\documentclass[PGL,biber,plain]{nowfnt} 

\usepackage[utf8]{inputenc}

\DeclareUnicodeCharacter{0308}{}
\DeclareUnicodeCharacter{0301}{}

\usepackage{graphicx}
\usepackage{xspace}
\usepackage{listings}
\usepackage{comment}
\usepackage{amsmath}
\usepackage{lscape} 
\usepackage{subcaption}
\usepackage{enumitem}
\makeatletter
\newcommand{\enlabel}[2]{#2\def\@currentlabel{#2}\label{#1}}
\makeatother

\newcommand{\tablelbl}[1]{\textbf{#1}}

\newcommand{\macrop}[1]{macroprogramming\xspace{}}
\newcommand{\Macrop}[1]{Macroprogramming\xspace{}}
\newcommand{\MacroP}[1]{MacroProgramming\xspace{}}

\newcommand{\meta}[1]{\textcolor{blue}{#1}}

\newcommand{\revision}[1]{{\color{darkgreen}#1}}
\renewcommand{\revision}[1]{#1}

\usepackage{tabularx}
\usepackage{longtable}
\usepackage{makecell}
\usepackage{tikz}
\usetikzlibrary{arrows,shapes,positioning,shadows,trees}
\usepackage{cleveref}

\definecolor{darkgreen}{rgb}{0,0.5,0}

\lstdefinelanguage{scafi}{
  basicstyle=\footnotesize\ttfamily\lst@ifdisplaystyle\scriptsize\fi,
	frame=single,
	escapechar=\%,
  keywords={abstract,case,catch,class,def,%
    do,else,extends,final,finally,%
    for,if,implicit,import,match,mixin,%
    new,null,object,override,package,%
    private,protected,requires,return,sealed,%
    super,this,throw,trait,try,lazy,%
    type,val,var,while,with,yield,forSome},
  otherkeywords={=>,<-,<\%,<:,>:,\#},
  keywordstyle=\color{red}\textbf,
  keywordstyle=[2]\color{blue},
  keywords=[2]{exchange,branch,@@},
  keywordstyle=[3]\color{violet},
  keywords=[3]{fold,mux,collect,pair,broadcast,foldSum,distanceTo,channel,
  timestamp},
  keywordstyle=[4]\color{Emerald},
  keywordstyle=[5]\color{Brown},
  sensitive=true,
  morecomment=[l]{//},
  morecomment=[n]{/*}{*/},
  commentstyle=\color{darkgreen},
  morestring=[b]",
  morestring=[b]',
  morestring=[b]"""
} 

\sloppypar

\title{Macroprogramming: Concepts, State of the Art, and Opportunities of Macroscopic Behaviour Modelling}

\maintitleauthorlist{
Roberto Casadei\\
Alma Mater Studiorum---Universit{\`a} di Bologna \\
roby.casadei@unibo.it
}

\issuesetup
{%
 copyrightowner={R.~Casadei},
 volume        = xx,
 issue         = xx,
 pubyear       = 2022,
 isbn          = xxx-x-xxxxx-xxx-x,
 eisbn         = xxx-x-xxxxx-xxx-x,
 doi           = xx.xxxx/XXXXXXXXX,
 firstpage     = 1, 
 lastpage      = 78
 }

\usepackage{mwe}

\author[1]{Casadei,Roberto}

\affil[1]{Alma Mater Studiorum---Universit{\`a} di Bologna; roby.casadei@unibo.it}

\articledatabox{\nowfntstandardcitation}

\renewcommand{\cite}[1]{\citep{#1}}

\begin{document}

\makeabstracttitle

\begin{abstract}
\revision{
\emph{\Macrop} 
 refers to the theory and 
 practice of
 conveniently 
 expressing the macro(scopic) behaviour of a system
 using a single program.
}
\Macrop{} approaches are motivated by the need of effectively capturing \emph{global/system-level} aspects
 and the \emph{collective behaviour} of a set of interacting components,
 \revision{while abstracting over low-level details}.
In the past,
 this style of programming has been primarily adopted 
 to describe the data-processing logic
 in \emph{wireless sensor networks};
 recently,
 research forums on \emph{spatial computing},
 \emph{collective adaptive systems},
   and \emph{Internet-of-Things}
  have provided renewed interest in macro-approaches.
However,
 related contributions
 are still fragmented    and lacking conceptual consistency.  Therefore, to foster principled research,
  an integrated view
 of the field is provided,
 together with opportunities and challenges.
\end{abstract}

\renewcommand{\thesection}{\arabic{section}}

\section{Introduction}
\label{s:intro}

\revision{
\emph{\Macrop} 
 refers to the theory and 
 practice of
 conveniently 
 expressing the macro(scopic) behaviour of a system
 using a single program,
 often leveraging macro-level abstractions
 (e.g., collective state, group, or spatiotemporal abstractions).
}
This is not to be confused with the use of macros (abbreviation for \emph{macroinstructions}), mechanisms for compile-time 
 substitution of program pieces,
 available in programming languages ranging from C and Common Lisp to Scala and Rust.
\Macrop{} is a paradigm 
 driven by the need of designers and application developers
 to capture \emph{system-level behaviour} while
 abstracting, in part, the behaviour \revision{and interaction} of the individual components involved.
It can be framed as a \emph{paradigm} since it embodies a (\emph{systemic}) view or perspective of programming,
 and accordingly provide \emph{lenses} to the programmer
 for understanding and working on particular aspects of systems---especially those related to collective behaviour, interaction,  and global, distributed properties.

In the past,
 this style of programming has been primarily adopted 
 to describe the behaviour of \emph{wireless sensor networks (WSN)}~\cite{mottola2011programming-wsn},
 where data gathered from sensors are to be 
 processed, aggregated, and possibly moved across different parts or regions
 of the network in order to be consolidated into useful, actionable information.
More recently,
 certain research trends and niches 
 have provided renewed interest in macro approaches.
Research in the contexts of
 \emph{Internet of Things (IoT)} and 
 \emph{cyber-physical systems (CPS)}
 has proposed \macrop{} approaches
 (cf. \cite{mizzi2018dartagnan,Azzara2014pyot-macroprogramming-iot})
 to simplify the development of systems
 involving a multitude of connected sensors, actuators, 
 and smart devices. 
In the \emph{spatial computing} thread~\cite{beal2012organizing-the-aggregate},
 space can represent both a means and a goal
 for \macrop{}.
Indeed, declaring \emph{what} has to be done \emph{in} a spatiotemporal region allows systems to self-organise to effectively carry out the task at hand, dynamically adapting to the specifics of the current deployment and spatial positions of the components involved.
Similarly, one can program a system, such as a drone fleet, in a high-level fashion to make it seek and maintain certain shapes and connectivity topologies.
Indeed, swarm-level programming models have been proposed in robotics research~\cite{pinciroli2016buzz-swarm-programming}.
In \emph{distributed artificial intelligence (DAI)} and \emph{multi-agent systems (MAS)} research~\cite{DBLP:journals/aim/Adams01},
 an important distinction is made between the \emph{micro} level of individual agents
 and the \emph{macro} level of an ``agent society'',
 sometimes explicitly addressed by \emph{organisation-oriented} programming approaches~\cite{DBLP:journals/scp/BoissierBHRS13}.
In the field of \emph{collective adaptive systems (CAS)} engineering~\cite{Ferscha2015cas,DBLP:journals/sttt/NicolaJW20}, \macrop{} abstractions can promote collective behaviour exhibiting self-* properties (e.g., self-organising, self-healing, self-configuring)~\cite{DBLP:journals/computer/KephartC03,DBLP:conf/dagstuhl/LemosGMSALSTVVWBBBBCDDEGGGGIKKLMMMMMNPPSSSSTWW10}.
In software-defined networking (SDN),
 the logically centralised view of the control plane
 has promoted a way of programming the network
 as \emph{``one big switch''}~\cite{DBLP:conf/conext/KangLRW13}.

This work draws motivation from a profusion of \macrop{} approaches and languages 
 that have been proposed in the last two decades,
 aiming to capture the aggregate behaviour of certain classes of distributed systems.
However, contributions are sparse, isolated in research niches, and tend to be domain-specific as well as technological in nature.
This survey aims to consolidate the state of the art,
 provide a map of the field,
 and foster research  on \macrop{}.

This article is organised as follows.
\Cref{sec:method} covers the method adopted for carrying out the survey.
\Cref{sec:domains} provides an overview of the research fields where \macrop{} techniques have been proposed. This overview helps to trace a historical development and the motivations for the approach.
\Cref{sec:framework} defines a conceptual framework 
 and taxonomy for \macrop{}.
\Cref{sec:survey} is the core of the survey: it classifies and presents 
the selected primary studies.
\Cref{s:opp-ch} provides an analysis of the surveyed approaches
 and discusses opportunities and challenges of \macrop{}.
\Cref{s:rw} covers related work, discussing the contributions of other secondary studies.
Finally, \Cref{s:wrap-up} provides a wrap-up.

\section{Survey Method}
\label{sec:method}

This section briefly describes
 how the survey has been carried out.
It focusses on motivation, research questions, data sources, presentation of results, and terminology.

\subsection{Survey Method}
\label{survey-method}

Though this is not a systematic literature review (SLR),
 parts of its development process have been inspired by
 guides for conducting SLRs in software engineering, such as~\cite{keele2007guidelines}.
More details follow.

\subsubsection{Review motivation}\label{review-motivation}
As anticipated in \Cref{s:intro},
 the survey draws motivation by 
 the emergence of a number of works 
 that more or less explicitly identify themselves
 as \macrop{} approaches.
Related secondary studies have been carried out in the past: 
 they are reviewed in \Cref{s:rw}.
However, they focus on particular perspectives or domains (e.g., spatial computing, or WSN programming),
 are a bit outdated,
 and consider \macrop{} as a particular class of approaches
 in their custom scope.
Critically, \emph{\macrop{} has never been investigated as a field per se}, yet.
Another major motivation lies in the \emph{fragmentation} of \macrop{}-related works across disparate research fields and domains.
Therefore, a goal of this very survey is to provide a \emph{map} of \macrop{}-related literature,
 promoting interaction between research communities
 and development of the field.
More motivation is given by the urge of the following research questions.

\subsubsection{Research goals and questions}
The goal of this article is to explore the literature on \macrop{} \emph{in breadth},
 synthesise the major contributions,
 and provide a basis for further research.
The focus is on the \emph{programming} perspective,
 rather than e.g. modelling formalisms for analysis and prediction;
 namely, the contribution can be framed in \emph{language-based software engineering}~\cite{DBLP:journals/scp/Gupta15}.
To better structure the investigation, we focus on the following research questions, inspired by the ``six honest serving men''~\cite{kipling1902six-honest-serving-men} as e.g., in~\cite{flood1994keep}.
\begin{enumerate}[label=\textbf{S.\arabic*}]
\item[\enlabel{rq-why}{RQ0})] \emph{Why, where, and for who is \macrop{} most needed?}
\item[\enlabel{rq-what}{RQ1})] \emph{What is \macrop{} and, especially, what is not? }
\item[\enlabel{rq-how}{RQ2})] \emph{How is \macrop{} implemented? Namely, what are the main \macrop{} approaches and abstractions?}
\item[\enlabel{rq-opp}{RQ3})] \emph{What opportunities can arise from research on \macrop{}?}
\item[\enlabel{rq-challenges}{RQ4})] \emph{What are the key challenges in \macrop{} systems?}
\end{enumerate}
\ref{rq-why} is addressed in \Cref{sec:domains}.
\ref{rq-what} is addressed in \Cref{sec:framework}.
\ref{rq-how} is addressed in \Cref{sec:survey}.
Finally, \ref{rq-opp} and \ref{rq-challenges} are addressed in \Cref{s:opp-ch}.

\subsubsection{Identification, selection, and quality assessment of primary research}
Primary research studies have been identified by searching literature databases (such as Google Scholar, DBLP, IEEEXplore, ACM DL) for keywords such as ``macroprogramming'', ``global-level programming'', ``network-wide programming'', and ``swarm programming'',
Terminology is fully covered and discussed in \Cref{sec:terminology}.
Additional sources include other secondary and primary studies, which are surveyed in \Cref{s:rw} and \Cref{sec:survey}, respectively.

The survey scope is wide and includes PhD theses, technical reports, and papers presented at workshops, conferences, and journals as well as across different domains and research communities.
Works that are deemed too preliminary (e.g., position papers), not enough ``macro'' (refer to \Cref{sec:framework}), or neglecting the ``programming'' aspects (e.g., describing a middleware but no programming language) have been excluded, after being manually inspected.

\subsubsection{Data extraction, synthesis, and dissemination}
For each primary study, 
 notes are taken regarding
 its \emph{self-positioning} (i.e., how the authors define their contribution),
 its \emph{programming model} (i.e., what main abstractions are provided),
 its \emph{implementation} (i.e., how macro-abstractions are mapped to micro-level operations),
 and \emph{source-code examples}.
The data is synthesises using the conceptual framework introduced in \Cref{sec:framework}.
When covering and summarising the primary works in the survey  (~\Cref{sec:survey}),
 we tend to keep and report the terminology originally used in the referenced papers,
 possibly explained and compared with the terminology used in this manuscript.
This should help to preserve the richness and nuances of each work while the common perspective is ensured by proper selection and emphasis of the information included in the descriptions.
Examples -- adapted from those already included in the primary studies or created anew from composing code snippets described in those papers -- are provided when they are reasonably ``effective'' or ``diverse'' from those already presented: i.e., they are brief and simple in transmitting how the reviewed approach looks like and works.

\subsection{A Note on Terminology}
\label{sec:terminology}

A first issue in \macrop{} research is 
 the fragmentation and ambiguity of terminology,
 which -- together with domain fragmentation (see \Cref{sec:domains}) -- leads to 
 (i) difficulty when searching for related work,
 and 
 (ii) obstacles in the formation of a common understanding.
Across literature, multiple terms such as macroprogramming, system-level programming, and global-level programming are used to refer to the same or similar concepts:
 this does not promote a unified view of the field
 and hinders progress by preventing the spread of related ideas.
At the same time,
 there is a problem of usage
 of both over- and under-specific terms.
Overly general terms 
 both witness the lack and prevent the formation 
 of a common ground.
On the other hand,
 overly specific terms,
 mainly due to domain specificity of research endeavours,
 fail at recognising the general contributions
 or at advertising the effort in the context of a bigger picture.
 
In the following, we list some terms that have been used 
 (or might be used)
 -- with more or less good reason -- 
 when referring to \macrop{},
 and analyse their semantic precision (by reasoning on their etymology and other common uses)
 as well as alternative meanings
 in literature
 (for conflicts with more or less widespread acceptations).

\paragraph{Macroprogramming, macro-programming, macro programming, macro-level programming}
These are the premier terms
 for the subject of this article 
 and may indeed refer to \revision{\emph{programming macroscopic aspects of systems} (often, by leveraging macro-level abstractions)}.
However, these terms are sometimes also used in other computer programming-related contexts.
The potentially ambiguity stems from 
 word ``macro'',
 which is and can be used to abbreviate both term
 ``macroscopic'' 
 and term ``macroinstructions''---often used in the sense of \emph{macros},
 i.e., the well-known programming language mechanism 
 for compile-time substitutions of program pieces.
Indeed, it is common to say that macros are written
 using a macro (programming) language.
The result is that searching for these terms 
 leads to a mix of results from both worlds.
Unfortunately, being macros a very common mechanism~\cite{lilis2020csur-metaprogramming-survey}, macroscopic programming-related entries remain relatively little visible in search results, unless other keywords are used to narrow the context scope---but then, only a fragment of the corpus can be located.
 
\paragraph{System programming, system-level programming, system-oriented programming}
All these terms are also ambiguous.
Indeed, they strongly and traditionally refer
 to \emph{low-level programming}, i.e.,
 programming performed at a level 
 close to the (computer) system (i.e., to the machine)~\cite{appelbe1985survey-system-programming}.
System programming languages include, e.g., C, C++, Rust, and Go.
A better name for these would probably be, 
 as suggested by Dijkstra, \emph{machine-oriented} languages,
 but such a ``system'' acceptation is a sediment of the field by now.
The scarce accuracy of the term was also somewhat acknowledged
 by researchers in the object-oriented programming community~\cite{Nygaard1997ecoop-invited-talk}.
However, in some cases, system-level programming
 is contrasted with device-level programming, to mean approaches that address ``a system as a whole''~\cite{DBLP:journals/csur/LiangCLL16}.

\paragraph{Centralised programming}
This term~\cite{DBLP:journals/ccr/GudeKPPCMS08,DBLP:conf/otm/LimaCBF06}
 commonly refers
 to programming a distributed system
 through a single program
 where distribution is (partially~\cite{waldo1996note-on-distributed-computing}) abstracted away, i.e.,
 like if the distributed system were a centralised system, 
 namely a software system on a single computer deployment.
\revision{An example of centralised programming is \emph{multi-tier programming}~\cite{DBLP:journals/csur/WeisenburgerWS20}.}
This notion is certainly related to \macrop{},
 since a ``centralised perspective''
 where several distributed components can be \revision{addressed} at once
 is a macroscopic perspective.
However, as discussed in \Cref{sec:framework},
 programming the macro level 
 often implies \emph{more} than programming the individual components
 from a centralised perspective.

\paragraph{High-level programming}
This term, identifying a style of programming that abstracts many details of the underlying platform, lacks of precision.
\Macrop{} is a form of high-level programming,
but not all the high-level programming is \macrop{}
(for a conceptual framework for \macrop{}, refer to \Cref{sec:framework}).

\paragraph{Global programming, global-level programming, global computation}
\revision{These terms may be related to macroprogramming because, in general, a global view is also a macroscopic view (though the reverse is not always true).}
Global computation~\cite{cardelli1997global-comp,bent2001global-comps} refers to 
 the computation performed by ``global computers''
 made of ``global communication and computation infrastructure''---essentially, distributed systems on the Internet.
\revision{A ``global computer'' may be a target of \macrop{}. Moreover,} in computer science,
 terms ``global programming'' and ``global-level programming'' \revision{also} sometimes \cite{carbone2007structured-global-programming}  refer to \emph{choreographic programming}~\cite{felipe2020choreographic-programming},
 i.e., a form of programming addressing the interaction between services 
 in service compositions by a global perspective, 
 through so-called \emph{choreographies}.
Outside computer science, these more commonly refer to planning in organisational management. 

\paragraph{Domain-specific or alternative terminology: network-wide programming, organisational programming, swarm programming, aggregate programming, ensemble programming, global-to-local programming, team-level programming, organisation-oriented programming etc.}
These terms will be explained and properly organised in the following sections.
From this list of terms, however,
 it is already possible to get a sense of
 (i) an intimate need, from different research communities,
  to linguistically emphasise a focus on macroscopic aspects of systems,
  and
 (ii) the urge for a common conceptual framework
  where such disparate contributions can be framed.

\section{Historical Development and Scope of Macroprogramming}
\label{sec:domains}

In this section, we provide an overview of 
 the main research fields and application domains 
 where \macrop{} techniques have been proposed,
 also tracking elements of historical development of the paradigm.

\subsection{Wireless Sensor and Actuator Networks (WSAN)}
WSANs are networks of embedded units capable of processing, communication, and sensing and/or acting~\cite{mottola2011programming-wsn}.
They are a technology providing relatively low-cost 
 monitoring and control of physical environments.
Given the large number of involved devices,
 and the reasonable levels of heterogeneity and dynamicity
 for a given application,
 it became apparent 
 that a benefit could be provided 
 by high-level programming models
 abstracting from a series of low-level network details
 while still seeking to preserve efficiency.
When a system consists of a large number of rather homogeneous entities, 
 individuals tend to become less important
 to the functionality
 (while may well contribute to non-functional aspects):
 a WSN with 50 devices might perform worse than a 100-devices network,
 but these two networks can be programmed the same.
Additionally, 
 developers and researchers started realising that the individual sensors 
 are actually a \emph{proxy} or a \emph{probe} for 
 more important application abstractions
 such as information, streams, and events.
At a next step,
 those abstractions started to become more high-level,
 and to address larger portions of the system beyond individual sensors,
 such as neighbourhoods~\cite{hood}, or regions~\cite{welsh2004abstract-regions};
 accordingly, abstractions related to those more coarse-grained entities emerged, denoting 
 contexts, aggregate views, fields---increasingly non-local abstractions.
Among the high-level approaches,
 languages providing a \emph{centralised view} of the WSN emerged;
 then, the step to \macrop{} was short.
This is, indeed, one of the first domains 
 where \macrop{} was introduced.

Early works like TinyDB~\cite{tinydb}, Pieces~\cite{liu2003state-centric-program-wsan-pieces}, Abstract Regions~\cite{welsh2004abstract-regions}, and Regiment~\cite{regiment}
 are among the first contributions 
 explicitly defining themselves as \macrop{}.
A survey on \macrop{} for WSNs can be found in~\cite{mottola2011programming-wsn}.

\subsection{Spatial Computing}

Space is generally important in ICT systems.
This has been especially
 motivated and investigated in the \emph{Computing Media and Languages for Space-Oriented Computation} seminar in Dagstuhl~\cite{dagstuhl2006seminar-space-oriented-computation},
  where three key issues  are found to be recurrent 
  in many computer-based applications:
  \emph{(i) coping with space}, for efficiency in computation;
  \emph{(ii) embedding in space}, as in embedded and pervasive computing;
  and \emph{(iii) representing space}, for spatial awareness.
What became apparent, also from WSN programming research,
 is that devices situated in space
 can become \emph{representatives} of the spatial region
 they occupy and of the corresponding context.
In this view, distributed systems and networks
 can be seen as \emph{discrete approximations
 of continuous space-time regions and behaviours}~\cite{DBLP:journals/nca/BachrachBM10}.
Therefore,
 \macrop{} abstractions
 may abstract individual devices
 and rather focus 
 on spatial patterns 
 that such devices
 should cooperatively (re-)create---e.g., for morphogenesis~\cite{DBLP:journals/tsmc/JinM11}.
In general, dealing with \emph{situated systems}~\cite{DBLP:journals/adb/LindblomZ03} -- namely, systems where components have a location in and coupling with (logical or physical) space, with typical corresponding consequences such as partial observability and local (inter)action --
 is simplified when recurring to spatial abstractions
 such as, e.g., computational fields~\cite{DBLP:journals/pervasive/MameiZL04}.

Spatial computing approaches are extensively surveyed in~\cite{beal2012organizing-the-aggregate} (see~\Cref{s:rw} for details on the study) and include examplars of \macrop{}
such as Regiment~\cite{regiment} and MacroLab~\cite{hnat2008macrolab}.

\subsection{Internet of Things, Cyber-Physical Systems, Edge-Fog-Cloud Computing Systems}
The \emph{Internet of Things (IoT)}~\cite{DBLP:journals/cn/AtzoriIM10}
 refers to a paradigm and set of technologies
 supporting interconnection of smart devices
 and the bridging of computational systems with physical systems---the latter element being emphasised also through term \emph{Cyber-Physical Systems (CPS)}~\cite{DBLP:journals/computer/Serpanos18}.
IoT systems share many commonalities with WSANs,
 so it is not surprising that contributions from the latter field
 have been extended to address IoT application development.
Actually, the IoT can be considered as a superset of WSANs,
 with additional complexity 
 due to the exacerbation of issues like heterogeneity, mobility, topology, dynamicity, infrastructural complexity, as well as functional and non-functional requirements.
However, an IoT system 
can still be considered as a collective of interconnected smart devices,
 amenable to being considered by a macroscopic perspective.

Moreover, IoT systems 
 tend to be more heterogeneous 
 and infrastructurally rich,
 comprising edge, fog, and cloud computing layers~\cite{DBLP:journals/jsa/YousefpourFNKJN19}
 to support various requirements including low-latency and low-bandwidth consumption.
Interestingly, also the edge, the fog, and the cloud
 can be considered as computational (eco-)systems
 programmable at the macro-level~\cite{DBLP:journals/fgcs/PianiniCVN21}.
This idea also underlies orchestration approaches
 based on Infrastructure-as-Code~\cite{infrastructure-as-code},
 which can be considered a form of centralised, declarative programming.

Examples of IoT/CPS \macrop{} approaches
 include PyoT~\cite{Azzara2014pyot-macroprogramming-iot},
 DDFlow~\cite{Noor2019ddflow-visual-macroprogram-iot},
 and MacroLab~\cite{hnat2008macrolab},
 whereas
 preliminary approaches also considering edge/fog/cloud 
 comprise ThingNet~\cite{qiao2018thingnet-macro-iot-edge}.

\subsection{Swarm robotics}

A set of interacting robots can work as a collective,
 also known as a \emph{swarm}.
In this case, the focus of external observers tends
 to shift from the activity of individual robots
 to the activity of the swarm as a whole.
Various tasks make sense at such a macro-perspective.
For instance, we could ask a swarm to:
 move in flock formation 
 towards a destination;
 split and later merge 
 for avoiding a large obstacle;
 use, in a coordinated way, the sensing capabilities 
 to estimate physical quantities
 (e.g., the mean temperature in a certain area)
 or other indicators (e.g., the risk of fire in a forest); or
 use, in a coordinated way, sensing and actuation capabilities
 to efficiently perform actions and tasks
 (e.g., quickly collecting toxic waste in industrial plants)
 possibly going beyond individual capabilities
 (e.g., moving heavy objects).
Another prominent sub-field in robotics
 with emphasis on macroscopic features 
 is modular, morphogenetic robotics~\cite{DBLP:journals/tsmc/JinM11,DBLP:journals/trob/ZykovMDL07},
  which considers collections of building-block modules 
  that should dynamically self-reconfigure into functional shapes
  in order to address tasks, change, or damage.
Indeed, the overall morphology of a modular swarm 
 is a macro-level structure
 that must be  dynamically sought
 through activity and cooperation
 of the individual robots.
The traditional question is: how can the individual robots be programmed such that the desired overall shape is produced?
By a \macrop{} perspective, this question turns into:
 how can a swarm \emph{as a whole} be programmed such that the overall shape is produced?
Of course, this ultimately entails a definition of the behaviour of the individuals as well;
 however, the idea is to encapsulate the complexity of such a collective behaviour at the middleware level, behind proper macroscopic abstractions.

Examples of \macrop{} languages for swarm robotics include 
Meld~\cite{Meld} (for modular robotics),
Voltron~\cite{Mottola2014voltron} (for drone teams),
Buzz~\cite{pinciroli2016buzz-swarm-programming},
TeCoLa~\cite{Koutsoubelias2016tecola},
and WOSP~\cite{varughese2020swarm-wosp} (for elementary robots).

\subsection{Complex and Collective Adaptive Systems}
Complex and collective adaptive systems (CAS)
 are collectives (i.e., collections of individuals)
 exhibiting a non-chaotic behaviour
 that is adaptive to the environment
 and cannot be (easily) reduced to the behaviour
 of the individuals,
 but that rather \emph{emerges} from complex networks
 of situated interactions.
These kinds of systems were originally observed in nature,
 but researchers 
 have tried to bring those principles and ideas
 for development artificial, ICT-based CASs~\cite{Ferscha2015cas,DBLP:journals/sttt/NicolaJW20}.
The field of CAS engineering
 emerges 
 from  
 swarm computational intelligence~\cite{DBLP:books/daglib/p/Kennedy06} 
 and
 autonomic, self-adaptive computing~\cite{DBLP:journals/computer/KephartC03,DBLP:conf/dagstuhl/LemosGMSALSTVVWBBBBCDDEGGGGIKKLMMMMMNPPSSSSTWW10}.
The goal of CAS programming
 is to program the collective adaptive behaviour
 of a system.
In general, two approaches are possible:
 \emph{local-to-global}, 
  where local behaviour is specified
  in order to promote emergence of a target global behaviour;
  or
 \emph{global-to-local},
 where the idea is to specify the intended global behaviour
 and come up with a mechanism to synthesise the 
 corresponding local behaviour.

Since the notion of a \emph{collective} (also known as \emph{ensemble}) is per se 
 a macro-level abstraction,
 it is natural to adopt \macrop{} techniques.
Examples are provided in \Cref{sec:survey}
 and include 
 ensemble-based approaches such as DEECo~\cite{bures2013deeco}  and SCEL~\cite{DBLP:journals/taas/NicolaLPT14},
 and aggregate programming~\cite{DBLP:journals/computer/BealPV15}.

\subsection{Other domains}

In the following domains,
 \macrop{} has not actually been proposed explicitly,
 but similar needs can be perceived
 and very related ideas have indeed been considered.

\subsubsection{Software-defined networking}
Software-defined networking (SDN)~\cite{DBLP:journals/pieee/KreutzRVRAU15}
 is an approach for the management of computer networks
 based on the idea of separating the data plane (forwarding) and the control plane (routing).
Thanks to this separation, network devices become just entities responsible for forwarding, whereas control logic can be logically centralised in a single component.
This logical centralisation directly leads to centralised programming (cf. \Cref{sec:terminology})
 and hence to a macroprogramming viewpoint.
This is also visible in the editorial note~\cite{Beckett2019},
 which provides a brief historical reflection on
 the development of such a vision, also known as \emph{network-centric} or \emph{network-wide programming}~\cite{martins2017network-wide-programming}.

Examples of network-wide programming include
 NetKAT~\cite{Anderson2014netkat} and SNAP~\cite{Arashloo2016snap}. 
 
\subsubsection{Parallel Programming and High-Performance Computing (HPC)}
Literature on parallel programming
 includes some germs of \macrop{} ideas
 as well,
 even though the focus on performance and low-level system programming
 arguably has been hindering adoption of high-level abstractions.
However, these can be found in parallel, \emph{global-view} languages,
 such as those implementing the Partitioned Global Address Space (PGAS) model~\cite{DBLP:journals/csur/WaelMFCM15},
 where, e.g., directives 
 have been proposed to represent
 \emph{``high-level expressions of data distributions, parallel data movement, processor arrangements and processor groups''}.
Indeed, addressing the behaviour of multiple processors 
 in terms of macroscopic patterns
 rather than in terms of micro-instructions
 could simplify programmability
 and still reach good performance 
 through smart global-to-local mapping.

Other elements of similarities 
 can be traced
 between Valiant's \emph{Bulk Synchronous Parallel (BSP)} model~\cite{DBLP:journals/cacm/Valiant90}
 and the execution model of
 \macrop{} approaches such as aggregate computing~\cite{DBLP:journals/computer/BealPV15},
 where multiple parallel processors work 
 in \emph{supersteps} involving communication and computation
 as specified by a single global program.
Moreover, 
 this tendency towards programming by a macroscopic perspective
 has been witnessed by some BSP-based models.
For instance, in the domain of graph-processing,
 as discussed in the paper \emph{From ``Think Like a Vertex'' to ``Think Like a Graph''}~\cite{giraphpp},
 the Giraph++ framework has been proposed
 by replacing the vertex-centric model of Giraph
 with a \emph{graph-centric model}
 to provide efficiency benefits 
 by directly exposing graph partitions and optimising communications.

\revision{
\subsection{Final remarks}\label{scope-macroprogramming-remarks}
It is evident from the domains covered in this section
 that the target of \macrop{}
 is often a \emph{collective},
 namely a (largely homogeneous, usually~\cite{brodaric2020pluralities-collectives-composites}) collection
 of physical or computational entities---e.g., 
 a collection of sensors,
 a collection of drones,
 a collection of routers in a network,
 a collection of partially-autonomous agents.
In these contexts, the goal of \macrop{} 
 is usually that 
 of defining 
 how such collectives should behave as a whole,
 while abstracting from certain low-level details,
 for instance: 
 how sensor data is to be processed and diffused
 by a WSN, without specifying routing details~\cite{gummadi2005macroprogramming-wsn-kairos};
 how a fleet of drones should organise into teams
 to explore an environment, without specifying how the different spatial locations are assigned to individual robots and the specific fleet-aware movements of individual robots~\cite{Mottola2014voltron};
  how a set of network nodes should monitor or control network activity,
  without specifying micro-level node decisions
  such as
  where to place, how to distribute, and optimise access to network state variables~\cite{Arashloo2016snap}; and 
 how to specify the self-organisation logic of a collection of agents, without  specifying how activity is scheduled,
 how individuals interact in terms of message-passing, and how agents fully behave~\cite{Viroli-et-al:JLAMP-2019}.

The aforementioned examples
 show both commonalities and specific traits.
First of all, 
 we always have a (possibly dynamic)
 \emph{collection} of entities.
Concerning autonomy, 
 the target entities may range from
fully passive (like routers, which act only upon reception of packets)
 to fully autonomous (like agents, which encapsulate control).
Concerning interaction,
 the entities of the system
 may interact using diverse mechanisms
 like message-passing
 or
 shared state
 (including stigmergy~\cite{pinciroli2016virtual-stigmergy-swarm}).
Concerning system structure,
 the system may have a
 static, regular network topology (as in a WSN)
 or 
 a dynamic, ad-hoc network topology (as in a system of mobile agents).
 
Regarding the application tasks
 commonly addressed by \macrop{},
 we observe that 
 they are typically tasks amenable to macroscopic evaluation
 (e.g., moving a fleet of drones to cover a spatial area),
 often considering a multiplicity of inputs and outputs
 (e.g., the starting and final configurations of the fleet),
 or
 tasks that require a collaboration of several entities to be carried out
 (e.g., computing the routing path for a given packet request sent to an individual network node).
For instance, in network applications,
 it is not the routing decisions
 made by an individual network node that matter,
 but rather how such individual routing decision
 relates with those of other nodes,
 that make for a ``good'' overall network activity.
While traditional approaches
 address the issue by 
 providing each entity of the system
 with different programs,
 or with a same program assuming the individual perspective (in case of homogeneous entities),
 \macrop{} adopts a change of perspective
 whereby a system behaviour or its outputs can be specified as a whole, by a conceptually centralised point of view.
Indeed, regarding the goals of \macrop{},
 we can find three recurrent general objectives:
 (i) \emph{abstraction}, namely expressing a certain global behaviour in a convenient way, while keeping low overhead;
 (ii) \emph{optimisation}, namely moving the problem of realising efficient micro-level activity from the programmer to the runtime system; and
 (iii) \emph{adaptivity}, namely promoting collective adaptation to change in the environment.
}

\section{A Conceptual Framework and Taxonomy}
\label{sec:framework}

In this section,
 after some preliminaries (\Cref{def-preliminaries}),
 we define \macrop{},
 describe its essential elements and concepts (\Cref{sec:macrop:def}),
 characterise it in terms of abstraction, 
 and distinguish it from other related notions like
 \emph{centralised programming} (\Cref{s:macro-not}).
Then, we propose a taxonomy and conceptual framework
 (\Cref{sec:taxonomy})
 for classifying and studying the \macrop{} approaches surveyed in \Cref{sec:survey}.

\begin{table}
\newcommand{\enfq}[1]{#1} 
\scriptsize
  \renewcommand{\arraystretch}{1.5}%
\begin{tabularx}{\textwidth}{ |p{2cm}|X| }
\hline
\tablelbl{Ref.} & \tablelbl{Definition} 
\\\hline
\cite{DBLP:conf/iwdc/BakshiP05} &
\emph{``The objective of macroprogramming is to allow the programmer to write a distributed sensing application \enfq{without explicitly managing control, coordination,
and state maintenance at the individual node level}. Macroprogramming languages provide abstractions that can specify \enfq{aggregate behaviors} that are \enfq{automatically synthesized} into software for each \enfq{node} in the target deployment.
The structure of the \enfq{underlying runtime system} will depend on the particular programming model.''}
\\\hline
\cite{whitehouse2006semantic-streams} & \emph{``Macroprogramming is a term often used to refer to the process of writing a program that specifies \enfq{global network behavior} as opposed
to the \enfq{behavior of individual nodes}.''}
\\\hline 
\cite{DBLP:conf/hicss/WadaBS08} & \emph{``Macroprogramming is an emerging \enfq{programming paradigm} for wireless sensor networks (WSNs). It allows developers to implement each WSN application from a \enfq{global viewpoint} as a \enfq{whole} rather than a viewpoint of sensor nodes as \enfq{individuals}. A \enfq{macro-program} specifies an application's \enfq{global behavior}. It is \enfq{transformed} to \enfq{node-level (micro) programs}, and the micro-programs are deployed on individual
nodes. Macroprogramming aims to increase the simplicity
and productivity in WSN application programming.''}
\\\hline
\cite{DBLP:journals/taas/Mamei11} &
\emph{``Macro programming [...] is the ability to specify application tasks at a \enfq{global level} while relying on compiler-like software to translate the \enfq{global tasks} into the \enfq{individual} component activities.''}
\\\hline
\cite{awan2007cosmos} &
\emph{`` Macroprogramming specifies \enfq{aggregate system behavior}, as opposed to device-specific programs that code
\enfq{distributed behavior} using explicit messaging. [...]  Composing applications with reusable components allows the macroprogrammer to focus on application
specification rather than \enfq{low-level details or inter-node messaging}.''}
\\\hline
\cite{DBLP:journals/tosn/SugiharaG08} & \emph{``Network-level abstractions, or equivalently macroprogramming, share the approach
with \enfq{group-level abstractions}, but go further by \enfq{treating the whole network
as a single abstract machine}.''}
\\\hline
\cite{bai2009wasp-archetype-based} &
\emph{``\enfq{Network-level} programming languages, also called macro-programming languages, treat the whole network as a single machine. Lower-level details such as routing and communication are hidden from programmers.''}
\\\hline
\cite{sookoor2009macrodebugger} & \emph{``Macroprogramming provides the user with the illusion
of programming a \enfq{single machine} by \enfq{abstracting away the low-level details of message
passing and distributed computation}.''}
\\\hline
\cite{DBLP:conf/inss/HnatW10} & \emph{``Macroprogramming systems addresses the difficult problem of how to \enfq{program a system of devices} to perform a
\enfq{global task} without forcing the programmer to develop device-specific implementations''}
\\\hline
\cite{DBLP:series/eatcs/PathakP11} &
\emph{``In macroprogramming, abstractions are provided to specify the \enfq{high-level collaborative behavior} at the \enfq{system level}, while \enfq{intentionally hiding most of the low-level details concerning state maintenance or message passing} from the programmer''}
\\\hline
\cite{martins2017network-wide-programming} & \emph{``[...] the behavior of a CPS is better understood at a global system level. In order to reflect this from a programming language abstraction standpoint we rely on \enfq{network-wide programming}, or macroprogramming [...] This paradigm of programming allows the system developer to write \enfq{one piece of code for the network, specifying the application at a global semantic level}. The task of the compiler is then to not only produce a machine translation of the code, but also decide how to \enfq{split this code into several images to run on many devices}. These images should set up the necessary communication channels, buffering and orchestration between processing devices. In this way, the role of the compiler is not to produce an executable but to produce a set of deployable software images, along with their deployment requirements.''}
\\\hline
\end{tabularx}
\caption{Some descriptions of \macrop{} from the literature.}
\label{table:defs}
\end{table}

\subsection{Preliminaries}\label{def-preliminaries}

Consider the problem of programming the behaviour of a computational system $\mathcal{S}$
 composed of multiple computational entities \revision{(namely Turing-equivalent machines able to process information and possibly interact with other entities)~\cite{DBLP:journals/corr/HorsmanSWK13}}.
Let $A$ and $B$ be two different entities of that system. \revision{
We have the following main \emph{modes}
 for \emph{affecting}
 their behaviour
 in order to affect the behaviour or properties ascribable to the overall system $\mathcal{S}$ (which, as we will shortly see, is essentially the goal of \macrop{}).
 \begin{enumerate}
\item \emph{Change their context (e.g., inputs).} The entities will be indirectly influenced by the different context. For instance, if $A$ is a sensor, it might sense a different value, which may in turn affect $B$ and so on.

\item \emph{Interaction (e.g., trigger/orchestrate their behaviour).}
 For instance, if $A$ is an actuator, it might be commanded to act upon the environment, which may in turn affect $B$ and so on. 

\item \emph{Set their behaviour.} Part of the behaviour of $A$ and $B$ may be set or changed such that, when activated (e.g., in a reactive or proactive way), certain global outcomes will be produced.
\end{enumerate}
Let us use term \emph{program}
 to mean an (abstract) description that can be \emph{executed} by some (abstract) computational entity.
Notice that the modes (1) and (2)
 allow a program
 to affect $A$ or $B$, and hence $\mathcal{S}$,
 by having it executed by another entity, 
 say $C$,
 that is assumed to be external to the arbitrary boundary of $\mathcal{S}$.

}

\begin{figure}
\centering
\includegraphics[width=0.8\textwidth]{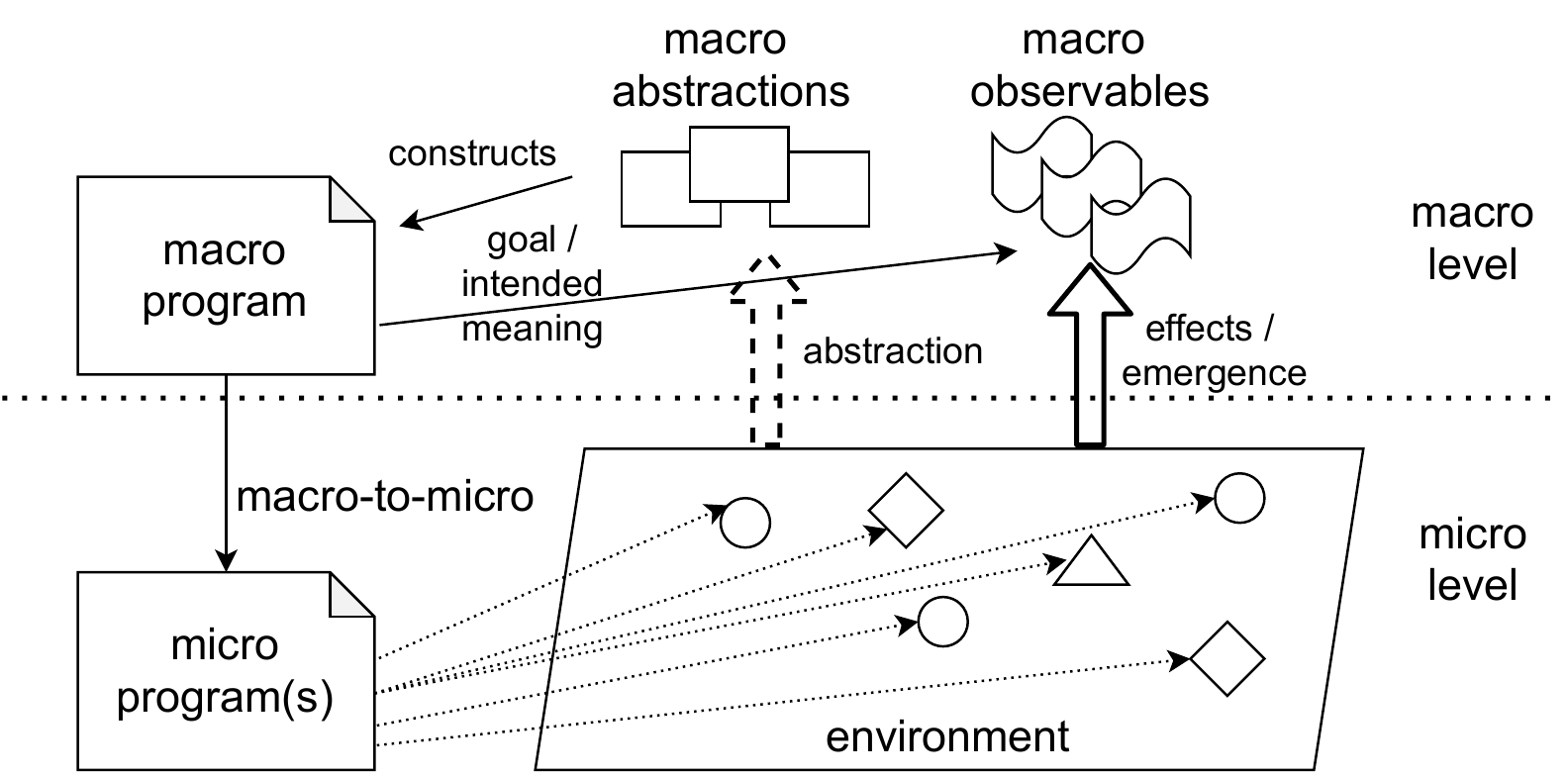}
\caption{\revision{The general idea of macroprogramming.}}
\label{img:macroprogramming}
\end{figure}

\subsection{\Macrop{}: Definition and Basic Concepts}
\label{sec:macrop:def}

We define macroprogramming 
 as \emph{an abstract paradigm for
 \revision{programming the (macro)scopic behaviour of systems
 of computational entities\footnote{\revision{Possibly corresponding to physical devices through a notion of \emph{digital twin}~\cite{DBLP:journals/access/RasheedSK20} or \emph{physical computation}~\cite{DBLP:journals/corr/HorsmanSWK13}.}}}}.
As a paradigm \revision{(see \Cref{macro-as-paradigm} for a discussion on this)},
 it is ``an approach to programming 
 based on a mathematical theory \textbf{or a coherent set of principles}''~\cite{vanroy2009programming-paradigms} \revision{(bold is added)}.
Macroprogramming is based on the following principles,
 which can be partially extracted from the various definitions given in literature (cf. \Cref{table:defs}):
\begin{itemize}
\item[P1] \emph{Micro-macro distinction.} Two main levels of a system are considered:
 a \emph{macro} level (of global  structures, \revision{of state,} of behaviour) and a \emph{micro} level \revision{(of computational entities).}
 \item[P2] \emph{Macroscopic perspective.} The programming activity tends to focus on macroscopic \revision{aspects of a \emph{system},
 which may include summary observations
 and views whereby} micro-level entities 
 are considered by a \emph{global} (or \emph{non-local})
 and conceptually centralised perspective. 
\item[P3] \emph{Macroprogram.} The output of the \macrop{} activity is a program 
  that is conceptually executed by the system as a whole
 \revision{and whose intended meaning adopts the macroscopic perspective}.
 \item[P4] \emph{Macro-to-micro mapping.} 
 A \macrop{} implementation
 has to define \emph{how}
 a macro-program is executed,
 by the system as a whole,
 which entails defining a \emph{macro-to-micro mapping} logic---sometimes also known as \emph{global-to-local} mapping~\cite{DBLP:series/csm/Hamann10}.
  In other words, from a macroprogram,
  micro-level programs or behaviours
  are derived or affected \revision{(see \Cref{def-preliminaries})}.
\end{itemize}
\revision{
\Cref{img:macroprogramming} shows the general idea of the approach, graphically. 
The following sections detail on the above principles.

}

\subsubsection{\revision{On micro-macro and local-global distinction}}
The micro-macro levels and the local-global scales usually used as equivalent concepts 
 to distinguish smaller elements/scopes 
 and larger elements/scopes
 somewhat ``containing'' or ``being implied by'' the former.
The micro-macro distinction~\cite{alexander1987micro-macro-link} (sometimes also space out by an intermediate, or \emph{meso} level) is typical in many scientific areas including social sciences, systemics, and distributed artificial intelligence~\cite{DBLP:conf/mabs/SchilloFK00} (cf. multi-agent systems~\cite{DBLP:books/daglib/0023784}).
For the sake of programming, 
 just like a system (as an ontological and epistemological element) can be \emph{defined} according to a boundary condition~\cite{mobus2014principles-of-systems-science},
 the distinction between two dimensions, micro and macro,
 is similarly made through a design-oriented boundary or membership decision defining what belongs to one level or the other.

\revision{
The intended meaning of macroprograms,
 and hence the ultimate goal of \macrop{},
 seems to be related to the notion of \emph{emergence}~\cite{holland1998emergence,DBLP:conf/atal/WolfH04,gignoux2017emergence,DBLP:journals/jocec/KalantariNM20}.
In \cite{gignoux2017emergence},
 the authors
 use graph theory to provide formal definitions of macroscopic states and microscopic states,
 and characterise emergence by analysing
 the general relationships 
 between microscopic and macroscopic states.

What can we say, in general, 
 about 
 the entities 
 at the micro and macro levels
 in \macrop{}?
Micro entities 
 have a computational behaviour,
 which may be autonomous (proactive), 
 active, or reactive;
 and may or may not interact 
 with other micro entities.
So, for instance, 
 data elements do not make for micro entities
 (they have no behaviour),
 while agents, actors, objects, and microservices do\footnote{\revision{Possibly, even humans and other physical entities~\cite{DBLP:journals/corr/HorsmanSWK13}.}}.

Regarding the macro level,
 we can distinguish
 between
 macro-level observables
 and 
 macro-level constructs.
A \emph{macro-level observable}
 is a high-level observation of the system behaviour,
 i.e., a macro state as defined in \cite{gignoux2017emergence},
  which is associated to the system as a whole
  and might be difficult to derive from micro state (the set of observations about the micro-level entities).
The intended meaning, or goal, of a macroprogram,
 is generally a function of macro-level observables 
 over some notion of time.
A \emph{macro-level construct} or \emph{abstraction}
 is, instead, a description that can be mapped
 down to affect the behaviour of 
 two or more micro-level entities (cf. \Cref{def-preliminaries}).
The problem of implementing the logic 
 for such a mapping
   is the macro-to-micro problem of \macrop{}.
}

\subsubsection{\revision{On collectives}}
\Macrop{} usually targets so-called \emph{collectives}\revision{---see \Cref{sec:domains}}.
Term ``collective'' derives from Latin 
\emph{colligere}, which means ``to gather together''. 
Typically~\cite{brodaric2020pluralities-collectives-composites}, a collective is an entity that gathers multiple \emph{congeneric} elements together \revision{by some notion of \emph{membership}}. 
``Congeneric'' means ``belonging to the same genus'',
namely, of related nature.
In other words, a collective is a group of
similar individuals or entities that share something (e.g., a trait, a goal, a plan, a reason for unity, an environment, an interface)
 which justifies seeing them as a collective, overall. 
A group of co-located workers, a swarm of drones, 
 the cells of an organ
are examples of collectives, whereas a gathering of radically different or unrelated entities such as cells, rivers,
and monkeys is not, intuitively.
Being congeneric, the elements of a collective 
 generally share goals
 and mechanisms for interaction and hence collaboration.
The differences among the elements, 
 often promoting larger collective capabilities by collaboration,
 may be due to genetic factors, individual historical developments, 
 and the current environmental contexts driving diverse responses on similar inputs.

\revision{
Heterogeneous collectives also exist
 (e.g., aggregates involving humans, autonomous robots, and sensors)
 and can be addressed by macroprogramming~\cite{scekic2017hybridcas}.
However, heterogeneity 
  tends to complicate \macrop{}
  by posing more importance on individuals' perspectives
  or widening the macro-to-micro gap---see \Cref{challenge-heterogeneity} for a discussion.
Finally, 
 we observe that a 
 precise characterisation of 
 groups and collections of entities
 is subject to research~\cite{brodaric2020pluralities-collectives-composites},
 in philosophical and mathematical fields like
 applied ontology (the study of being in general),
 and mereology (the study of parts and the wholes they form).
}

\subsubsection{\revision{On declarativity}}
A typical aspect of \macrop{} is \emph{declarativity}.
\emph{Declarative programming}~\cite{loyid1994declarative-programming}
 is a    \revision{paradigm 
 which focusses on expressing 
 \emph{what} the goal of computation is
 rather than \emph{how} it must be achieved}.
Common and concrete aspects of a computation that can be abstracted away
 include 
 the order of function evaluation (cf. functional programming),
 proving theorems from facts (cf. logic programming), and
 the specifics of data access (cf. query plans in databases and SQL).
The general idea is to provide high-level abstractions
 capturing system-wide concerns 
 by making assumptions promoting convenient mapping to 
 component-level concerns.
As such assumptions tend to be specific to an application domain,
 macroprogramming languages typically
 take the form of \emph{domain-specific languages (DSLs)}~\cite{beal2012organizing-the-aggregate}.

\subsection{What \Macrop{} Is (Not)}
\label{s:macro-not}

Programming
 essentially always deals with multiple interacting software elements,
 be them functions, objects, actors, or agents.
Even though paradigms 
 are more a matter of \emph{mindset} and \emph{abstractions},
 rather than a matter of strict demarcation,
 a \emph{demarcation issue} may be considered to better
 delineate a (nevertheless, fuzzy) boundary of macroprogramming.

\revision{
Macroprogramming is \revision{often} 
 centred around \emph{macro-abstractions}: informally, constructs that involve, in some abstract way,
 (the context, state, or activity of)
 two or more micro-level entities.
For instance:
\begin{itemize}
\item \emph{macro-statements} (or \emph{macro-instructions}), for imperative \macrop{} languages (e.g., ``move the entire swarm to that target location'',
 or ``update the WSN state history to record the current temperature of the area'');
\item \emph{macro-expressions}, evaluating to a macro-value
 (e.g., ``the direction vector of the swarm towards the target location'', ``the mean temperature of the area covered by the network'');
\end{itemize}
Other examples of macro-abstractions can be found in \Cref{s:technical-analysis}.
}

Consider the following artificial Scala program:
\begin{lstlisting}[language=scala]
// Library code (non-macroprogramming)
object swarm {
  def robots = // ...
  def move(target: Pos): Unit = robots.foreach(robot => robot.move(target))
  def energyLevel(): Double = robots.map(_.energyLevel).sum / robots.size
  def positions(): Set[Pos] = robots.map(_.position)
  def monitor(area: Area): Unit = // ...
  // ...
}

// User code (macroprogramming)
if(swarm.energyLevel() < WARNING_ENERGY_LEVEL){
  swarm.move(rechargingStation())
} else { 
  swarm.monitor(targetArea())
}
\end{lstlisting}
\revision{
The \texttt{swarm} object provides a macro-abstraction
 over the set of underlying \texttt{robots}.
}
Indeed, such a code might be written to abstract from a series of low-level details:
 the obstacle avoidance behaviour of individual robots;
 the fact that robots of the swarm move collectively in flock formation;
 the way sensors and actuators perceive distances to other robots,
 obstacles, and acceleration, to control stability and speed of each moving robot.
\revision{
The intended meaning of the program 
 may refer to macro-observables
 that may or not may accessible by the program (cf. side-effects).
The library code provides an implementation
 of the \macrop{} system.
It maps the expressions of the user macro-program 
 down to micro-level behaviour.
Here, the macro-to-micro approach may be interpreted as an interaction mode -- it is the running thread
 that interacts with the micro-level entities
 through the program control flow --
 or an execution mode -- the macro-program is 
 executed by the micro-level entities.
This simplified example shows a \macrop{} language
 as an library/API within an existing host language (Scala), also called an internal DSL; actual examples of internal \macrop{} DSLs include Chronus~\cite{wadaa2010chronus-spatiotemporal-macroprogramming-wsn} and ScaFi~\cite{scafi}.

Doing \macrop{} is very much a matter of perspective.
}
If the micro-macro distinction we are considering is robots vs. a swarm, then the library code (Lines~1-9), individually addressing each robot of the swarm with a specific instruction, is not \macrop{}, properly;
vice versa, the user code (Lines~11-16), addressing the swarm as a whole, does represent an example of \macrop{}.
However, the library code could be considered \macrop{}
 under a micro-macro viewpoint of sensors/actuators vs. a \texttt{robot}.

\subsubsection{\revision{Weak vs. strong macroprogramming}}
In a nutshell,
 the central idea of \macrop{}
 is considering \emph{the entire system
 as the abstract machine} for the operations.
Notice that adopting a \emph{centralised perspective} to programming,
 where a centralised program
 has access to all the individual entities,
 is not \revision{generally sufficient for effective \macrop{}}:
there should typically be \emph{at least one \revision{intermediate} level of indirection\footnote{\revision{Informally, indirection refers to the ability to reference some object through another object; it can be interpreted, e.g., based on static or dynamic scope.}}},
 where macro-operations
 turn into micro-operations.
\revision{
In the example above, 
 while the library code can directly access the individual robots,
 the user code indirectly accesses them 
 through the \texttt{swarm} macro-abstraction.
}

Essentially, \emph{directly} feeding micro-operations to the micro-level entities
or specifying the individual behaviours of the parts
 breaks the \macrop{} abstraction, or makes it \emph{leaky}~\cite{spolsky2004leaky-abstractions,kiczales1992on-abstraction}.
This is one reason 
 (in addition to limited emphasis on behaviour)
 for which, e.g.,
 formalisms for concurrent systems such as process-algebraic approaches~\cite{DBLP:journals/tcs/Baeten05},
 certain component-based approaches, \revision{
 and multi-tier programming~\cite{DBLP:journals/csur/WeisenburgerWS20}}
 are not generally considered \macrop{}.
However, 
 several approaches in literature
 defined themselves as \macrop{}
 despite basically embodying merely a form of centralised programming.
Some of these may provide some \macrop{} abstractions
 (e.g., an object from which individual entities can be \revision{dynamically} retrieved),
 but would nevertheless appear as a \emph{weak} form of \macrop{}.
\revision{
We may consider 
 the \emph{macroscopic stance} as a degree,
 and hence
 define \emph{strong} \macrop{} approaches
 those where \emph{only} macro-abstractions are provided.
}
For demarcation purposes,   we propose to call those centralised programming approaches
 that inherently adopt a macro-level, global perspective
 but directly address individuals through micro-level instructions
 as \emph{weak \macrop{}} or \emph{meso-programming}.
\revision{
Considering the macro perspective as a continuum,
 and hence admitting that languages can be ``more macro''
 or ``less macro'',
 somewhat allow us to avoid defining clear boundaries
 and be comprehensive (which may be important at these early stages, as well as for the sake of this survey).
For sure, however, we can state that writing separate programs
 for different entities of the system 
 from their individual perspectives
 is \emph{not} \macrop{}.
}

\subsubsection{\Macrop{} as a Paradigm}
\label{macro-as-paradigm}

\cite{vanroy2009programming-paradigms}
 defines a \emph{programming paradigm}
 as \emph{``an approach to programming a computer[-based system] based on a mathematical theory \textbf{or a coherent set of principles}''} \revision{(bold is added)}.
Van Roy classifies paradigms 
 according to (i) whether or not they can express observable nondeterminism
 and 
 (ii) how strongly they support state
  (e.g., according to whether it is named, deterministic, and concurrent).
Also interesting is Van Roy's view
 of computer programming
 as a way to deal with complexity (e.g., number of interacting components)
 and randomness (non-determinism)
 to
 make
 aggregates (unorganised complexity)
 and 
 machines (organised simplicity)
 into systems
 (organised complexity).
Macroprogramming effectively deals with aggregates,
 turning them into programmable systems.

We argue the principles outlined in this section form sufficient ground for \macrop{} to be considered a paradigm, and hence aggregate multiple approaches under its umbrella.
\revision{
It is a paradigm in a way similar to \emph{declarative programming}~\cite{loyid1994declarative-programming}, 
 which is ``concerned with writing down \emph{what} should be computed and much less with \emph{how} it should be computed''~\cite{DBLP:journals/tcs/FinkelsteinFL03}.
Then, paradigms like functional and logic programming 
 are considered as more specific forms of declarative programming.
As shown in \Cref{sec:survey},
 also concrete macroprogramming languages 
 can adopt a specific paradigm (e.g., functional, logic, or object-oriented).
The interpretation of \macrop{} as a property degree
 is also coherent with the property degree of ``declarativity'': a language may be more or less declarative according to the amount of details it allows to omit for a same semantic element.
}

The notion itself of a paradigm 
 has sometimes been criticised in teaching programming~\cite{krishnamurthi2019programming-paradigms-and-beyond}
 for its fuzziness and coarse grain,
 preferring epistemological devices like notional machines~\cite{DBLP:conf/iticse/FincherJMDBHHHL20}.
However, our stance is that the notion of a paradigm may still be useful 
 as a lens or perspective
 for observing, comparing, and relating
 several concrete programming approaches,
 and as a core notion around which researchers
 on disparate topics can self-identify and connect
 through shared terms and ideas.

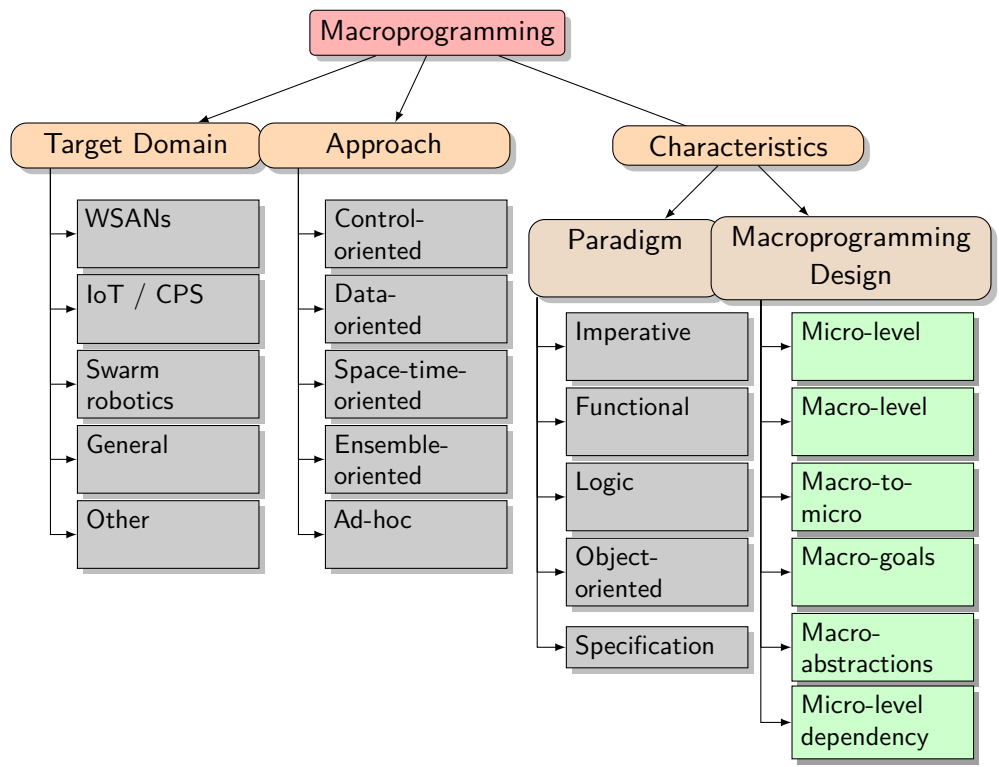
\begin{figure}
\centering
\tikzset{
  basic/.style  = {draw, drop shadow, font=\sffamily, rectangle},   root/.style   = {basic, rounded corners=2pt, thin, align=center, fill=red!30},
  level 2/.style = {basic, rounded corners=6pt, thin,align=center, fill=orange!30, text width=8em},
  level 3/.style = {basic, thin, align=left, fill=gray!40, text width=6.2em},
  level 4/.style = {basic, thin, align=left, fill=gray!40, text width=6em, fill=brown!30,
  align=center},
  trait/.style = {level 3, fill = green!20},
}
\begin{tikzpicture}[
  level 1/.style={sibling distance=40mm},
  edge from parent/.style={->,draw},
  >=latex
]
\newcommand{\xsft}[0]{12pt}
\newcommand{\ysft}[0]{-5pt}
\node[root] {Macroprogramming}
  child {node[level 2] (c2) {Target Domain}}
  child {node[level 2,xshift=-20pt] (c1) {Approach}}
  child {
   [sibling distance=30mm] node[level 2] (c3) {Characteristics}
   child { node[level 4] (c3a) {Paradigm \\ ~} } 
   child { node[level 4,text width=9em] (c3b) {\Macrop{} Design} }   
  };

\begin{scope}[every node/.style={level 3}]

\small

\node [below of = c1, xshift={\xsft},
yshift={\ysft}] (c11) {Control-\\oriented};
\node [below of = c11] (c12) {Data-\\oriented};
\node [below of = c12] (c13) {Space-time-oriented};
\node [below of = c13] (c14) {Ensemble-oriented};
\node [below of = c14] (c15) {Ad-hoc\\~};

\node [below of = c2,xshift={\xsft},yshift={\ysft}] (c21) {WSANs\\~};
\node [below of = c21] (c22) {IoT / CPS\\~};
\node [below of = c22] (c23) {Swarm\\robotics};
\node [below of = c23] (c24) {General\\~};
\node [below of = c24] (c25) {Other\\~};

\node [below of = c3a,xshift={\xsft},yshift={\ysft}] (c31) {Imperative\\~};
\node [below of = c31] (c32) {Functional\\~};
\node [below of = c32] (c33) {Logic\\~};
\node [below of = c33] (c34) {Object-oriented};
\node [below of = c34] (c35) {Specification};

\node [trait,below of = c3b,xshift={\xsft},yshift={\ysft}] (c3b1) {Micro-level\\~};
\node [trait,below of = c3b1] (c3b2) {Macro-level\\~};
\node [trait,below of = c3b2] (c3b3) {Macro-to-micro};
\node [trait,below of = c3b3] (c3b4) {Macro-goals\\~};
\node [trait,below of = c3b4] (c3b5) {Macro-abstractions};
\node [trait,below of = c3b5] (c3b6) {Micro-level dependency}; 
\end{scope}

\foreach \value in {1,...,5}
  \draw[->] (c1.195) |- (c1\value.west);

\foreach \value in {1,...,5}
  \draw[->] (c2.195) |- (c2\value.west);

\foreach \value in {1,...,5}
  \draw[->] (c3a.204) |- (c3\value.west);

\foreach \value in {1,...,6}
  \draw[->] (c3b.205) |- (c3b\value.west);
\end{tikzpicture}
\caption{Taxonomy}
\label{taxonomy}
\end{figure}

\subsection{Taxonomy}
\label{sec:taxonomy}

We propose to classify and analyse 
 \macrop{} approaches
 according to the following elements,
 succinctly represented in \Cref{taxonomy}.
\begin{itemize}
\item[1)] \emph{Target domain.} It refers to the application domain explicitly addressed by a \macrop{} approach. This is relevant since domain-specific abstractions and assumptions are typically leveraged to properly deal with the abstraction gap induced by declarativity.
Label ``General'' is used to indicate that an approach 
 addresses distributed systems in general,
 whereas ``Other'' means that the approach 
 addresses a specific domain different from the others.
\item[2)] \emph{Approach.} We propose to classify \macrop{} languages according to the main approach they follow.
	\begin{itemize}
	\item \emph{Control-oriented}. Emphasis is on specification of control flow and instructions for the system.
	\item \emph{Data-oriented.} Emphasis is on specification of data and data flow.
	\item \emph{Space-time-oriented.} Emphasis is on specification spatial, geometric, or topological aspects and their evolution over time.
	\item \emph{Ensemble-oriented.} Emphasis is on specification of organisational structures as well as tasks and interaction between groups of components.
	\item \emph{Ad-hoc.} The followed approach is peculiar and cannot be easily related with the previous ones.
	\end{itemize}
\item[3)] \emph{Characteristics.}
\item[3a)] \emph{Paradigm.} The paradigm upon which \macrop{} abstractions are supported (the main one in case of multi-paradigm languages).
\item[3b)] \emph{Macroprogramming design.} Elements characterising a particular \macrop{} language.
	\begin{itemize}
	\item \emph{Micro-level}: the individual components and aspects that collectively make up the system.
	\item \emph{Macro-level}: the system as a whole and its macroscopic aspects.
	\item \emph{Macro-to-micro}: \revision{the approach followed by macro-programs to affect micro-level behaviour. 
	We distinguish four main modalities based
	on the discussion in \Cref{def-preliminaries}:
	(i) \emph{context}, where global state, inputs, or node parameters are set;
	(ii) \emph{interaction}, where a process is used to orchestrate micro-level entities;
	(iii) \emph{compilation}, where the macroprogram
	 is translated into the micro-programs;
	(iv) \emph{execution}, where the macro-program is executed by the micro-level entities
	according to some (ad-hoc) execution model.
	}.
		\item \emph{Macro-goals}: the objectives that macro-programs are meant to reach \revision{(typically, abstraction, flexibility, and optimisability---as a result of declarativity)}.
	\item \emph{Macro-abstractions}: the abstractions provided by a \macrop{} approach that are instrumental for achieving or capturing macroscopic aspects or goals of the system.
	\item \emph{Micro-level dependency}: the extent to which the \macrop{} language depends on micro-level components or aspects.
	We consider three levels:
	(i) \textbf{D}ependent (if micro-level elements are always visible),
	(ii) \textbf{I}ndependent (if micro-level elements are abstracted away),
	or
	(iii) \textbf{S}calable (if micro-level elements can be abstracted away as well as accessed, in case).
	\end{itemize}
\end{itemize}
Elements of this taxonomy
 integrate and are partially inspired by some perspectives 
 of previous work covered in \Cref{s:rw}.

{
\newcolumntype{L}[1]{>{\raggedright\let\newline\\\arraybackslash\hspace{0pt}}m{#1}}
\newcolumntype{C}[1]{>{\centering\let\newline\\\arraybackslash\hspace{0pt}}m{#1}}
\scriptsize

\begin{landscape}
\begin{longtable}{|L{0.01\textwidth}|L{0.17\textwidth}|L{0.10\textwidth}|L{0.13\textwidth}|L{0.13\textwidth}|L{0.1\textwidth}|L{0.1\textwidth}|L{0.15\textwidth}|L{0.22\textwidth}|L{0.15\textwidth}|}
\caption{\revision{Summary of the surveyed \macrop{} approaches.
The first column indicates whether the manuscript explicitly advertises the approach as \macrop{}.}
\label{table:results-summary}}
\\\hline
\textbf{} & \textbf{Name/Ref.} & \textbf{Domain} & \textbf{Approach} & \textbf{Paradigm} & \textbf{Micro-level} & \textbf{Macro-level} & \textbf{Macro-to-micro} & \textbf{Macro-abstractions} & \textbf{Micro-dependency}
\\\hline\endhead
$\bullet$ & Market-Based Macroprogramming~\cite{Mainland2004mbm} & WSANs & ad-hoc & specification & nodes & WSN & context & virtual markets; good price & independent\\\hline
$\bullet$ & BNM~\cite{DBLP:journals/taas/Mamei11} & General & ad-hoc & specification & nodes & network & execution & Bayesian network tasks & dependent\\\hline
 & Graph-centric programming, Giraph++~\cite{giraphpp} & Other & ad-hoc & imperative & nodes & graph & execution & subgraph & dependent\\\hline
 & NetKAT~\cite{Anderson2014netkat}; SNAP~\cite{Arashloo2016snap} & Other & ad-hoc & imperative & switches & network & compilation & network state; network slices & dependent\\\hline
 & WOSP~\cite{varughese2020swarm-wosp} & Robotics & ad-hoc & specification & robots & swarm & execution & common behaviour & dependent\\\hline
 & PIECES~\cite{liu2003state-centric-program-wsan-pieces} & Other & control-oriented & object-oriented & (sensor) nodes & WSAN & interaction & global state; state pieces; groups & dependent\\\hline
$\bullet$ & Kairos~\cite{gummadi2005macroprogramming-wsn-kairos} & WSANs & control-oriented & imperative & nodes & WSN & compilation & centralised view; node iterators; neighbourhoods; remote data access & dependent\\\hline
$\bullet$ & PyoT~\cite{Azzara2014pyot-macroprogramming-iot} & IoT/CPS & control-oriented & object-oriented & resources & IoT system & interaction & resource groups & dependent\\\hline
 & Buzz~\cite{pinciroli2016buzz-swarm-programming} & Robotics & control-oriented & imperative & robots & swarm & execution & swarm; neighbourhood; virtual stigmergy & dependent\\\hline
 & Dolphin~\cite{lima2018dolphin} & Robotics & control-oriented & imperative & vehicles & vehicle network & interaction & vehicle sets; vehicle selection expressions & dependent\\\hline
$\bullet$ & makeSense mPL~\cite{makesense} & WSANs & control-oriented & object-oriented & nodes & WSN & compilation & distributed actions (report, tell, collective actions) & dependent\\\hline
 & Warble~\cite{saputra2019warble} & IoT/CPS & control-oriented & object-oriented & things & IoT system & interaction & things selectors; bindings & dependent\\\hline
 & TinyDB~\cite{tinydb} & WSANs & data-oriented & specification & nodes & WSN & compilation & database & independent\\\hline
$\bullet$ & ATaG~\cite{bakshi2005atag} & WSANs & data-oriented & specification & nodes & WSN & compilation & data flow graph & independent\\\hline
$\bullet$ & Semantic Streams~\cite{whitehouse2006semantic-streams} & WSANs & data-oriented & logic & nodes & WSN & execution & event streams; semantic services; inference units; regions & independent\\\hline
$\bullet$ & Regiment~\cite{regiment} & WSANs & data-oriented & functional & nodes & WSN & compilation & time-varying signals; regions & scalable\\\hline
$\bullet$ & COSMOS~\cite{awan2007cosmos} & WSANs & data-oriented & specification & nodes & WSN & compilation & data flow graph & independent\\\hline
$\bullet$ & Flask~\cite{flask} & WSANs & data-oriented & functional & nodes & WSN & execution & nfold macroprogramming combinator & independent\\\hline
$\bullet$ & SOSNA~\cite{karpinski2008stream-macro-wsan-sosna} & WSANs & data-oriented & functional & nodes & WSAN & execution & streams of spatial values & scalable\\\hline
$\bullet$ & MacroLab~\cite{hnat2008macrolab} & IoT/CPS & data-oriented & imperative & nodes & CPS & compilation & macrovector; neighbourhoods & scalable\\\hline
$\bullet$ & Nano-CF~\cite{Gupta2011nano-cf} & WSANs & data-oriented & specification & nodes & WSN & execution & services; jobs & dependent\\\hline
$\bullet$ & Pico-MP~\cite{dulay18pico-mp-pubsub-macroprogramming} & WSANs & data-oriented & logic & nodes & WSAN & compilation & global formula on network data & dependent\\\hline
$\bullet$ & D'Artagnan~\cite{mizzi2018dartagnan}; Porthos~\cite{mizzi2019porthos-blockchain-macroprogramming} & IoT/CPS & data-oriented & functional & IoT devices & IoT system & compilation & data streams & independent\\\hline
 & DDFlow~\cite{Noor2019ddflow-visual-macroprogram-iot} & IoT/CPS & data-oriented & specification & IoT devices & IoT system & execution & data flow graph & independent\\\hline
 & MOISE~\cite{DBLP:journals/ijaose/HubnerSB07} & General & ensemble-oriented & specification & agents & multi-agent system & execution & organisations; roles; groups; missions & dependent\\\hline
 & Scopes~\cite{jacobi2008structuring-wsn-scopes} & WSANs & ensemble-oriented & specification & nodes & WSN & execution & scope (ensemble); scope membership & independent\\\hline
$\bullet$ & EcoCast~\cite{tu2011ecocast} & WSANs & ensemble-oriented & object-oriented & nodes & WSN & compilation & group handles; group-wide operations & dependent\\\hline
 & DEECo~\cite{bures2013deeco} & General & ensemble-oriented & specification & components & distributed system & execution & ensemble & dependent\\\hline
 & SCEL~\cite{DBLP:journals/taas/NicolaLPT14}; AbC~\cite{alrahman2015abc-calculus}; CARMA~\cite{carma}; AErlang~\cite{denicola2018aerlang} & General & ensemble-oriented & specification & components & self-* system & execution & ensembles; group-oriented communication & dependent\\\hline
$\bullet$ & Voltron~\cite{Mottola2014voltron} & Robotics & ensemble-oriented & imperative & drones & swarm & compilation & teams; spatially situated tasks & dependent\\\hline
 & Comingle~\cite{lam2015comingle-distributed-logic-programming-ensemble} & General & ensemble-oriented & logic & app nodes & distributed system & compilation & collective information; system state evolution & dependent\\\hline
 & TECOLA~\cite{Koutsoubelias2016tecola} & Robotics & ensemble-oriented & object-oriented & robots & robotic team & execution & team-level services; mission groups; membership rules & dependent\\\hline
 & PaROS~\cite{paros} & Robotics & ensemble-oriented & object-oriented & robots & swarm & execution & abstract swarms; path planning; task partitioning & dependent\\\hline
 & Aggregate Programming~\cite{Viroli-et-al:JLAMP-2019}; Proto~\cite{proto06a}; Protelis~\cite{protelis}; ScaFi~\cite{scafi} & General & ensemble-oriented & functional & devices & distributed system & execution & computational fields; neighbourhoods; macro-behaviour functions & scalable\\\hline
 & SmartSociety~\cite{scekic2017hybridcas} & General & ensemble-oriented & object-oriented & human and machine peers & socio-technical system & interaction & collectives; collective-based tasks & independent\\\hline
 & Abstract Regions~\cite{welsh2004abstract-regions} & WSANs & space-time-oriented & imperative & nodes & WSN & compilation & regions; region-aware data access & independent\\\hline
 & SpatialViews~\cite{ni2005manetspatialview} & General & space-time-oriented & imperative & devices & MANET & interaction & spatial views (virtual networks) & dependent\\\hline
$\bullet$ & STOP \cite{wada2007spacetime-oriented-macroprog-stop}; Chronus~\cite{wadaa2010chronus-spatiotemporal-macroprogramming-wsn} & WSANs & space-time-oriented & object-oriented & nodes & WSN & interaction & space-time slices & independent\\\hline
 & Meld \cite{Meld} & Robotics & space-time-oriented & logic & modular robots & robot ensemble & compilation & collective information; collective deduction & independent\\\hline
$\bullet$ & Sense2P \cite{choochaisri2012logic-macroprogramming-wsn-sense2p} & WSANs & space-time-oriented & logic & nodes & WSN & execution & logical rule & independent\\\hline
 & PLEIADES \cite{bouget2018pleiades} & General & space-time-oriented & functional & nodes & distributed system & execution & shape templates & dependent\\\hline
\end{longtable}
\end{landscape}

\restoregeometry

}

\section{\Macrop{} Approaches}
\label{sec:survey}

This section provides a survey of \macrop{} languages,
 which are analysed as per the conceptual framework of \Cref{sec:framework}.
The contributions are classified and organised as per the approach classes proposed in \Cref{sec:taxonomy}.
A summary of the survey is provided in \Cref{table:results-summary}.

\subsection{Control-oriented approaches}
\label{sec:survey:control}

Control-oriented approaches
 emphasise an \emph{imperative} \macrop{} style 
 where control flow is specified 
 \revision{and/or explicitly controlled}
 for the system
 and instructions are issued to 
 query or act on system components.
\revision{
This constrasts with data-driven approaches
 where control flow is a consequence 
 of relationships among data.
With control orientation, implicit or explicit sequences, conditionals, and loops 
 may be used to describe what the macro-system or its components have to perform.
}

\paragraph{Kairos~\cite{gummadi2005macroprogramming-wsn-kairos}}
 It is a procedural macroprogramming language for WSNs 
 that assumes loose synchrony and leverages eventual consistency to keep low overhead.
The approach is \emph{control-driven} and \emph{node-dependent}---i.e.,
 nodes and node state 
 are explicitly manipulated at the programming level.
In Kairos, the programmer writes a centralised program expressing the global specification of a distributed computation, which is compiled to a node-specific program.
Kairos exposes three main abstractions:
 addressing of arbitrary nodes (e.g., by names or iterators like \lstinline|node_list|),
 inspection of one-hop neighbour nodes
 (e.g., via function \lstinline|get_neighbors|), 
 and 
 remote data access at nodes (e.g., with expressions \lstinline|variable@node|).
As an example, consider a simple self-healing hop-gradient computation, i.e., 
 an algorithm that makes each node in the system yield the corresponding hop-by-hop distance towards a root node~\cite{DBLP:conf/saso/AudritoCDV17}.
\begin{lstlisting}[language={c}]
node_list nodes = get_available_nodes();
int dist;

for(node n = get_first(nodes); n!=NULL; n=get_next(nodes)){
  // Initialisation
  if(n==root){ dist = 0 } else { dist = INF };
  
  // Event loop
  for(;;){
    sleep(sleep_interval);
    node_list nbrs = get_neighbors(n);
    for(node nbr = get_first(nbrs); nbr!=NULL; nbr=get_next(nbrs){
      if(dist@nbr+1 < dist){ dist = dist@nbr+1; }
    }
  }
}
\end{lstlisting}
\revision{
Concerning macro-to-micro mechanics and implementation,
 during the translation 
 of the macro-program into node-level programs,
 references to remote data are expanded into calls
 to the \emph{Kairos runtime}, a software component which is assumed to be available in every node of the system.
Specifically, the Kairos runtime deals with \emph{managed objects} (objects owned by a node that are to be made available to remote notes) and \emph{cached objects}
 (local views of managed objects owned by remote nodes),
 through asynchronous hop-by-hop communication---contrast this with synchronous data access calls in Kairos programs.
Issues at the middleware level include supporting end-to-end reliable routing and management of dynamic topologies.
}

\paragraph{PyoT~\cite{Azzara2014pyot-macroprogramming-iot}}
PyoT is defined as a Python-based macroprogramming framework for the IoT and WSNs.
A simple example adjusted from the paper is the following.
\begin{lstlisting}[language={Python}]
temperatures = Resource.objects.filter(title='temp')
results = [temp.GET() for temp in temperatures]
avg = sum (results) / len(results)
TEMP_THRESHOLD = 24
if avg > TEMP_THRESHOLD:
    Resource.objects.get(title='fan').PUT('on')
\end{lstlisting}
This is just a script 
 that collects values from temperature sensors,
 computes the mean of the temperature,
 and turns the fan device on if the mean exceeds a certain threshold.
Notice the imperative approach
 and the global perspective by which
 resources are accessed.
The only relevant abstraction is that of a \emph{resource},
 inherited by its RESTful design
 which is typical in IoT platforms.
\revision{
Architecturally, PyoT has one or more worker nodes
 managing corresponding sets of sensors (i.e., entire IoT systems or WSNs)
 and executing tasks issued from Pyot programs
 by the users through shells or virtual control rooms.
}

\paragraph{Buzz~\cite{pinciroli2016buzz-swarm-programming}} Buzz is an imperative swarm-oriented macroprogramming language and system.
In Buzz, a \emph{swarm} consists of a set of robots equipped with the Buzz virtual machine and running the same Buzz script in a step-by-step fashion.
In each step, a robot (i) collects sensor readings and incoming messages;
(ii) executes a portion of the Buzz script;
(iii) sends output messages; and
(iv) applies actuators on actuator values hold in the state.
Robots can share information through \emph{virtual stigmergy}~\cite{pinciroli2016virtual-stigmergy-swarm} 
 (i.e., communication via distributed tuples spaces, inspired by environment-mediated interaction of social insects)
or 
by querying neighbours.
The following example of Buzz code shows how swarm behaviour is programmed imperatively at the swarm-level.
\begin{lstlisting}[language=Python,morekeywords={function}]
function init(){ # this function is run for initialisation
  s1 = swarm.create(1) # a newly created, empty swarm with ID=1
  s2 = swarm.create(2) # another swarm
  s1.select(id % 2 == 0) # join the swarm based on robot's id
  s2.join # every robot joins the swarm 2 unconditionally
  s3 = swarm.difference(3, s2, s1) # a new swarm with robots in s2 but not in s1
}

function step(){ # this function is run at each time step
  # ...
  s3.exec( function() { ... }) # every robot in swarm s3 runs the given function
  if(...){ s1.leave() } # conditionally leaving a swarm
  # ...
  n = neighbors.count() # number of neighbours
  nbrs = neighbors.kin() # neighbours in the same swarm
  temperatures = nbrs.map( function(robotId,data){
      return data.temperature # data provides access to the attributes of a robot
  }) # map every neighbour robot in nbrs with the corresp. temperature
  # ...
}
\end{lstlisting}
Notice the language comprises both single-robot and swarm-based primitives.
Within the \lstinline|step| function,
 the point of view is of an individual robot;
 however, swarm abstractions enable selective addressing of individuals and multiple dispatch of operations,
 and neighbourhood abstractions promote local coordination.
\revision{
In the implementation, each robot keeps track 
 of memberships in swarms
 and data from neighbours;
 optimisations are applied to reduce 
 the communication overhead.
}

\paragraph{Dolphin~\cite{lima2018dolphin}}
Dolphin is an open-source Groovy-based task-oriented macroprogramming language per autonomous vehicle networks.
Lima et al. describe it as an ``extensible task orchestration language [...] delegating platform-dependent networked operations to the platform''.
The macro-level abstraction in Dolphin is the \emph{vehicle set}, a dynamic group of vehicles which can be manipulated through set operations and by pick/release operators.
An example of a Dolphin program, slightly adapted from the paper and commented, for coordinating three unmanned underwater vehicles (UUVs) surveying a region and one unmanned aerial vehicle (UAV) for collecting those surveys is as follows.
\begin{lstlisting}[language=java]
// (1) configuration
r = ask 'Radius of operation area? (km)' // ask radius to user
APDL = (location 41.18500, -8.70620) ^ r.km // geo-referenced area
// (2) Vehicle selection
UUVs = pick { count 3; type 'UUV'; payload 'DVL', 'Sidescan' ; region APDL }
UAV = pick { type 'UAV'; region APDL }
// (3) function yielding UUV task i
def UUVTask(i) {
    imcPlan('survey' +i) >>     // the actual survey task
    action { post ready:i } >>  // signal readiness for rendez-vous
    imcPlan { planName 'sk' + i; skeeping duration: 600 }
}
// (4) Execute tasks
execute UUVs: UUVTask(1) | UUVTask(2) | UUVTask(3), // concurrent composition
        UAV: allOf {
            when { consume ready: 1 } then imcPlan('rv1')
            when { consume ready: 2 } then imcPlan('rv2')
            when { consume ready: 3 } then imcPlan('rv3')
        }
// (5) End
release UUVs + UAV // + is for union of two vehicle sets
\end{lstlisting}
Single-vehicle tasks are expressed through \emph{IMC plans}, i.e., task specifications based on a message-based interoperability protocol called \emph{Inter-Module Communications (IMC)}~\cite{pinto2013lsts}.
The program asks a radius \lstinline|r| to the user,
 defines a geographical region \lstinline|APDL|,
 then defines two vehicle sets, \lstinline|UUVs| (with three UUVs) and \lstinline|UAV| (a singleton set with a single UAV), which are tasked via \lstinline|execute| and finally \lstinline|release|d.
In summary, macroprogramming in Dolphin is concerned with ``grouping'' and ``tasking''; however, the different vehicles are given specific micro programs (IMC plans) not obtained from the global specification, which is only responsible for orchestrating the system.

\paragraph{Warble~\cite{saputra2019warble}}
Warble is another macroprogramming framework for the IoT.
The following example adapted from the paper shows 
 how to select the three light or thermostat things 
  closest to \lstinline|myLocation|
  and perform a one-time operation on them.
\begin{lstlisting}[language={java}]
Warble warble = new Warble(); // for discovery of things
List<Selector> template = new ArrayList<Selector>();
template.add(new TypeSelector(LIGHT, THERMOSTAT));
template.add(new NearestThingSelector(myLocation));
List<Thing> things = warble.fetch(template, 3);
for(Thing thing : things){
  if(thing instanceof Light) ((Light) thing).on();
  if(thing instanceof Thermostat) ((Thermostat) thing).setTemperature(24);
}
\end{lstlisting}
Warble also support ``dynamic binding'' 
 for continuous discovery and operations.
\begin{lstlisting}[language={java}]
Plan plan = new Plan();
plan.set(Plan.Key.LIGHTING_ON, true); 
plan.set(Plan.Key.AMBIENT_TEMPERATURE, 30);

DBinding dBind = warble.dynamicBind(template, 3);
dBind.bind(plan); // start binding based on plan 
\end{lstlisting}
Warble is quite similar to PyoT; essentially,
 differences are mainly in the API
 and in the architecture and implementation,
 and therefore in the non-functional properties of the system.

\paragraph{makeSense m{\scriptsize{}PL}~\cite{makesense}}
makeSense is a framework for WSN application development.
It comprises a two-step compilation process
 where (i) 
 a Business Process Modelling Notation (BPMN)~\cite{DBLP:journals/bpmj/ZarourBGD20} model is compiled into a macro-program expressed in a \macrop{} language called m{\scriptsize{}PL};
 and
 (ii) the m{\scriptsize{}PL} program is then compiled 
 to the target underlying WSN platform such as TinyOS~\cite{DBLP:books/sp/05/LevisMPSWWGHWBC05} or Contiki~\cite{DBLP:conf/lcn/DunkelsGV04}.
The m{\scriptsize{}PL} language comprises the following meta-abstractions:
 \emph{local} actions that affect a single device
 and \emph{distributed} actions that affect multiple devices
 and include \emph{report} actions (modelling many-to-one interactions),
 \emph{tell} actions (modelling one-to-many interactions),
 and 
 \emph{collective} actions (modelling many-to-many interactions).
These meta-abstractions

\subsection{Data-oriented and database abstraction approaches}

Data-oriented approaches 
 define the macro-level behaviour of a system
 in terms of goals and activities of data gathering and processing.
Sometimes, this is taken to the extreme,
 considering the system as a kind of distributed database
 keeping spatiotemporal or aggregated data.

\paragraph{TinyDB~\cite{tinydb}}
It is a query processing system that considers a WSAN as a database.
TinyDB supports an SQL-like language
 for expressing queries and actuations.
A query looks like the following:
\begin{lstlisting}[keywords={SELECT,WHERE,FROM,SAMPLE,PERIOD}]
SELECT nodeId, temperature  WHERE temperature > k FROM sensors 
SAMPLE PERIOD 5 minutes
\end{lstlisting}
Therefore, the approach is fully declarative 
 and the system must find itself a strategy to
 map the global goal to local behaviour of the sensor nodes.
\revision{
We remark that
 the \emph{behaviour} of the individual nodes
 is driven partly by the query-like macroprogram
 and partly by a basic ``execution protocol'' (providing a structure for the emergence of global behaviour) which is the same for all the nodes.
Nodes work in \emph{epochs}, corresponding to sampling periods, in a synchronised fashion.
They sleep for most of the time; they wake up
 to sample sensors,
 gather neighbour data,
 process data,
 and send results to their parent node.
This execution protocol is very similar to those
 used by other \macrop{} approaches, such as aggregate computing~\cite{Viroli-et-al:JLAMP-2019} which is a paradigm for self-organising systems of agents.
}

\paragraph{ATaG (Abstract Task Graph)~\cite{bakshi2005atag}}
ATaG is a data-driven macroprogramming approach for sensor networks 
 where macro-programs take the form of annotated dataflow graphs.
In these graphs,
 abstract channels 
 connect abstract data 
 with abstract tasks.
Then, the graph is augmented with annotations specifying 
 (i) how tasks are to be instantiated on the network nodes (e.g., on a specific node, anywhere, once every $N$ nodes, on every node in a spatial area),
 (ii) how tasks are to be scheduled
 (e.g., periodically, or when any or all of the inputs are available),
 and
 (iii) information flow patterns 
 (e.g., through particular kinds of channels
 modelling neighbourhood interaction, parent-child interaction, and task-wide broadcast).
For instance, a temperature monitoring application
 could be designed as follows.

\begin{center}
\footnotesize
\tikzset{
  n/.style  = {draw, font=\sffamily\footnotesize
, rectangle, auto},
lbl/.style  = {draw, fill=black!15, font=\sffamily\tiny, rectangle, auto, text width=2.8cm},
elbl/.style  = {draw, fill=black!15, font=\sffamily\tiny, rectangle, auto, midway}
}
\begin{tikzpicture}[node distance=1.5cm and 2cm]

\node[n,ellipse] (monitor) [] {Monitor};
\node[n,ellipse] (corr) [right=of monitor] {Corroborator};

\node[n] (temp) [below = of monitor] {Temperature};
\node[n] (alarm1) [right= of temp] {Local Alarm};
\node[n] (alarm2) [right= 1cm of alarm1] {Global Alarm};

\node[lbl] (mlbl) [above = 0cm of monitor] {\texttt{[nodes-per-instance: 1]}\newline
\texttt{[periodic: 10 || any-data]}
};

\node[lbl] (mlbl) [above = 0cm of corr] {\texttt{[nodes-per-instance: 1]}\newline
\texttt{[any-data]}
};

\draw[->] (monitor.south west) -- (temp.north west) node [elbl,left] {\texttt{1-hop \&\& $\lnot$local}};
\draw[->] (temp) -- (monitor) node [elbl,right] {\texttt{local}};
\draw[->] (temp) -- (corr) node [elbl,near start,below] {\texttt{10m:pull}};
\draw[->] (monitor) -- (alarm1) node [elbl,near start] {\texttt{local}};
\draw[->] (alarm1) -- (corr) node [elbl] {\texttt{local}};
\draw[->] (corr) -- (alarm2) node [elbl] {\texttt{local}};

\end{tikzpicture}
\end{center}

\paragraph{Semantic Streams~\cite{whitehouse2006semantic-streams}}
Semantic Streams is a logic-based, declarative language 
 for expressing semantic queries over WSN data.
It builds on two main abstractions: \emph{event streams} and \emph{inference units} (processes on event streams).
For instance, the following program
\begin{lstlisting}[language={prolog}]
stream (Y), isa (Y, histogram),  % histogram events
property (Y , X, stream),        % a histogram event has a stream X
property (Y , time, property),   %   and a time property
stream (X), isa(X, objectDetected),  %  stream X consists of object detection events
property (X, [[0,0,0],[50,50,0]], region). % ... within a given region
\end{lstlisting}
can be used to query for and plot \lstinline|objectDetected| events 
 in a given area across time.
\revision{
The \macrop{} system implementation
 is based on service composition and embedding.
The query planner builds a task graph to be deployed to individual nodes,
 which will dynamically instantiate services,
 resolve conflicts between tasks and resources,
 and  
 execute the queries.
}

\paragraph{COSMOS~\cite{awan2007cosmos}}
COSMOS is a macroprogramming system for heterogeneous sensor networks.
It consists of a \emph{dataflow} macroprogramming language, called \emph{mPL}, and an operating system, called \emph{mOS}.
Macroprograms specify the aggregate behaviour of a sensor network
 in terms of typed \emph{functional components} (mapping inputs into outputs)
 and \emph{interaction assignments}
 describing how functional components are connected
 to form an asynchronous dataflow graph.
Constraints can be used to affect instantiation of components in the physical nodes.
Constructs called \emph{contracts} can be applied to dataflow paths 
 to provide higher-level abstractions such as region-scoped/neighbourhood broadcasts or 
 load-aware resource management.
An adapted excerpt of mPL code from the paper follows.
\begin{lstlisting}[language=C,otherkeywords={IA}]
// Functional Component (FC) declarations
  mcap = MCAP_FAST_CPU,  // execution is possible only in nodes with fast CPU
  fcid = FCIS_FFT,       // ID of the FC
  in[craw_t],            // input type of the FC
  out[freq_t]            // output type of the FC
}

// Logical instances
fft_fc : fft;
device : source;
device : sink;

// Interaction Assignment (IA)
IA {
  source -> fft[0];
  fft[0] -> sink;
  ctrl[0] --> thresh[1]; // first output of ctrl is connected to the second input of thresh
}

// FC implementations (in C)
cp_ret_t fft_fc(cp_param_t*param, bq_node_t*pbqn, num_conn_t ind) { /*...*/ }
// ...
\end{lstlisting}
The COSMOS compiler takes a macro-program as input
 and produces an annotated dataflow graph
 which is communicated on the network nodes hosting mOS.

\paragraph{MacroLab~\cite{hnat2008macrolab}}
MacroLab is a vector-based macroprogramming framework
 where the global behaviour of a CPS
 is specified through a macro-program consisting of Matlab-like operations.
It exposes a \emph{macrovector} abstraction
 which is an unordered data structure 
 where each element is associated with a system node.
As an example, consider a MacroLab program inspecting a collection of temperature sensor nodes.
\begin{lstlisting}[language={matlab}]
RTS = RunTimeSystem();
temperatureSensors = SensorVector('Temperature', 'uint16');
temperatureValues = Macrovector('uint16');
neighborTemperatureValues = neighborReflection(temperatureValues)
CRITICAL_TEMP = uint8(50);

every(1000) {
  temperatureValues = sense(temperatureSensors);
  maxTemperatureOverall = max(temperatureValues);
  meanTemperatureInNbrhood = smean(neighborTemperatureValues);
  criticalNodes = find(meanTemperatureInNbrhood > CRITICAL_TEMP);
  % ...
}
\end{lstlisting}
As we see, the program consists of a loop 
 where temperature sensors are queried to
 populate a macro-vector \lstinline|temperatureValues|;
 then, vector operations are used to get the maximum temperature in the sensor network,
 to aggregate neighbourhood-wide temperatures in a mean value,
 and to find the node where that mean temperature is higher of a threshold.
Both synchronised (e.g., \lstinline|smean|---with the \lstinline|s| prefix) and unsynchronised (e.g., \lstinline|max|, \lstinline|find|) operations are supported, to enable various semantics and optimisations.
Interestingly, MacroLab also provides a \emph{deployment-specific code decomposition} approach 
 where 
 macro-programs can be decomposed 
 into different sets of micro-programs
 supporting different deployment scenarios.
The MacroLab decomposer works by 
 choosing a macrovector representation
 and then using rules to map vector operations
 into micro-level network operations.
Three main representations include:
 a \emph{decentralised} representation
 where vector values for a node are stored in that node;
 a \emph{centralised} representation
 where vectors are centrally stored in a single node (e.g., a base station);
 or
 a \emph{replicated} representation
 where vectors are replicated in every node.

\paragraph{Flask~\cite{flask}}
Flask is a functional \macrop{} DSL,
 implemented in Haskell,
 that supports a dataflow-oriented design of 
 the behaviour of sensor networks.
It provides a \emph{\macrop{} combinator}, called \texttt{nfold},
 that combines local with other nodes' signals together;
 this is implemented by the whole network
 by aggregating values on a spanning tree.
Flask has been used to program an implementation of TinyDB, called FlaskDB.

\paragraph{Sense2P~\cite{choochaisri2012logic-macroprogramming-wsn-sense2p}}
 It is a \emph{logic macroprogramming system} for WSNs,
 based on \emph{LogicQ}~\cite{choochaisri2008logicq-wsn-as-globally-deductive-db},
 a system that abstracts sensor networks as relational databases
 and supports collecting data and spreading logic queries.
 In Sense2P, programs are expressed in a Prolog-like language: 
 \emph{facts} represent sensor data; 
 \emph{rules} enable generation of new facts;
  and
 \emph{queries} enable checking and retrieving for (derived) facts.
The macroprogrammers think like the facts were in a centralised database: the actual process of data retrieval is abstracted away.
An example from the paper follows: it gets a set of all ``hot objects'' in a spatial area denoted by atom \lstinline|area70|.
\begin{lstlisting}[language=Prolog]
hotObject(Obj, AreaID):- detect(Obj, AreaID), temperature(Obj,T), T > 50.
?- hotObject(X, area70).
\end{lstlisting}
Operationally, the subqueries (\lstinline|detect|, \lstinline|temperature|)
 are propagated in the network and evaluated
 only in the nodes that support them.
So, only sensors detecting any object in \lstinline|area70| are considered.
Missing facts/rules or false conditions cause the evaluation on the node to be suppressed.
\revision{
The system architecture of Sense2P
 has two main components:
 (i) a query processing engine,
 with a compiler that translates Sense2P macro-programs
 into compiled code,
 and a run-time processing unit 
 which executes the compiled program
 and submits queries to be disseminated
 across the network;
 and 
 (ii) a data-gathering engine 
 which is responsible of providing results to queries
 by exploiting a routing tree connectign sensors to the WSN base station.
}

\paragraph{PICO-MP~\cite{dulay18pico-mp-pubsub-macroprogramming}}
 It is a publish/subscribe system for WSNs
 where subscriptions are expressed 
 through \emph{global FOL formulae}.
Subscriptions are checked in a decentralised way
 against the published events to produce notifications bubbling up in a tree overlay network of brokers.
This is done by having brokers perform local checks
 as described by \emph{projections} of the global formulae.

\paragraph{D'Artagnan~\cite{mizzi2018dartagnan}}
D'Artagnan is a stream-based, functional, macroprogramming DSL embedded in Haskell for IoT systems.
Its core idea is to specify stream processing functions 
 that periodically execute on sensor data.
For instance,
\begin{lstlisting}[language={Haskell}]
average :: [Stream Float] -> Stream Float
average ss = sum ss ./. consStream (length ss)
\end{lstlisting}
is a generic function 
 that computes a stream of averages
 from a list of streams of values from sensors:
\begin{lstlisting}
let input1 = input (device 1) (sensor 1)
    input2 = input (device 2) (sensor 1)
    input3 = input (device 2) (sensor 2)
in average [input1, input2, input3]
\end{lstlisting}
The language also provides two communication operators: \lstinline|pull| to perform a request-response interaction,
and \lstinline|push| for a fire-and-forget action.
For instance, in expression
\begin{lstlisting}
input1 .+. (push (device 2) (input2 .*. input3))
\end{lstlisting}
the multiplication is performed on \lstinline|device 2|
 before sending the partial result in \lstinline|device 1| (cf. the definition of \lstinline|input1|),
 hence avoiding transmission of both \lstinline|input2| and \lstinline|input3| with separate messages.

\paragraph{Nano-CF~\cite{Gupta2011nano-cf}}
Nano-CF is a macroprogramming framework for WSNs.
Architecturally,
 a Nano-CF system consists of three layers:
 (i) the \emph{runtime environment} provides support for executing local actions above the sensor node OS;
 (ii) the \emph{integration layer} deals with packet delivery, data aggregation, scheduling, and batching tasks;
 (iii) the \emph{coordinated programming environment (CPE)}
 allows users to compile and send programs to the sensor nodes as well as to receive data from the WSN.
A Nano-CF macro-program consists of a number of \emph{service} definitions (expressing the local behaviour of a sensor node)
and a set of \emph{job descriptors}
 each describing in which nodes a service must be executed, 
 at which frequency (and tolerable deviation),
 and how output data must be gathered
 (through aggregation functions such as \lstinline|min|, \lstinline|max|, \lstinline|sum|, \lstinline|count|, and \lstinline|noaggr| for keeping all data instances).
For instance,
 consider the following program for counting the number of rooms occupied in a building and getting the temperatures of all the rooms (adapted and simplified from~\cite{Gupta2011nano-cf}).
{
\lstset{language={},morecomment=[s][\color{black!40!green}]{/*}{*/}}
\begin{lstlisting}[
morekeywords={JOB,END,SUM,NOAGG,SERVICE,uint8,return,if,endif,
else,gets}]
JOB:
  occupancy_monitor <L1,L2,...> <20s,5s> SUM
  temperature_collection_service <L1,L2,...> <50s,0s> NOAGG
END

SERVICE: occupancy_monitor uint8  
  if(/* determine if occupied e.g. using local sensors */) return 1; else return 0; endif
END

SERVICE: temperature_collection_service uint16
  return gets(TEMP);
END
\end{lstlisting}
}
Notice the macro-level viewpoint in the \lstinline|JOB| block,
 where service are mapped to multiple nodes
 and overall service data is aggregated.

\paragraph{DDFlow~\cite{Noor2019ddflow-visual-macroprogram-iot}}
DDFlow is a graphical language for programming IoT systems
 through specification of declarative dataflow graphs.
A \emph{dataflow graph} describes an application
 as a set of interrelated action \emph{nodes}. 
Nodes represent stateful macro-actions.
The actual deployment of tasks onto the devices of the system
 can be configured by specifying a (possibly dynamic) spatial region
 or a set of target devices 
 through corresponding attributes in the dataflow node itself.
Nodes are connected through \emph{wires}, 
 which can be of three main types: one-to-one (stream), one-to-many (broadcast), or many-to-one (unite).
The following is an example of object tracking (adapted from the paper).
\begin{center}
\footnotesize
\tikzset{
  n/.style  = {draw, font=\sffamily\footnotesize
, rectangle}, auto
}
\begin{tikzpicture}[node distance=0.2cm and 1.4cm]
\node[n,text width=2.3cm] (t1genframe) [] {{ \textbf{Generate frame} \newline \scriptsize\texttt{Region:(lat,long,r)}}};
\node[n] (t2classify) [right=of t1genframe] { \textbf{Classify} };
\node[n,text width=1.8cm] (t3filter) [right=of t2classify] { \textbf{Filter} \newline \scriptsize\texttt{``Target object''}};
\node[n,text width=1.6cm] (t4alert) [above right= of t3filter] { \textbf{Sound Alert} \newline \scriptsize\texttt{DEVICE: <ip>} };
\node[n,text width=2.5cm] (t5follow) [below right= of t3filter] { \textbf{Follow} \tiny\newline \texttt{LOCATION: Frame.loc} \newline \texttt{Target: ``Target Object''}};

\draw[->] (t1genframe) -- (t2classify) node [text width=0.5cm,midway] {Stream};
\draw[->] (t2classify) -- (t3filter) node [midway] {Stream};
\draw[->] (t3filter) -- (t4alert) node [midway] {Unite} ;
\draw[->] (t3filter) -- (t5follow) node [midway,below,xshift=-0.2cm,yshift=-0.1cm] {Broadcast};
\end{tikzpicture}
\end{center}
Clearly, the graph provides a macro-view of the system
 that abstracts from the underlying IoT nodes.

\subsection{Space-time oriented approaches}

Space-time-oriented \macrop{} approaches
 are those that 
 leverage spatial and temporal abstractions 
 to organise the behaviour of a system.

\paragraph{Pieces (Programming and Interaction Environment for Collaborative Embedded Systems)~\cite{liu2003state-centric-program-wsan-pieces}}
Pieces is a \emph{state-centric} programming model for WSANs
where
\begin{quote}
programmers think in terms of dividing the global state of physical phenomena into a hierarchical set of independently updatable pieces with one computational entity (called a \emph{principal}) maintaining each piece.~\cite{liu2003state-centric-program-wsan-pieces}
\end{quote}
That is, a principal is an agent that interacts with other principals to update its piece of state, and that may move across WSAN nodes.
Pieces leverages the notion of \emph{collaboration group}, i.e., a scoped set of principals playing different roles that collaborate to a state update,
  to abstract communication and resource allocation patterns.
Examples of groups include \emph{geographically constrained} groups (a set of nodes located in some geographical region),
 \emph{n-hop neighbourhood} groups (a set of nodes within $n$ hops from a given anchor node), and
 \emph{publish/subscribe} groups (a set of consumer and producer nodes on certain topics).
\begin{lstlisting}[language=java]
public class SomePrincipal extends MobilePrincipal {
  private InputPortAgent someAgent;
  private BeliefState someBelief;
  private UtilityFunction someUtility = new InformationUtility();
  private GeoConstrainedLeaderGroup someGroup; 
  
  public void initialize() {
    someAgent = new SomeAgent(this);
    someGroup = new GeoConstraintedLeaderGroup(geometricExtent, someAgent, ...);
  }
  
  // The principal is awakened every 1 second by a time trigger.
  public void react(WakeupEvent event) {
    updateState(); // Compute a new target belief state
    moveTo(someGroup, someUtility); // Choose next host for this principal
  }
  
  // Function to update the state of the principal.
  public synchronized void updateState() {
    if(someAgent.isInputReady())
      someBelief = someStateUpdateFunction(someBelief, someAgent.getInputData());
  }
}
\end{lstlisting}
The above code shows that principals are defined 
 individually,
 but these somewhat abstract 
 from the underlying WSAN nodes.
\revision{In other words, principals are executors
 different than the micro-level entities 
 of the system (sensors): 
 they orchestrate the activity of global state management.}

\paragraph{Abstract Regions~\cite{welsh2004abstract-regions}} 
It provides a \emph{``region-based collective communication interface [...] to hide the details of data dissemination and aggregation within regions''}~\cite{welsh2004abstract-regions}.
Supported classes of operators include those for neighbour discovery, enumeration of nodes in a region, 
data sharing,
and data aggregation (or reduction).

\paragraph{Regiment~\cite{newton2004region-streams-regiment,regiment}}
Regiment is a functional reactive spatiotemporal macroprogramming language.
It is based on the abstractions
 of \emph{time-varying signals}
 (to model, e.g., the values produced by a temperature sensor)
  and \emph{regions},
  modelling dynamic collections of spatially distributed signals
  (i.e., regions are essentially the same concept as computational fields).
Signals and regions can be manipulated through typical functional operations such as map, filter, and fold.
These global operations abstract data acquisition, storage, and communication:
 it is the job of the compiler to map these to local operations on the network nodes
 (through a process called \emph{deglobalisation}).  
New regions can be constructed based on spatial and topological relationships between nodes through \emph{region formation primitives}, which are grouped into two categories:
 (i) functions for \emph{growing regions} from ``source'' nodes called \emph{anchors} (implemented using \emph{spanning trees})
 and 
 (ii) \emph{gossip-based functions} (based on one-hop broadcasts).
A sample Regiment program computing the average temperature across an entire sensor network (denoted via region \lstinline|world|) is as follows (adapted from~\cite{regiment}).
{
\lstset{language={},morecomment=[l][\color{black!40!green}]{\%}}
\begin{lstlisting}[morekeywords={fun,::,->},emph={rmap,rfold,smap}]
doSum :: float (float, int) -> (float, int);
doSum(temperature, (sum, count)) { (sum+temperature, count+1) }

temperatureRegion = rmap(fun(node){ sense("temperature", node) }, world);
sumSignal = rfold(doSum, (0.0, 0), temperatureRegion)
avgSignal = smap(fun((sum,count)){ sum / count }, sumSignal)

BASE <- avgSignal % move such information to the base station
\end{lstlisting}
}
As in functional reactive programming,
 change in signals 
 is propagated to dependent signals.
So, as temperature sensors yield new values,
 as sensors fail, disconnect, or enter the network,
 the signals and regions depending on them are updated,
 ultimately adjusting the average temperature value,
 which is finally transmitted to the base station.
Notice how this macro-program abstracts 
 low-level details such as network communications.

\paragraph{SpatialViews~\cite{ni2005manetspatialview}}
 This approach works by abstracting a MANET into \emph{spatial views} (i.e., collections of \emph{virtual nodes}) of a configurable space-time granularity, that can be iterated on to visit nodes and request services. 
In detail, the model is as follows.
A physical network consists of physical nodes. A physical node has a spatio-temporal location and a set of provided services.
A virtual node is the digital twin of a physical node: its programming abstraction.
A spatial view defines a virtual network over the physical network
 which is discovered and instantiated when iterated.
Operationally, the system works by migratory execution of the program during iteration.
The SpatialViews language is implemented as an extension to Java.
\begin{lstlisting}[language={java},otherkeywords={spatialview,{@},{\%},visiteach}]
// Spatial views are collections of virtual nodes
spatialview sv1 = Camera @ BuildingC.Floor3;
spatialview sv2 = TemperatureSensor @ CampusB % 50; // 50 meters space granularity

// Discover virtual nodes in a spatial view
visiteach x : sv1 every 5 forever { x.getPicture().upload(); }

// Take average of temperatures
sumreduction float s = 0;
sumreduction int n = 0;
visiteach y : sv2 { s += y.read(); n++; }
float avg = s/n;
\end{lstlisting}
Space-time granularities are used to distinguish virtual nodes, which are visited once per iteration; instead,
the underlying physical nodes might be visited more than once (e.g., because of mobility or after a quantum of time granularity).
We remark that this work did not use any ``macroprogramming''-like term to label SpatialViews,
 though clearly embracing the paradigm.

\paragraph{SpaceTime Oriented Programming (STOP)~\cite{wada2007spacetime-oriented-macroprog-stop},
 a.k.a. \emph{Chronus}~\cite{wadaa2010chronus-spatiotemporal-macroprogramming-wsn}}
 This WSN macroprogramming system
 exposes a spacetime abstraction
 to support collection and processing of past or future data
 in arbitrary spatio-temporal resolutions.
Architecturally, it consists of a network of 
battery-powered sensors (where data is gathered)
and base stations (where data is processed) 
linked to a gateway connected to the STOP server,
which holds network data in the so-called \emph{spatiotemporal database}.
Operationally, the system is implemented through mobile agents
 carrying data to the STOP server, which in turn updates the database:
 \emph{event agents} detect events 
 and replicate themselves to move hop-by-hop
 towards a base station, where they finally \emph{push} data;
 by contrast, \emph{query agents} move across a spatial region
 in order to \emph{pull} relevant data.
The STOP/Chronus language is an object-oriented, Ruby DSL 
 enabling on-command and on-demand (event-driven) data collection and processing.
An example, selected and adapted from~\cite{wada2007spacetime-oriented-macroprog-stop}, is the following.
\begin{lstlisting}[language=Ruby]
sp = Spacetime.new(Polygon.new(points), RelativePeriod.new(NOW, Hr-1))

spaces = sp.get_spaces_every(Min 5, Sec 10, 80)

values = spaces.collect { |space|
  space.get_data('f-spectrum', MAX, Min 2){
    |event_type, value, space, time |
    # ...
  }
}
\end{lstlisting}
This program queries data in space-time ``slices'' 
 that abstract the data generation activity of the underlying collection of sensor nodes.
Indeed, it focusses on a macroscopic perspective.

\paragraph{SOSNA~\cite{karpinski2008stream-macro-wsan-sosna}}
 SOSNA is a stream-based, macroprogramming language for WSANs
 where programs operate on streams of spatial values.
 Spatial values are essentially like the regions in Regiment and are called \emph{field streams}.
The other kind of spatial value is given by \emph{cluster streams}, which are spatially-limited fields with a singleton node (\emph{cluster head}) holding cluster field data.
A cluster stream is built via cluster operators from a field stream
 which internally drive a
 \emph{leader election} (according to the used operator and local data)
 and a
 \emph{clustering} process of nodes through spanning trees of bounded height (based on a compilation parameter and possibly resulting in multi-hops paths).
Other operators allow for moving data from cluster members to cluster heads (\lstinline|fold|) and vice versa (\lstinline|unfold|) as well as for evolving state (through \lstinline|pre x|, which refers to the stream \lstinline|x| at the previous round)
 and aggregating data from neighbour nodes (\lstinline|foldnbrs|).
 Execution of SOSNA programs is round-wise and synchronous:
 a \emph{round} consists of an application-specific number of \emph{steps};
 in each step, neighbours exchange a \emph{protocol packet}.
Any network operator requires a fixed number of execution steps,
and the compiler can statically infer the maximum number of steps of each round.
Consider this example from the paper:
\begin{lstlisting}[keywords={where,clmax,fold}]
object = where (sensor > THRESH) clmax sensor 
totalMass = fold (+) sensor object 
objX = (fold (+) (posX*sensor) object) / totalMass 
objY = (fold (+) (posY*sensor) object) / totalMass
\end{lstlisting}
It is a simple object tracking program. The cluster stream \lstinline|object| is defined by 
the \lstinline|clmax| (cluster-max) operator 
on the field stream \lstinline|sensor| filtered for values greather than \lstinline|THRESH|. Then, the local values \lstinline|sensor| of all the cluster members are accumulated into \lstinline|totalMass|, and then the coordinates of the object are computed by applying the formula of centre-of-mass.
In summary, SOSNA is similar to Proto~\cite{proto06a},
 but requires synchronisation 
 and is tailored to WSANs. 

\paragraph{Karma~\cite{karma}}
Karma is a framework for programming swarms of micro-aerial vehicles.
It proposes a \emph{hive-drone} model 
 where the user submits tasks to the hive
 which is responsible for orchestrating the drones
 based on a central data store.
The programmer writes a set of process definitions
 that include 
 an \emph{activation predicate}
 (used by the hive to determine how to allocate tasks to the drones)
 and a \emph{progress function}
 (used to determine progress towards the goal).
During a mission or sortie, a drone writes data to a \emph{scratchpad}, which is ultimately flushed to the data store at the hive.

\paragraph{Pleiades~\cite{bouget2018pleiades}}
It is a topology programming framework leveraging
self-organising overlays and assembly-based modularity~\cite{bruneton2006fractal-assembly-based}
to construct and enforce self-stabilising structural invariants
in large-scale distributed systems.
Shapes are described through templates specifying positions and neighbours for nodes; configurations of shapes are disseminated in the system and used by nodes for joining shapes; shape formation is regulated through protocols.
However, these features are not captured linguistically.
A simple example from the paper is a naive self-stabilising ring.
\resizebox{\textwidth}{!}{
\begin{minipage}{\textwidth}
\centering
\begin{align*}
E_{ring} & = [0,1[;  & \text{position space}
\\
f_{ring}(n) &= rand([0,1[); & \text{projection function (assigns nodes to positions)}
\\
d_{ring}(x,y) & = min(|x-y|,1-|x-y|); & \text{ranking function}
\\
k_{ring} & = 2 & \text{number of neighbours per node}
\end{align*}
\end{minipage}
}

\subsection{Collective adaptive systems and ensemble-based approaches}\label{approaches-ensemble}

Macroprogramming is also popular 
in the field of multi-agent (MAS)~\cite{DBLP:books/daglib/0023784} and collective adaptive systems (CAS)~\cite{Ferscha2015cas} engineering.
CASs approaches are quite related to spatiotemporal approaches
 since CASs are often situated and space represents a foundational structure for coordination.
In these approaches, it is common to consider large, dynamic groups of devices
 as first-class abstractions,
 which are commonly referred to as 
 \emph{ensembles}, \emph{collectives}, or \emph{aggregates}.
The general idea is to support interaction between (sub-)groups of devices
 by abstracting certain details away (e.g., membership, connections, concurrency, failure).
With respect to the network abstraction and other macroprogramming approaches,
 the works focus more on 
 addressing the specification of dynamic ensembles,
 do not take an explicit, spatial space
 or are not limited to data gathering and processing.

\paragraph{Aggregate programming~\cite{Viroli-et-al:JLAMP-2019}}
Aggregate programming is a \macrop{} paradigm,
 founded on field calculi~\cite{Viroli-et-al:JLAMP-2019},
 for expressing the decentralised, self-organising behaviour
 of a (spatiotemporally situated) distributed system.
It builds on the \emph{computational field} abstraction,
 a conceptually distributed data structure
 that maps any device of a system to a value, over time.
Then, macroscopic behaviour can be expressed
 in terms of a single program
 which manipulates fields
 through constructs for state management,
 neighbourhood-based interaction,
 and domain partitioning (i.e., the ability to run a computation on a subset of the system nodes).
Aggregate programming is supported by languages
 such as the Scala-internal DSL ScaFi~\cite{scafi}
 and the standalone DSL Protelis~\cite{protelis}.
For instance, the problem of counting, in any device,
 the number of neighbour devices 
 experiencing a high temperature can be expressed in ScaFi as follows:
\begin{lstlisting}[language={scafi}]
foldhood(0)(_+_)(if(nbr(sense("temperature"))) 1 else 0)
\end{lstlisting}
where \lstinline|foldhood(init)(acc)(f)| folds over the neighbourhood
of each device
 by aggregating the neighbours' evaluation of \lstinline|f|
 through accumulation function \lstinline|acc|, starting with \lstinline|init|.
The interesting aspect about aggregate programming
 is that it is possible to 
 capture collective behaviour into reusable \emph{functions}
 (from which libraries of domain-specific features can be defined)
 and 
 \emph{compose} functions ``from fields to fields''
 to define 
 increasingly complex behaviour.
For instance, the following \lstinline|channel| functionality
 reuses functions provided by the ScaFi library 
 to build a minimum-width path field 
 from a source to a destination device,
 which is -- crucially -- able to self-adapt to input changes (i.e., different source or destination) and topology changes (e.g., as devices move or leave the system).
\begin{lstlisting}[language={scafi}]
// source: input Boolean field (true only in the source device)
// target: input Boolean field (true only in the target device)
// width: input floating-point field for enlarging the channel
def channel(source: Boolean, target: Boolean, width: Double): Boolean = {
  distanceTo(source)+distanceTo(target) <= distanceBetween(source,target)+width
} // output: true if the device belongs to the channel, false otherwise
\end{lstlisting}%
Notice how this program abstracts from the individual devices at the micro-level:
 such a \lstinline|channel| function
 denotes a macro-level structure 
 that is sustained by repeated computation and interaction
 from the underlying network of devices.
In virtue of this flexibility, aggregate programming
 can be deemed a \emph{scalable} \macrop{} approach
 as it retains the ability to address individual devices
 but provides tools for raising the abstraction level.

\paragraph{Distributed Emergent Ensembles of Components (DEECo)~\cite{bures2013deeco}}
 DEECo is a CAS development model
 where components can only communicate
 by dynamically binding together through ensembles.
A DEECo component
 is an autonomous entity
 made of \emph{knowledge} (i.e., state),
 exposed to external world through \emph{interfaces} (providing a partial view of the state),
 and \emph{processes} (i.e., behaviour)
 which manipulate local knowledge, possibly perform side-effects, and are scheduled periodically or on-demand.
A DEECo ensemble 
 dynamically binds components 
 according to a \emph{membership condition} and
 consists of one \emph{coordinator} component 
 and multiple \emph{member} components
 interacting by implicit \emph{knowledge exchange}.
A DEECo application has the following structure:
{
\lstset{language={},morecomment=[l][\color{black!40!green}]{//}}
\begin{lstlisting}[alsoletter={-,.},morekeywords={component,interface,features,knowledge,process,in,out,
function,scheduling,knowledge-exchange,member,membership,
coordinator,ensemble}]
interface I1: field1, field2
interface I2: field3

component C1 features I1:
  knowledge:
    field1 = 77
    field2 = [ "a", "b" ]
  process P:
    in knowledge_field2, out knowledge_field1 function: // ...
    scheduling: triggered(changed(knowledge_field2))

component C2 features I2: // ...

ensemble E:
  coordinator: I1
  member: I2
  membership: f(coordinator.field1) && g(member.field3)
  knowledge-exchange: coordinator.field2 <- // ...
  scheduling: periodic (5000ms)
\end{lstlisting}
}
DEECo has also a Java implementation called jDEECo\footnote{\url{https://github.com/d3scomp/JDEECo}} 
 which enables the definition of components and ensembles
 through Java annotations.

\paragraph{Meld~\cite{Meld}}
Meld is a logic macroprogramming language for modular robotics, inspired by P2~\cite{loo2006p2-declarative-networking} (a declarative language for overlay networks).
It abstracts low-level coordination in robot ensembles
by taking a global perspective to programming:
macroprograms are compiled to microprograms
distributed to the individual robots.
It assumes robot interaction is only possible between immediate, in-contact neighbours.
In Meld, production rules are used to generate new facts from existing ones
 to possibly enable other rules (forward chaining);
 facts that are invalidated will be \emph{eventually} deleted;
 and, interestingly, \emph{aggregate rules} can be used to collapse multiple facts into one (e.g., by maximising/minimising or folding).
The runtime system at the robots 
 must deal with the sharing of facts via communication
 as well as the consistent deletion of facts.
An example, adapted from the paper, shows how to use Meld 
 to reach a destination using three robots where each robot is able to move only by rolling against other robots.
\begin{lstlisting}[language={Prolog}]
Nbr(a,b). Nbr(a,c). Nbr(b,c). 
At(a, (0,1)). At(b, (0,2)). At(c,(1,1.73)).
RobotRadius(<size>).

Nbr(A,B) :- Nbr(B,C), A=C, !. % reflexivity rule for Nbr

Dist(A, min<n>).
Dist(A,0) :- At(A,P), P = destination().
Dist(A,n+1) :- Neighbor(A,B), Dist(B,n). % NB: gets Dist for each neighbour!

Farther(S,T):- Neighbor(S,T), Dist(S,DS), Dist(T,DT), DS >= DT.

MoveAround(S,T,U):- Farther(S,T), Farther(S,U), U /= T.
\end{lstlisting}

\paragraph{Comingle~\cite{lam2015comingle-distributed-logic-programming-ensemble}}
Inspired by Meld, Comingle is a distributed, logic programming framework for systems of mobile devices.
In this model, devices are identified through a location
 and contribute to the system through \emph{located facts};
 the set of all located facts is called a \emph{rewriting state};
 \emph{rules} operate on rewriting states.
The rewriting semantics is global: it operates on a distributed data structure of located facts (contributed by the locations participating in the ensemble).

\paragraph{Scopes~\cite{jacobi2008structuring-wsn-scopes}}
Scopes is a macroprogramming language 
 for specifying the logical structure of WSNs.
It leverages the main abstraction of a \emph{scope},
 to represent a dynamic group of nodes (i.e., an ensemble).
Scopes can be created and deleted 
 through declarative, logical expressions. 
The middleware, then, is responsible for maintaining \emph{scope membership}.
The node that initiates the creation of a scope is called a \emph{root node}.
The scope supports bidirectional communication (according various routing algorithms supported by the framework) between the root node and the member nodes.
For instance, expression
\begin{lstlisting}[language={},morekeywords={CREATE,SCOPE,AND,
IN,SPHERE,EXISTS,SENSOR,AS}]
CREATE SCOPE temperatureNeighborhood AS 
 ((EXISTS SENSOR TEMPERATURE AND TEMPERATURE > 20)
   AND (IN SPHERE ( SPHERE (ROOT_NODE_POS, 30), NODE_POS)))
\end{lstlisting}
would create a scope selecting as members the nodes which have a temperature sensor providing a temperature of more than 20 degrees and are situated within a radius of 30 metres from the root node.

\paragraph{Attribute-based interaction in component ensembles}
A collection of formalisms have been proposed to address distributed adaptive systems through ensembles and attribute-based communication, i.e., 
 a style of interaction where recipient groups are dynamically determined via attributes.
Service Component Ensemble Language (SCEL)~\cite{DBLP:journals/taas/NicolaLPT14}
 is a kernel language to specify the behaviour of autonomic components, 
 the logic of ensemble formation,
 as well interaction through attribute-based communication
 (which enables implicit selection of a group of recipients).
A simpler process calculus inspired by SCEL is \emph{AbC (Attribute-based Communication)}~\cite{alrahman2015abc-calculus},
 capturing the essence of this interaction style.
AbC has been implemented for the Erlang programming language 
through the AErlang library~\cite{denicola2018aerlang}.
CARMA (Collective Adaptive Resource-sharing Markovian Agents)~\cite{carma} is a related stochastic process algebra and language that models collective of components
that may dynamically aggregate into ensembles,
 also using attribute-based communication (uni- or multi-cast) to
 implement broadcasts for coordinating large ensembles of devices.

\paragraph{TeCoLa~\cite{Koutsoubelias2016tecola}}
TeCoLa is described as a \emph{``programming framework for high-level coordination of robotic teams [...] with novel and unified abstractions for controlling individual robots as well as teams of robots [...] and with the bulk of team management work being performed behind the scenes''}.
TeCoLa has the following concepts:
 a \emph{node} has \emph{resources}
 and exposes \emph{services}, consisting of \emph{methods} and \emph{properties} and remotely invokable via \emph{proxies}.
A \emph{mission group} is a dynamic set of nodes participating in a mission, which is monitored and controlled by a \emph{coordinator} entity (e.g., a leader node or a command-control centre), and can be further split into \emph{teams} whose actual shape is automatically managed according to a \emph{membership rule} (e.g., on the set of services to be supported by team members).
When all the members of a team provide a certain service,
 that service is said to be \emph{promoted} at the team-level, meaning that it can be requested on the team itself, causing a corresponding service request on all the team members and yielding a vector of results.
TeCoLa has been implemented in Python, enabling macro-programs such as the following.
\begin{lstlisting}[language=python]
class TemperatureSvc(object):               # define a service
  __metaclass__ = Service
  def readTemperature(): # ...              # define a service operation
  def __activate(): # ...                   # hook method for service activation
  def __passivate(): # ...                  # hook method for service deactivation

co = Coordinator(...)                       # create coordinator

n1 = Node("TemperatureSensor")              # create node
n1 += TemperatureSvc()                      # configure node
# ...

co.group += n1                              # add node to mission group
# ...

t = co.defTeam(node=(any),                  # define a team based on membership conditions
               service="TemperatureSvc", property=(any))
ts = t.TemperatureSvc.readTemperature()     # invoke a team-promoted service (bulk operation)
for node, temp in temperatures.iteritems(): # do something
\end{lstlisting}
In this example,
 the set of collected temperatures depends on what nodes currently belong to team \lstinline|t|:
 these include nodes currently supporting the \lstinline|TemperatureSvc| service
 (which can in principle be dynamically de/activated according to conditions such as, e.g., the battery-level of the node).

\paragraph{Voltron~\cite{Mottola2014voltron}}
Voltron is defined as a \emph{team-level} programming model for drone systems.
It abstracts a set of individual drones
 through an abstraction of a \emph{team} of drones,
 which is tasked as a whole.
The specifics regarding what and when actions are performed by the individual drones is delegated to the platform system at runtime.
The programmer issues action commands to the drone team
 together with spatiotemporal constraints.
Indeed, tasks are actually associated to spatial locations, and the team self-organises to populate multisets of future values representing the eventual result of the task on a given location (i.e., it is similar to a computational field).
The paper provides an example of adaptive sampling of pictures in an archeological site, here reported with comments and minor adjustments.
\begin{lstlisting}[language=java,emph={spatial,foreachlocation,sceneChanged},
morekeywords={in}]
// definition of a spatial geometry as a filter of platform-generated Locations
boolean grid(Location loc) {
  if (loc.x % Drones.get("gridStep")==0 && loc.y % Drones.get("gridStep")==0) return true;
  else return false;
}

// spatial variable declarations
spatial Image tiles; // 1 or more Images, each associated to a different location

// loop program
do {  
  tilesOK = true; // loop flag
  tiles@grid = Drones.do("pic", Drones.cam, handleCam); // assign `tiles` according to grid locations
  foreachlocation (Location loc in tiles){  
    float aberr = abberationEstimate(tiles@loc);
    if (aberr > MAX_ABR && !sceneChanged) {
      Drones.stop(handleCam);
      int newStep = tune(Drones.get("gridStep"),aberr);
      Drones.set("gridStep",newStep);
      tilesOK = false;
      break;
    }
  }   
} while (!tilesOK);
\end{lstlisting}

\paragraph{PaROS (PROgramming Swarm)~\cite{paros}}
PaROS is a framework for programming
 swarms of robots.
It proposes an \emph{abstract swarm} abstraction,
 implemented through a Java API,
 to promote swarm orchestration and spatial organisation.
The API consists of functions for: 
 path planning,
 declaration of points of interest or spatial areas to be inspected,
 enumeration of the robots in the swarm,
 task partitioning,
 setting handlers for detection events or robot failure.
A program in PaROS looks like the following.
\begin{lstlisting}[language={Java}]
// Build a swarm from a set of drones
Swarm swarm = new Swarm(setOfDrones);
// Create flight plans by splitting an area and assigning sub-areas
swarm.areaDeclaration(targetArea);
// Define a collective task
swarm.setTask(Task.COVERAGE);
// Adds a handler for event detection
swarm.eventHandler((drone) -> { System.out.print("Event detected by " + drone); });
// Starts the mission: will run pathPlanning() and droneManipulation()
swarm.startMission();
// While the mission is running...
while(swarm.isMissionRunning()){
  for(Drone drone : swarm.getListOfDrones()){
    if(drone.isTaskComplete()){
      doSomethingWith(drone.getCameraImage(camera));
    }
  }
}
\end{lstlisting}
Many details regarding the coordination of the swarm 
 are abstracted away.
Therefore, PaROS promotes a multi-paradigm approach
 comprising elements from
 imperative, declarative, and event-driven programming.

\paragraph{SmartSociety platform~\cite{scekic2017hybridcas}}
This is
 a programming model of SmartSociety 
 for hybrid collaborative adaptive systems is proposed
 in which the designer specifies an environment where 
 collectives---i.e., persistent or transient teams of peers 
  (humans and machines)---are involved in collective tasks.
The approach can be used for applications
 involving crowdsourcing, human orchestration, and collective activities.
\begin{lstlisting}
TaskRequest r = new RideRequest(...);
CollectiveBasedTask cbt = CBTBBuilder.from(
    TaskFlowDefinition.usingContinuousOrchestration(...)
  ).withTaskRequest(r).build();
cbt.start();
\end{lstlisting}
The macro-abstraction of the \lstinline|CollectiveBasedTask|
 encapsulates team formation,
 plan execution and composition,
 as well as other collective activities like negotiation and incentivisation.

It is common for IoT programming frameworks and middlewares 
 provide a \emph{centralised view}
 of the entire IoT system
 and hence support a form of macroprogramming.

\paragraph{EcoCast~\cite{tu2011ecocast}}
EcoCast is an interactive, object-oriented/functional macroprogramming framework for Python.
Its basic idea is to \emph{``extend the concept of functional programming on lists of data to macroprogramming on groups of nodes''}~\cite{tu2011ecocast}.
It uses a particular kind of object, called a \emph{group handle},
 as a proxy for a group of nodes.
This group is static, i.e., it does not automatically deal with group membership like in ensemble-based approaches. An example of EcoCast code follows.
\begin{lstlisting}[language=Python,emph={ecFilter,ecMap,ecReduce}]
single_node = ecNode(77) # instantiates a node handle with ID=123
group = ecGroup([1,2,3,single_node]) # creates a group handle with nodes of IDs=1,2,3,77

dangerous_nodes = ecFilter(lambda x: x > TEMP_THRESHOLD, group, read_temperature)
ecMap(issue_warning_action, dangerous_nodes)
max_temperature = ecReduce(max, group, read_temperature)
\end{lstlisting}
From these lines, it is visible a ``node-'' and ``group-to-object mapping'' approach.
So, it follows a modern approach where object- and functional-orientation coexist to provide convenient APIs.
Operationally, EcoCast attempts to parallelise execution of actions and communications.

\paragraph{Organisation-oriented programming, MOISE~\cite{DBLP:journals/ijaose/HubnerSB07}}
Organisation-oriented programming
 is a  \macrop{} approach
 where an \emph{organisation} of agents
 is considered as a first-class entity.
In MOISE, a multi-agent system (MAS)
 is described through multiple specifications.
A structural specification defines the structure of a 
 MAS at the individual level (with \emph{roles}),
 at the social level (with \emph{links}, namely relationships between roles),
 and at the collective level (with \emph{groups}, defined in terms of roles, links, and constraints).
Similarly, 
 a MAS is functionally specified
 at the individual level (with \emph{missions}, namely goals assigned to roles)
 and at the collective level (with \emph{schemes}, namely trees of goals assigned to groups).
Such specifications are expressed through modelling languages.
During execution, mediated by the ``organisational platform'',
 agents join or leave groups,
 commit to missions in their groups' schemes,
 can reason about organisation entities,
 and are enforced to follow the specifications.

\subsection{Ad-hoc approaches}

\paragraph{Market-Based Macroprogramming (MBM)~\cite{Mainland2004mbm}}
In MBM, a sensor network is programmed as a \emph{virtual market}.
The nodes of the network
 follow a fixed behaviour protocol
 where they
 ``sell'' \emph{actions}
 to get a \emph{profit}.
They choose actions according to a local \emph{utility function} that expresses a trade-off between the profit and the \emph{cost} of performing the action.
For instance, the value of reading a sensor value or communicating with another agent could be in contrast
with time, energy or bandwidth expenses.
The macro-program defines the logic of updating the globally-advertised \emph{price} of the actions
to foster the desired global behaviour
by driving the choice of the actions useful for the current situation.

\label{sec:netkat-snap}
\paragraph{NetKAT~\cite{Anderson2014netkat}, SNAP (Stateful Network Abstractions for Packet processing)~\cite{Arashloo2016snap} etc.}
NetKAT and SNAP (which derives from NetKAT) are languages for software-defined networking (SDN)
 that are stateful
 and consider the network as \emph{``one big switch''}~\cite{DBLP:conf/conext/KangLRW13}.
The NetKAT language is based on KAT (Kleene Algebra with Tests) plus constructs for networking.
Conceptually, a macro-program in these languages
 is a function of a packet and network state (represented through global variables)
 that produces a set of packets and a new network state as output.
In practice, a program consists of the classical imperative constructs (assignment, conditionals, loops) which are however interpreted in the SDN domain.
The compiler translates the macro-program into micro-programs for the network devices
 dealing with \emph{traffic routing} and \emph{placement of state variables}.

\paragraph{Wave-Oriented Swarm Programming (WOSP)~\cite{varughese2020swarm-wosp}}
WOSP is an approach for swarm-level programming 
 that requires minimalistic communication,
 inspired by two biological mechanisms:
 (i) scroll waves in slime mould and 
  (ii) periodic light emission in fireflies.
Each robot of the swarm 
 follows a protocol where it is initially \emph{inactive}, 
 listening for incoming pings; 
 upon reception of a ping, it runs a ``relay code block''
 and goes into an \emph{active} state 
  where it emits a ping;
  after the emission of a ping,
  it goes in the \emph{refractory state},
  where it does nothing, being insensible to pings,
  and finally turns back to the inactive state
  after a refractory period.
On each state transition, the robot decrements an internal timer (randomly initialised) and performs the corresponding logic for the current state;
 after that, it checks if the timer has hit zero,
 and in case it runs an ``initialisation code block'' 
 for resetting the robot (state).
As an example, consider a simple ``leader election'' behaviour, whose pseudocode from the paper has been slightly adjusted.
\begin{lstlisting}[language={java}, morekeywords={function}]
function Initiate_Codeblock { 
  candidate <- true; 
  timer <- random(TIMER_MAX); 
}

function Relay_Codeblock { 
  timer <- null; // deactivate the timer
  candidate <- false; 
}
\end{lstlisting}
In this program, the idea is that any time a robot receives a ping before its internal timer has elapsed, it runs \lstinline|Relay_Codeblock| to withdraw from the election process.
In case no ping has been received,
 it can safely assume it has no rivals.
However, proper parameterisation is needed to ensure 
 a single leader remains after multiple election rounds.
To wrap up, though no evident macroprogramming abstraction is used at the code-level,
 the approach manages to steer collective behaviour
 by exploiting randomness, parameterisation, and simple local behaviours.

\paragraph{Bayesian Network-based Macroprogramming (BNM)~\cite{DBLP:journals/taas/Mamei11}}
In this work, Bayesian networks are used as macroprogramming language for spatially distributed systems.
The idea is to design Bayesian sub-networks 
 to capture functional requirements of the application
 and to deploy multiple copies of these Bayesian sub-networks
 to corresponding portions of the underlying network of devices.
Such a replication 
 results in the full Bayesian network
 being deployed in the distributed systems,
 hence supporting Bayesian inference in a distributed fashion (in turn enabling prediction, diagnosis, and anomaly detection).
The deployment is based on the connection of input and output \emph{ports} of the functional Bayesian subnetworks and of devices together.
Crucially, the details of distributed execution of Bayesian inference are abstracted by the macroprogram,
 which focusses only on the Bayesian subnetwork definitions (i.e., random variables and corresponding relationships) and logical connectivity (through ports).
As an example, took from the paper~\cite{DBLP:journals/taas/Mamei11}, consider the following functional Bayesian subnetwork for inferring the luminosity of an environment.
\begin{center}
\footnotesize
\tikzset{
  n/.style  = {draw, font=\sffamily\footnotesize
, rectangle}, auto
}
\begin{tikzpicture}[node distance=0.2cm and 1.4cm]
\node[n] (light) [label={\texttt{P(\textbf{Light} is \textbf{High})=0.3}}] {Light};
\node[n,fill=black!20] (nbr) [right=2cm of light, label=70:{\texttt{P(\textbf{Light}==\textbf{Neighbour Light})=0.9}}] {Neighbour Light};
\node[n] (measure) [below=0.6cm of light, label=below:{\texttt{P(\textbf{Light} is \textbf{High} given \textbf{Light} is \textbf{High})=0.8}}] {Measure};

\draw[->] (light) -- (nbr);
\draw[->] (light) -- (measure);

\end{tikzpicture}
\end{center}
In a concrete system, the above Bayesian sub-network will be deployed on each light sensor device:
the sensor will provide the light measurement as input to \texttt{Measure}, and the \texttt{Light} node will provide its output to the input of a neighbour's \texttt{Neighbour Light} node.

\paragraph{Graph-centric programming~\cite{giraphpp}}
Giraph++ is a framework for distributed graph processing
 that embodies a ``think like a graph'' programming model.
This is in contrast with traditional graph-processing approaches 
 following a ``think like a vertex'' paradigm.
Indeed, Giraph++ exposes the first-class concept of 
 a \emph{graph partition}
 that allows 
 to (i) obtain the vertices included in that partition,
 (ii) send data to or act over all the included vertices,
 and
 (iii) implement a user-defined function \texttt{compute()}
 operating on the whole graph partition.
According to the authors,
 this abstraction would enable various sorts of optimisations
 which were prevented by the vertex-centric approaches.

\section{Analysis and Outlook}
\label{s:opp-ch}

In this section, we 
 analyse some data from the survey (\Cref{s:data-trends}),
 the surveyed approaches by a technical point of view  (\Cref{s:technical-analysis}),
 and then
 review significant opportunities (\Cref{s:opportunities}) 
 and challenges (\Cref{s:challenges})
 related to \macrop{}.

\newcommand{\Nconsidered}[0]{66}
\newcommand{\Nincl}[0]{49}
\newcommand{\Ncore}[0]{39}
\newcommand{\NexplicitlyMacro}[0]{18}
\newcommand{\NwithRepo}[0]{18}

\subsection{Data and Trends}\label{s:data-trends}

In this survey,
 we have considered a total of \Nconsidered{} \emph{works},
 and have \emph{included} (i.e., considered as a \macrop{} approach, after manual analysis)
 \Nincl{} \emph{works}, of which \Ncore{} \emph{core} works have been identified
 (i.e., some approaches have been implemented through multiple published languages or DSLs) corresponding to the number of rows of \Cref{table:results-summary}.

\begin{figure}
\begin{minipage}{0.48\textwidth}
\includegraphics[width=\textwidth]{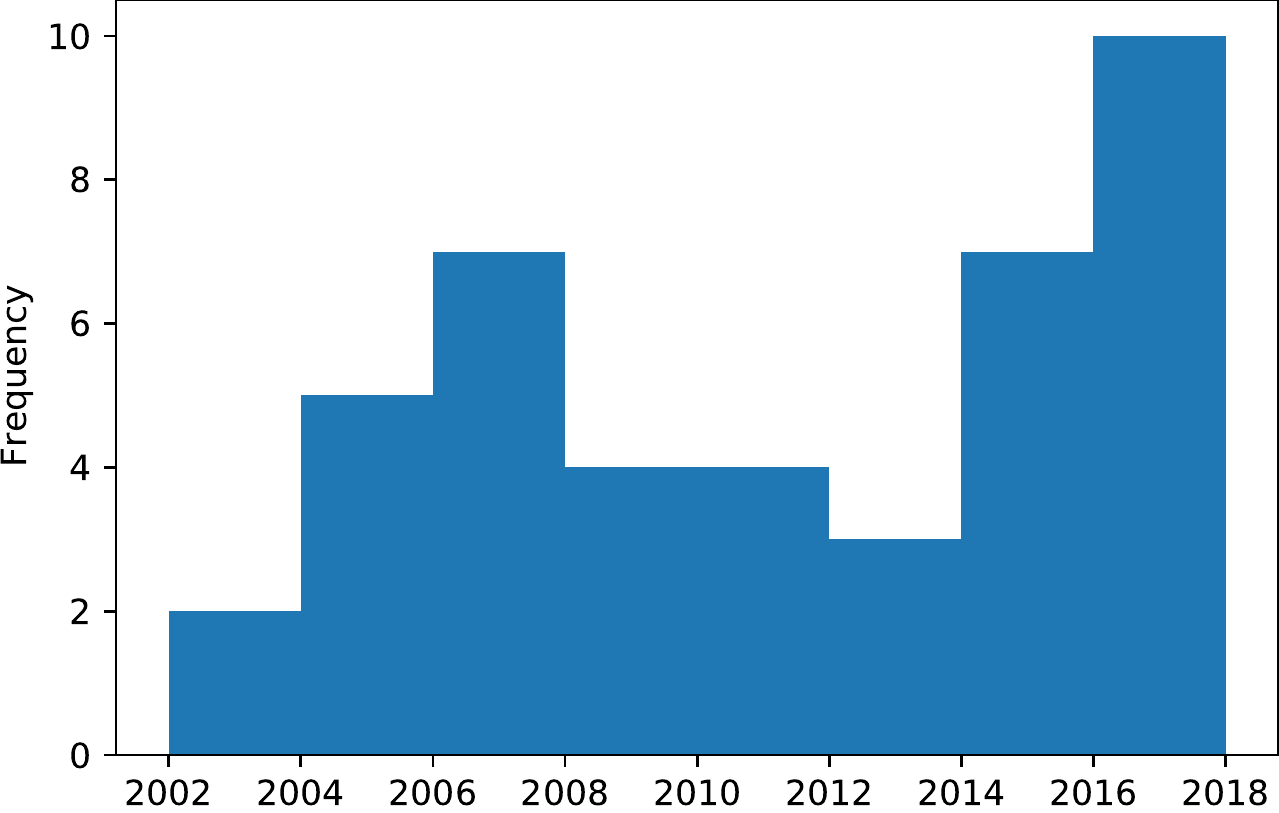}
\subcaption{\label{fig:year-histogram}Number of works per two-year periods.}
\end{minipage}\hfill
\begin{minipage}{0.48\textwidth}
\includegraphics[width=\textwidth]{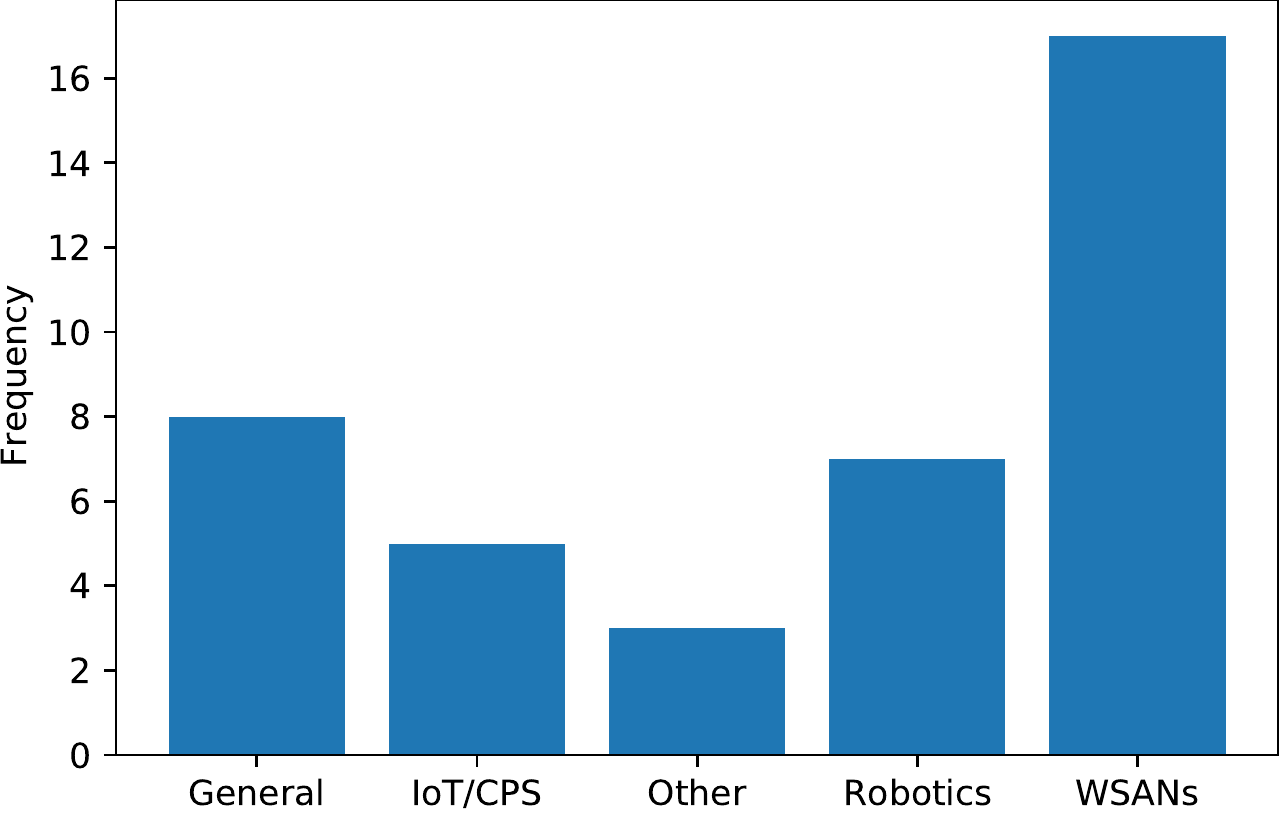}
\subcaption{\label{fig:domain-histogram}Number of core works per domain.}
\end{minipage}
\\[0.2cm]
\begin{minipage}{0.48\textwidth}
\includegraphics[width=\textwidth]{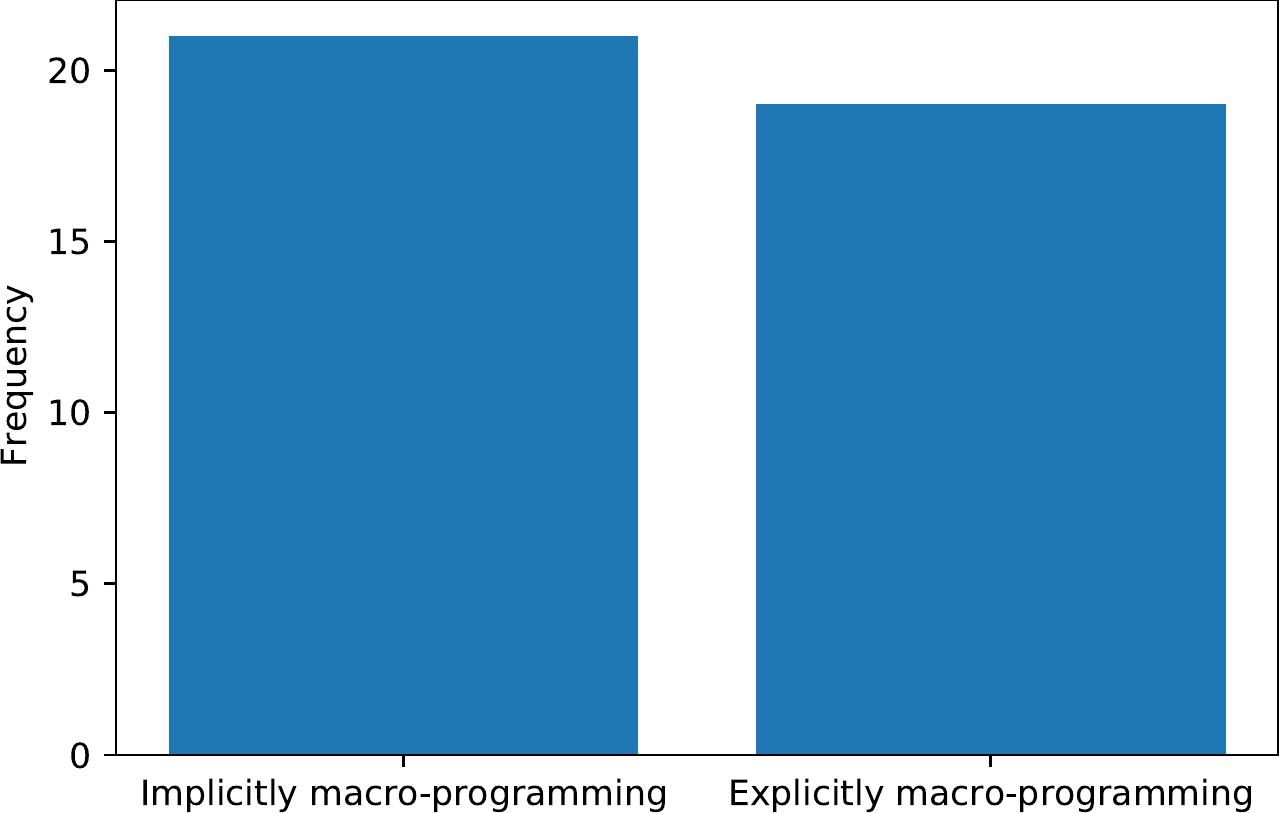}
\subcaption{\label{fig:explicitly-macrop}Number of core works that explicitly advertise themselves as \macrop{}.}
\end{minipage}\hfill
\begin{minipage}{0.48\textwidth}
\includegraphics[width=\textwidth]{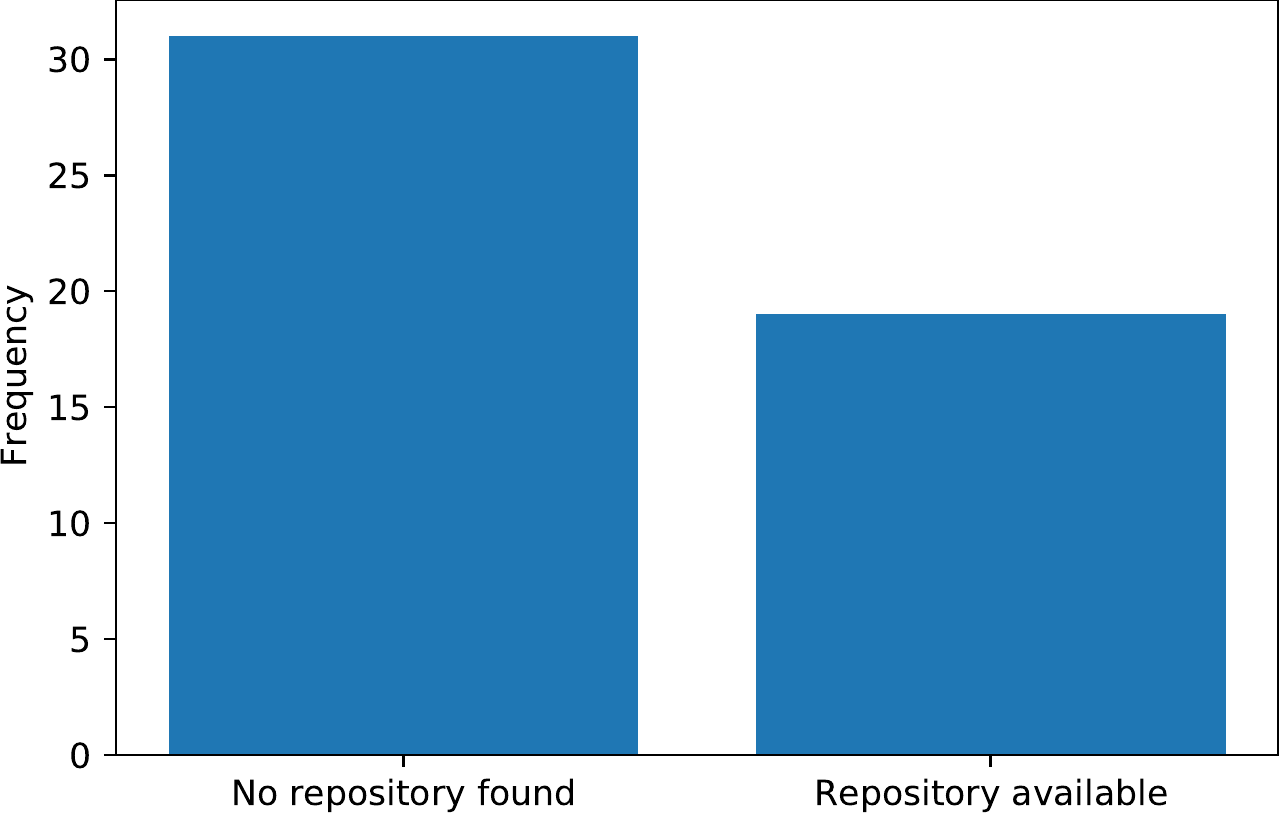}
\subcaption{\label{fig:accessibility-histogram}Number of works with publicly accessible software (e.g., through a website or repository).}
\end{minipage}
\caption{\label{fig:freqs}Collected data about non-technical aspects, from the survey.}
\end{figure}

The distribution of the included works
 by (publication) year is reported in \Cref{fig:year-histogram}.
From this histogram,
 we observe the rise of \macrop{}
 from WSN research in early 2000s,
 a loss of hype in early 2010s,
 and a new wave from 2014
 as a result of recent trends and developments
 in fields like the IoT, CPSs, and CASs
 (cf. \Cref{sec:domains}).
The distribution of works across domains 
 is shown in \Cref{fig:domain-histogram},
 where we observe a predominance of the WSN domain;
 the domain fragmentation seems to be a characteristic
 of the second wave of \macrop{}.
Another interesting datum
 is how many of the surveyed works
 explicitly advertise themselves as \macrop{}:
 according to \Cref{fig:explicitly-macrop},
 this is only the case for \NexplicitlyMacro{} out of \Ncore{} core works.
 
Another significant aspect concerns the availability of 
 accessible software for a \macrop{} language.
According to \Cref{fig:accessibility-histogram},
 the number of works for which
 a repository or website exists
 that provides access to software
 is \NwithRepo{} out of \Nincl{} works.
Arguably, this low score is partially due to 
 to the limited practice 
 of providing artifacts in early 2000s,
 as well as to the obsolescence of some of the proposed languages
 from those years.

\subsection{Analysis of \Macrop{} Approaches}
\label{s:technical-analysis}

\subsubsection{\revision{Paradigms}}
\begin{figure}
\centering
\begin{minipage}{0.48\textwidth}
\includegraphics[width=\textwidth]{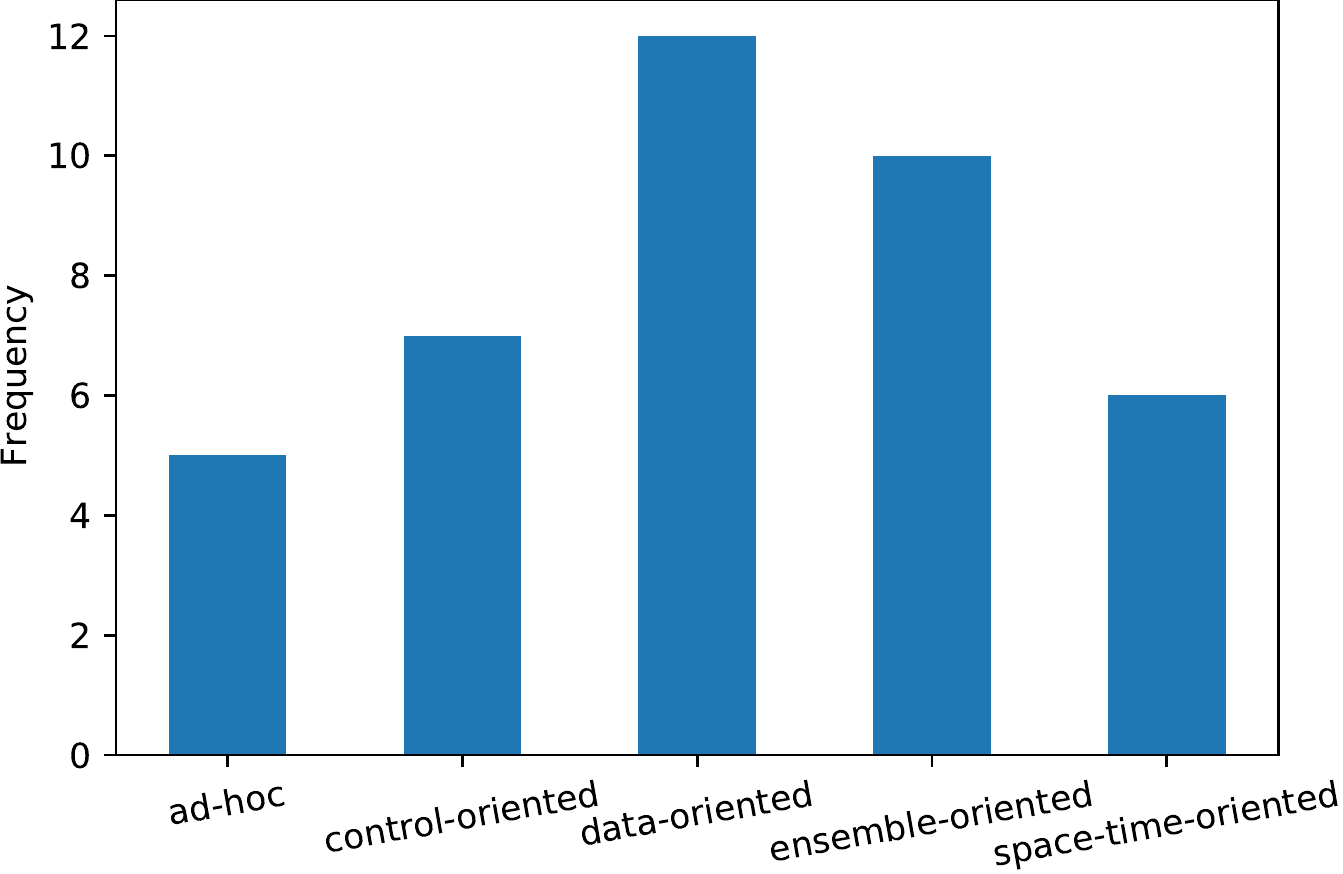}
\subcaption{\label{fig:approach-histogram}Number of core works per approach category.}
\end{minipage}\hfill
\begin{minipage}{0.48\textwidth}
\includegraphics[width=\textwidth]{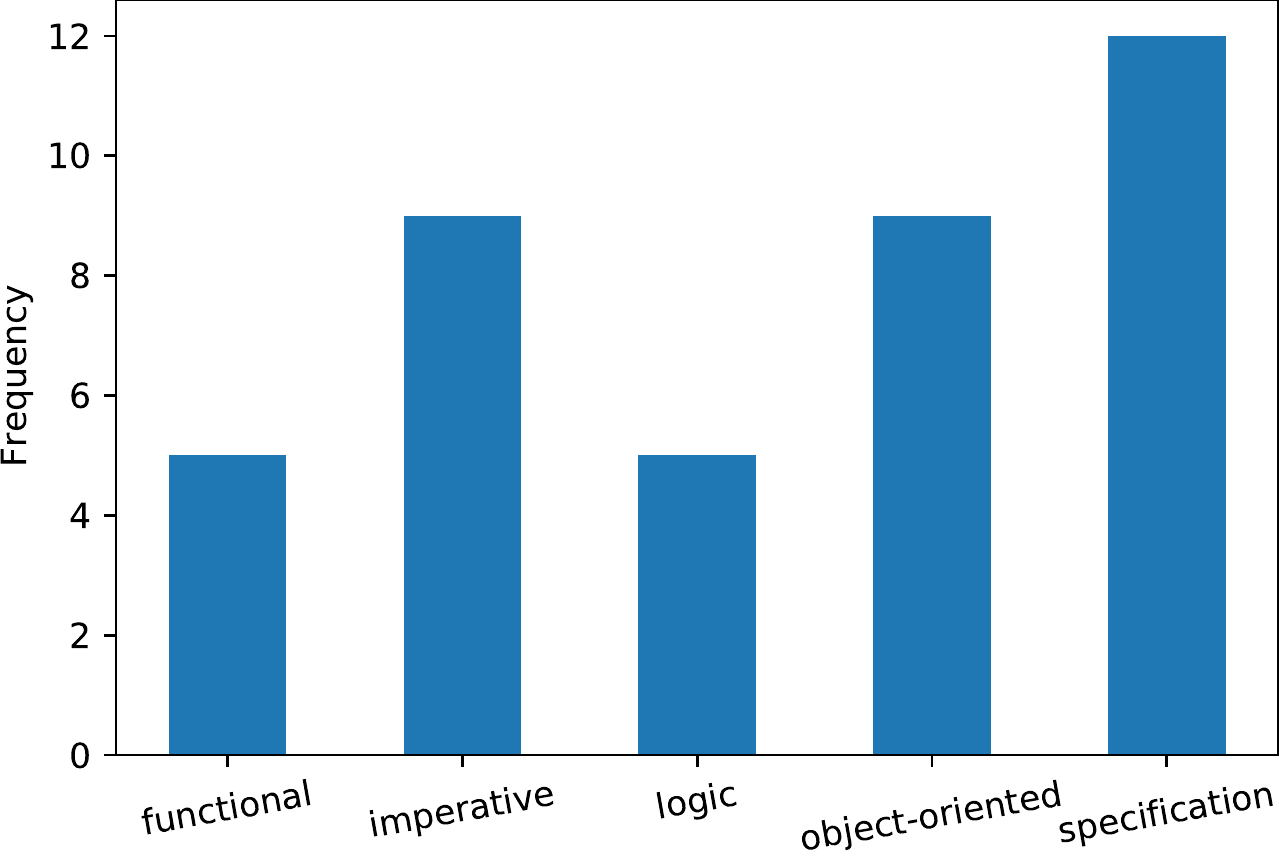}
\subcaption{\label{fig:paradigm-histogram}Number of core works per paradigm.}
\end{minipage}\\[0.2cm]
\begin{minipage}{0.48\textwidth}
\includegraphics[width=\textwidth]{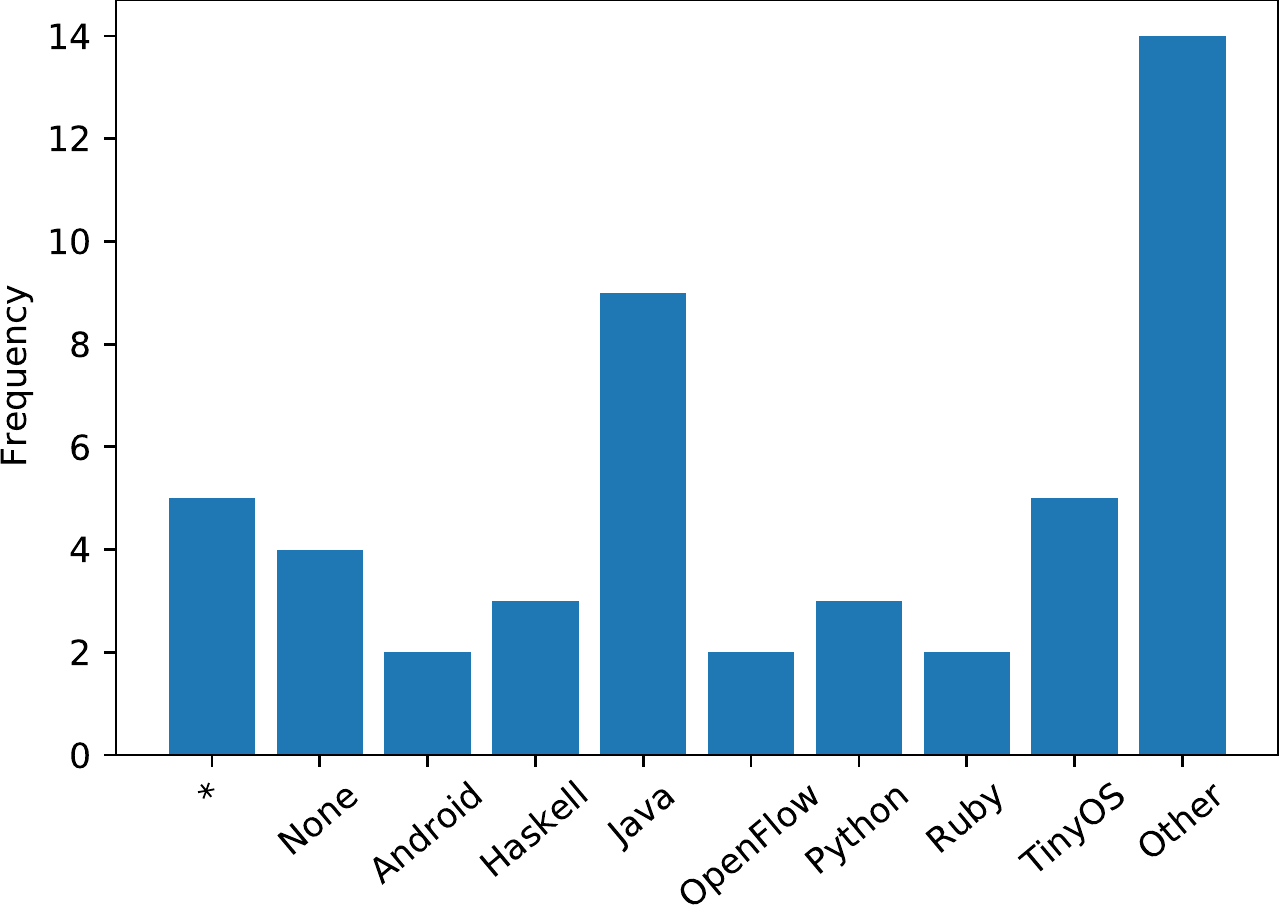}
\subcaption{\label{fig:host-histogram}Number of works per host platform/language. The {(*)} means that multiple languages are supported.}
\end{minipage}
\hfill
\begin{minipage}{0.48\textwidth}
\includegraphics[width=\textwidth]{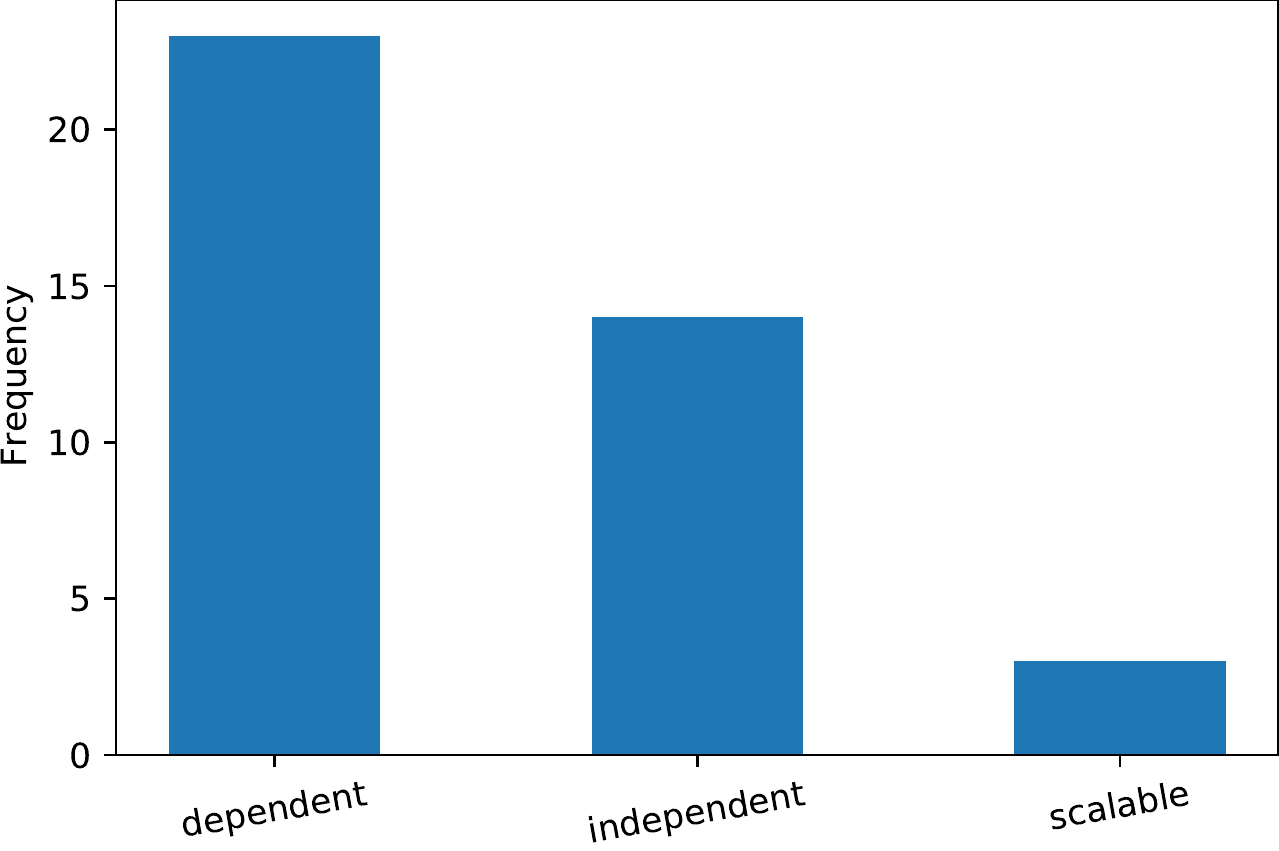}
\subcaption{\label{fig:microdep-histogram}Number of core works per micro-level dependency level.}
\end{minipage}
\\[0.2cm]
\begin{minipage}{0.48\textwidth}
\includegraphics[width=\textwidth]{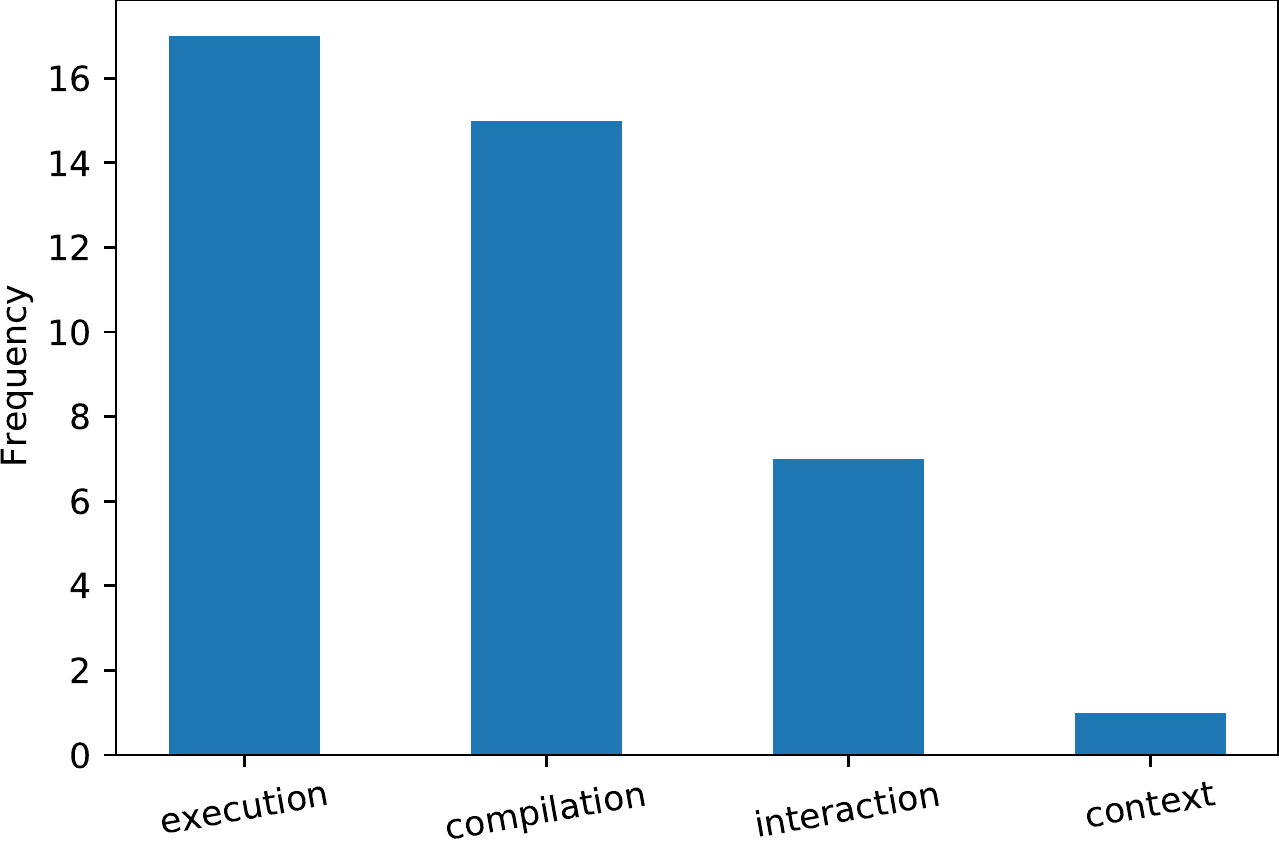}
\subcaption{\label{fig:macrotomicro-histogram}\revision{Number of core works per macro-to-micro mapping approach.}}
\end{minipage}
\caption{\label{fig:freqs}Collected data about technical aspects of the surveyed \macrop{} approaches.}
\end{figure}

The distribution of \macrop{} languages
 across the approach clusters and the basic paradigms (cf. \Cref{sec:taxonomy})
 is shown in \Cref{fig:approach-histogram}
 and \Cref{fig:paradigm-histogram}, respectively.
Apparently, the majority of works follow a data-oriented approach;
 arguably, this reflects the fact that 
 most of the works target the WSN domain,
 where the main goal is to extract data from the sensor network.
This is also coherent with the fact that 
 most of the \macrop{} languages 
 take a declarative, specification stance.
Beside that, all the main paradigms (logic, functional, imperative, object-oriented) have a discrete number of representatives---showing the orthogonality of the macro viewpoint to programming,
as well as the consequence of embedding .
On the other hand, 
 only a handful of works take an ad-hoc approach
 that could not be framed as either control-, data-, space-time-, or ensemble-oriented. 
 
\subsubsection{\revision{Underlying platforms and languages}}
In \Cref{fig:host-histogram},
 there is a hint about the underlying platforms or languages
 in which a given \macrop{} is supported or implemented.
We denote with ``*'' that an approach 
 supports multiple target platforms;
 with ``None'' that no implementation is provided or described;
 and we use a label ``Other'' to collect 
 platforms for which only a single occurrence exists
 (this is the case, e.g., of embedded platforms, simulators, or individual languages such as Embedded Matlab, PeerSim, or Groovy, respectively).
Several approaches found on the Java language and platform;
various approaches
 target TinyOS, an operating system for WSN motes;
however, it is most frequent to address specific platforms
 (as a reflection of a rather wide coverage of target domains and paradigms).

\subsubsection{\revision{Micro-level dependency}}
In \Cref{fig:microdep-histogram},
 we observe that the majority of works
 do depend on micro-level entities
 (e.g., because they need to be addressed individually);
 however, there are also many works that 
 abstract completely from the underlying set of components.
Not surprisingly, 
 only a few approaches
 are ``scalable'',
 allowing both to address individual nodes
 as well as to abstract from them entirely:
 these include Regiment~\cite{regiment} and
  aggregate programming~\cite{Viroli-et-al:JLAMP-2019}. 
Indeed, the main goal of the surveyed \macrop{} approaches 
 is to provide zero-cost abstractions
 to simplify the programming activity
 without renouncing to performance.
Moreover, in several cases,
 the programming models aim to provide 
 specific benefits:
 communication or execution efficiency (cf. WSN programming models),
 dynamic binding and architectures (cf. Scopes, DEECo, Warble, TECOLA),
 and self-organisation (cf. aggregate programming, Pleiades).
 
\subsubsection{\revision{Macro-to-micro mapping}}
\revision{As it can be observed from \Cref{fig:macrotomicro-histogram},}
The implementation of macro-to-micro mapping 
 is generally based on either \emph{compilation}
 of the macro-program into the programs for the individual nodes
 (also known as \emph{deglobalisation} or \emph{global-to-local compilation})
 or \emph{interpretation} of the macroprogram
 according to some \emph{execution protocol}
 (e.g., involving migratory execution of agents or
 orchestration of individuals).
\revision{
Quite frequent 
 is also the approach
 based on \emph{orchestration}, such as in Dolphin~\cite{lima2018dolphin} or SmartSociety~\cite{scekic2017hybridcas},
 or the definition of additional entities
 like \emph{mobile agents}
 which interact with micro-level entities 
 to promote desired emergents,
  as in PIECES~\cite{liu2003state-centric-program-wsan-pieces} or STOP/Chronus~\cite{wadaa2010chronus-spatiotemporal-macroprogramming-wsn}.
The less frequent mechanism 
 is ``context change'',
 e.g., parameter setting
 as found in WOSP~\cite{varughese2020swarm-wosp}
 or market-based macroprogramming~\cite{Mainland2004mbm},
 though these may not even be considered a ``programming approach'', strictly speaking.
}

\revision{
Unfortunately, the macro-to-micro mapping 
 is often not described formally (or even explicitly),
 which exceptions
 like SCEL~\cite{DBLP:journals/taas/NicolaLPT14},
 and aggregate programming~\cite{DBLP:journals/computer/BealPV15}.
For instance, in the latter approach, the core language -- namely the field calculus --
 has a macro-level denotational semantics
 linked 
 to the local operational semantics~\cite{Viroli-et-al:JLAMP-2019},
 using computational fields (global data structures) as bridging abstraction.
}
 
\subsubsection{\revision{\Macrop{} abstractions}}
Finally, we can observe that a number of abstractions 
 or features
 recur in \macrop{} approaches. These include:
\begin{itemize}
\item \emph{first-class groups}---the ability to directly express and manipulate groups of individuals (cf. group handles in EcoCast, swarms in Buzz or PaROS);
\item \emph{group lifecycle management}---the ability to evolve groups dynamically (cf. dynamic binding in Warble);
\item \emph{group addressing}---the ability to address a group, e.g., in terms of the individuals found in a certain spatial region
or that share certain capabilities (cf. Regiment, SpatialViews, STOP/Chronus);
\item \emph{distributed state}---the ability to address the state of a group of stateful entities (cf. fields in aggregate programming, state rewriting in Comingle);
\item \emph{group inspection}---the ability to inspect or iterate over the individuals of a group (cf. node iteration in Kairos, resources in PyoT);
\item \emph{group goal decomposition}---the ability to consider a global goal
 and ways to split it across multiple individuals (cf. task partitioning in PaROS, spatial decomposition in Karma);
\item \emph{group communication}---the ability to get data from or push data to a group (cf. report/tell/collective actions in makeSense, 
 or neighbourhood-based communication in COSMOS, and aggregate programming);
\item \emph{information flow patterns}---the ability to specify 
 how information should flow independently 
 of structure or concrete communication mechanisms
 (cf. ATaG);
\item \emph{group-level actions}---the ability to express \emph{what} a group should do (cf. functions in aggregate programming, activity nodes in DDFlow).
\end{itemize}
Sometimes, some of these aspects are abstracted away
 and implemented at the middleware layer:
 for instance,
 approaches that consider a WSN ``like a database''
 let the programmer express a query (global goal)
 and
 then handle its partitioning into micro-actions through underlying mechanisms and execution protocols.

\revision{
\subsubsection{On implementation and abstracted concerns}\label{ssec:impl-abstracted-concerns}
A major goal of \macrop{}
 is abstracting from 
 a series of low-level concerns.
This is also strikingly evident
 from the quotes reported in \Cref{table:defs},
 which suggest that the programmer
 can be relieved from
 ``explicitly managing control, coordination, and state maintenance at the individual node level''~\cite{DBLP:conf/iwdc/BakshiP05}
 in order to retrieve ``simplicity and productivity''~\cite{DBLP:conf/hicss/WadaBS08}
 through ``focus on application specification rather
than low-level details or inter-node messaging''~\cite{awan2007cosmos}.
Besides productivity,
 there is also the idea that
 low-level details 
 can be addressed efficiently or opportunistically
 at the middleware level---see \Cref{opportunity-middleware} for further considerations on this point.

The concerns that are abstracted may be classified 
 according to the fundamental dimensions of 
 structure, behaviour, and interaction.
\emph{Structural concerns}
 include connections between components
 and membership relationships.
As these elements tend to change dynamically,
 expressing them in a declarative fashion
 enables the underlying platform 
 to adopt flexible strategies for their reification.
For instance, in Buzz~\cite{pinciroli2016buzz-swarm-programming},
 each robot follows a protocol 
 to keep track of its membership in swarms,
 which further affects the set of its neighbours.
In ScaFi~\cite{DBLP:journals/eaai/CasadeiVAPD21},
 groups self-organise 
 by playing the logic expressed by the macroprogram
 in repeated sense-compute-interact rounds
 to continuously evaluate the ``spatiotemporal boundary''
 of the process/ensemble.
Therefore, we may conclude that often
 the macro-program is a piece of behaviour
 that is used to parameterize 
 a larger behaviour,
 supported by a proper runtime system or middleware,
 which provides the ``basic principle''
 for the collection of micro-level entities
 to act as a system.

\emph{Behavioural concerns} 
 that can be abstracted
 include specific decisions
 (e.g., what data must be stored or propagated),
 processing operations,
 and time aspects
 (e.g., when a certain behaviour is to be executed).
For instance, in SNAP~\cite{Arashloo2016snap},
 the individual switches 
 must determine how to route traffic and
 where the place state variables.
As another example, macro-programs in aggregate computing~\cite{DBLP:journals/computer/BealPV15},
 abstract from scheduling aspects,
 which enables dynamic tuning 
 of the frequency at which devices operate, making time a ``fluid'' notion in such systems~\cite{DBLP:journals/corr/abs-2012-13806}.
 
\emph{Interactional concerns}
 are also often abstracted.
In many cases, indeed, the details of communication,
 such as the specific format of the messages,
 the specific set of recipients,
 can be determine at runtime.
\Macrop{} approaches for WSN, for instance,
 generally provide abstractions 
 over routing
 and hop-by-hop information flows.

Among implementation strategies,
 a number of patterns recur
The macroprogram
 can, 
 as in PIECES~\cite{liu2003state-centric-program-wsan-pieces} or STOP/Chronus~\cite{wadaa2010chronus-spatiotemporal-macroprogramming-wsn},
 instruct mobile agents
 to move across the nodes of the network
 to access and process local state 
 to infer global information.
Orchestration -- cf. Dolphin~\cite{lima2018dolphin} and SmartSociety~\cite{scekic2017hybridcas} -- is similar
 but does not involve 
 moving agents.
Related is also 
 the approach, used for instance in Pyot~\cite{Azzara2014pyot-macroprogramming-iot},
 based on inferring tasks from the macroprogram
 and distributing them over the set of micro-level entities.
Round-based execution of macro-programs or projected micro-programs is also frequent and 
 can be found
 both in asynchronous variants,
 as in aggregate computing~\cite{DBLP:journals/computer/BealPV15},
 and in synchronous variants
 as in Giraph++~\cite{giraphpp}, SOSNA~\cite{karpinski2008stream-macro-wsan-sosna},
 and WOSP~\cite{varughese2020swarm-wosp}.
Such implementation strategies, 
 beside ``filling the abstraction gap'',
 are also aimed at optimising application-specific concerns,
 which may include 
 saving resources (e.g., energy or bandwidth)
 or promoting Quality of Service metrics like latency or reliability.
}

\subsection{Opportunities}
\label{s:opportunities}

\revision{
Research on \macrop{} 
 provides opportunities (with corresponding challenges, covered in~\Cref{s:challenges})
 in terms of synergies with related research fields
 and application domains.
}

\subsubsection{Model- and Language-based Software Engineering}
\label{lang-based-se}
Models, 
 as abstract representations of some aspect of a real or imagined object,
 play a key role in software engineering~\cite{DBLP:journals/ife/Ludewig04}.
Systems are generally described through multiple models
 covering different perspectives or viewpoints~\cite{DBLP:journals/ijseke/FinkelsteinKNFG92}.
As covered in~\Cref{sec:domains,sec:framework,sec:survey},
 \emph{macroscopic perspectives} 
 can be instrumental 
 for directly addressing 
 global properties, collective tasks, or system-level aspects.
Indeed, we can observe that prominent perspectives
 in software modelling
 (e.g., structure, behaviour, and interaction)
 can be considered at a microscopic or a macroscopic level. 
For the latter:
\begin{itemize}
\item The \emph{macro-structural view} considers the structural arrangement of multiple components of a system.
The creation of macro-structures is sometimes a goal of \macrop{} (cf. topology programming in Pleiades~\cite{bouget2018pleiades}).
\item The \emph{macro-behavioural view} considers behaviours emerging from multiple components of a system. This is generally the goal of \macrop{}: expressing the behaviour of a system as a whole (cf. swarm \macrop{} in Buzz~\cite{pinciroli2016buzz-swarm-programming}). 
\item The \emph{macro-interactional view} considers interaction and communication at increasingly non-local levels. This is often instrumental to drive macro-behaviour by expressing how information flows among several components or across large structures (cf. collective communication interfaces in Abstract Regions~\cite{welsh2004abstract-regions}).
\end{itemize}
Models have to be expressed in some language (also called a meta-model).
Languages exist for specification,
 design,
 implementation,
 and verification of software,
and contribute to a vision of \emph{language-based software engineering}~\cite{DBLP:journals/scp/Gupta15},
which promotes the use of high-level DSLs 
for building software.
Related notions such as
 \emph{goal-oriented}~\cite{DBLP:conf/sigopsE/Renesse98}
 or 
 \emph{declarative} programming~\cite{DBLP:conf/agp/Lloyd94,DBLP:conf/agp/BaldoniBMOT10}
 are used to denote a similar idea:
 the use of languages 
 to express an abstract model of a system
 emphasising \emph{what} has to be achieved rather than \emph{how}.
The benefit is that the complexity for efficiently mapping the \emph{what}
to the \emph{how} can be encapsulated in a middleware layer,
 while application developers can focus on domain abstractions and the business logic.
In this sense,
 \macrop{} can be considered as a particular domain of declarative programming;
 however, we think that research in this field
 can potentially provide insights 
 on the general principles and foundations of declarative programming.

\subsubsection{Intelligent middlewares}\label{opportunity-middleware}
Beside expressiveness, 
 the abstraction provided by \macrop{}
 can foster the implementation 
 of smart solutions at the middleware level.
In early \macrop{} approaches on WSNs,
 the goal was often simplifying the programming activity (i.e., productivity)
 while keeping performance overhead at acceptable levels.
In time, 
 the idea of actually \emph{improving} performance
 started to be considered as a research goal.
Indeed, overfitting solutions may not be able 
 to adequately generalise their performance
 to the various situations a system may experience in practice.
On the other hand, 
 more abstract architectures 
 could adapt to diverse situations
 and do that \emph{opportunistically}---by \emph{proactively} looking for opportunities of optimisation.
Of course, there is a trade-off between overfitting and underfitting models, and this revolves around a careful design of \macrop{} solutions in terms of (domain-specific) assumptions.

The middleware could be the part of the software system
 that implements the global-to-local mapping logic,
 possibly in a smart way.
Such a smartness could serve to
 avoid unnecessary computations or communications,
 change structure to promote functionality, or
 re-configure the application to improve performance or resiliency.
For instance, 
 in aggregate computing~\cite{DBLP:journals/fi/CasadeiPPVW20} and MacroLab~\cite{hnat2008macrolab},
 the logical macro-programmed system can be deployed
 variously on available infrastructure,
 where different deployments may result in different 
 non-functional trade-offs;
 moreover, their middleware can in principle adapt the deployment
 opportunistically as infrastructure, user preferences, or environmental conditions change.
Indeed, a key opportunity would be to leverage 
 recent advances in self-adaptive software 
 and autonomic computing
 as well as artificial intelligence and machine learning.

\revision{
Additionally, 
 it is interesting to note,
 in certain \macrop{} approaches, 
 the particular interplay between 
 the language (and hence the programs)
 and 
 the middleware.
For instance, in aggregate computing~\cite{Viroli-et-al:JLAMP-2019},
 the programs do not express control flow
 and the programmer rather assumes 
 the program to be collectively played by the system of devices
 through a 
 ``self-organisation-like execution protocol''.
As mentioned earlier, this provides opportunities 
 in terms of flexibility 
 in the implementation and deployment
 of the actual execution protocol~\cite{DBLP:journals/fi/CasadeiPPVW20}.
The point is how much these results can be generalised
 into design principles and patterns:
 this is a research opportunity (with corresponding challenges---see~\Cref{s:challenges})
 related to the (relationship between the) design of declarative programming languages
 and the design of intelligent middlewares for corresponding applications.
}

\subsubsection{Collective Intelligence, Soft Computing, Social Computing}
\revision{
The recent techno-scientific trends 
 and visions (cf. pervasive and ubiquitous computing, the IoT etc.)
 let us foresee an ever-increasing, world-wide deployment 
 of devices capable of computation and communication.
Reasoning just in terms of individual devices
 would hardly allow us to fully harness ``the power of the collective''.
On the other hand,
 adopting a vision of ``cyber-physical collectives and ecosystems''
 could provide a further perspective 
 for better addressing socio-technical 
 services and applications.
There are several systems
 that are amenable to be studied and engineered
 by a collective perspective,
 as well as several research fields that
 address aspects of such collective systems~\cite{tumer2004collectives}.
Works on \macrop{} are often found in such research areas (see \Cref{sec:domains} and \Cref{approaches-ensemble}),
 and might contribute (from its construction-oriented perspective) 
 to the overall research endeavour 
 about collective systems.
}

\emph{Computational collective intelligence (CCI)}
 is a sub-field of AI
 that focusses on
 ``the form of intelligence that emerges from the collaboration
and competition of many individuals (artificial and/or natural)''~\cite{szuba2001computational-collective}.
The affinity with \macrop{} is evident,
 as the latter generally provides a means
 for expressing \emph{what} collective intelligent behaviour
 should be like or work at,
 encapsulating the logic for building it
 in terms of rules of individual behaviour and interaction.
However, 
 the abstraction provided by strong \macrop{} languages
 tends to favour implementations 
 achieving approximated solutions
 in complex situations.
This is especially evident in \macrop{} languages
 for collective adaptive systems (\Cref{approaches-ensemble}),
 such as aggregate computing~\cite{Viroli-et-al:JLAMP-2019},
 where macro-programs express 
 global outcomes that are to be sought progressively
 in a self-organising fashion.
In this sense,
 \macrop{} promotes a language-based approach
 to \emph{soft computing}~\cite{DBLP:journals/soco/LiangH20}, namely
 the use of computing to approximately solve very complex problems
 despite uncertainty, perturbations, and partial knowledge.

A recent systematic literature review
 on ``collective intelligence''~\cite{suran2020frameworks-collective-intelligence},
 covers conceptual frameworks and models 
 for ``collaborative problem solving and decision making'', in the broad sense of \emph{social computing}~\cite{DBLP:journals/expert/WangCZM07}---namely the paradigm where humans, society, and computing technology integrate to promote information representation, processing, communication, and use.
The survey focusses on a high-level view 
 and purposefully abstracts from specific domains---not even mentioned, the programming viewpoint is completely neglected.
However, \macrop{} DSLs could work as inter-disciplinary artifacts 
 capturing relationship and behaviour 
 of groups and ecosystems.
Benefits could be obtained by addressing issues at the right perspective.

\subsection{Challenges}
\label{s:challenges}

\revision{
There are a number of challenges
 related to 
 the engineering of \macrop{} systems.
These include, e.g.,
 designing macro-level abstractions,
 bridging macro-level abstractions with micro-level activity,
 formalising the macro-to-micro mapping logic,
 providing formal guarantees about the correctness of such a mapping,
 and integrating \macrop{} systems with
 more traditional programming environments.
}

\subsubsection{Abstraction and global-to-local mapping}

A key challenge in \macrop{}
 is defining a good, coherent set of macro-level abstractions
 and identifying a proper way to map those 
 to micro-level activity while promoting 
 both functional and non-functional requirements.
This also includes finding a balance
 between over-fitting and under-fitting solutions:
 the former may hinder reusability
 and extensibility,
 while the latter,
 as an attempt to achieve a one-size-fits-all support,
 may complicate implementations.
As discussed previously,
 effective, highly-productive programming
 and smartness in implementations
 is where the most opportunities  arise
 and arguably the major concerns for any \macrop{} language.
The challenge revolves around
 ensuring that global-to-locally mapped behaviour
 results, when actually carried out, in local-to-global effects
 in a consistent (and possibly efficient) way.

\revision{
Implementing proper global-to-local mapping logic 
 is a key challenge in any \macrop{} system.
This is related to what Tumer and Wolpert call the \emph{inverse problem} in \emph{COIN (COllective Intelligence)}~\cite{tumer2004collectives}, i.e., configuring the laws of a system such that the desired collective behaviour is generated%
\footnote{\revision{However, the COIN approach aims to steer macro-behaviour by merely setting the local utility functions of reinforcement learning agents: this can hardly be seen as strong \macrop{} since no macro-level abstraction is used. According to the proposed terminology (\Cref{s:macro-not}), this can instead be seen as a weak form of \macrop{}, or meso-programming.}}.
The difficulty of this problem 
 is one of the reasons that
 make it hard to 
 find specific designs or solutions for \macrop{} in general:
 domain goals set peculiar global requirements, 
 and
 domain assumptions are often instrumental
 for effective mapping of abstractions
 down to the underlying ``platform''.
Still, the observation of regularities, namely ``patterns'',
 can provide for useful hints 
 to both the theory and practice of \macrop{}.
 
Moreover, some \macrop{} approaches
 such as, e.g., DEECo~\cite{bures2013deeco}, SCEL~\cite{DBLP:journals/taas/NicolaLPT14},
 and aggregate programming~\cite{DBLP:journals/computer/BealPV15},
 target \emph{complex}/\emph{collective adaptive systems}~\cite{Ferscha2015cas}---see \Cref{approaches-ensemble}.
Such systems feature complex networks of interactions
 that typically result in \emph{emergent properties (emergents)}~\cite{DBLP:conf/atal/WolfH04},
  namely macro-level properties that cannot be easily traced back to micro-level activity, because they are not -- by definition -- the result of mere summation of individual contributions (i.e., they are based on non-linear dynamics)~\cite{holland1998emergence}.
Due to its very nature,
 promoting desired emergents 
 is a challenge.
However, in some cases, emergence can be ``steered''.
Existing research~\cite{DBLP:journals/eaai/CasadeiVAPD21} seems to suggest that \macrop{}
 may provide a privileged perspective and approach
 for steering emergent behaviour 
 towards the desired emergents.
In a sense, 
 the development of a \macrop{} system
 might force its designers 
 to approach the problem
 by a mixed top-down/bottom-up strategy.
}

\subsubsection{Formal approaches to \macrop{}}
In software engineering,
 the use of formal methods 
 enables specification of non-ambiguous models of systems
 and promote their analysis and verification, possibly automated.
In \macrop{},
 languages backed by formal theories and calculi 
 may be analysed
 to verify qualitative or quantitative properties.
For instance, 
 in aggregate programming
 it has been possible,  by considering its core language (the field calculus),
 to prove Turing-like universality for space-time computations,
 identify language fragments supporting self-stabilising 
 and density-independent computations,
 prove optimality theorems for specific algorithms or encodings,
 and promote deployment-independence at the middleware level~\cite{Viroli-et-al:JLAMP-2019}.
In SCEL~\cite{DBLP:journals/taas/NicolaLPT14},
 statistical model checking tools can be used to
 verify reachability properties, i.e.,
 to compute the probability that a certain system configuration 
 (e.g., expressed as a predicate on collected information)
 is reached within a certain deadline.
Vice versa,
 several other \macrop{} languages
 focus mainly on providing a high-level API,
 simplifying the programming activity
 but providing little support for analysis and verification.
In some cases,
 the semantics of the DSL is not even specified formally.
Other approaches,
 such as WOSP~\cite{varughese2020swarm-wosp},
 provide certain properties (e.g., low communication overhead)
 by construction and
 use empirical methods (e.g., simulation) for verification.
Therefore,
 a challenge related to the identification of good abstractions
 and global-to-local mapping strategies
 is the definition of formal frameworks
 supporting both correct and efficient implementations
 as well as discovery of properties and results
 (both at the application and middleware level).
We note that this challenge (and opportunity)
 is also recognised by other fields of research
 including self-adaptive software 
 and robotics threads~\cite{DBLP:conf/c3s2e/WeynsIIA12,DBLP:conf/ifm/FarrellL018}.
 
\revision{
Besides applying formal methods
 for verification and analysis
 within specific \macrop{} systems,
 another challenge 
 lies in devising a 
 \emph{general, formal theory of \macrop{}}
 that abstracts from specific languages 
 and possibly even from concrete paradigms.
One possibility
 would be to rigorously identify
 a minimal but complete set
 of concepts or predicates
 applicable to programming systems
 to classify them as (a form of) \macrop{}.
The basic principles provided in \Cref{sec:macrop:def}
 could make for a starting point in this research.
The use of such a formal framework 
 could then be used to
 provide alternative, possibly more precise, classifications
 of \macrop{} approaches 
 with respect to the one provided in \Cref{sec:taxonomy}.
}

\subsubsection{Heterogeneity}\label{challenge-heterogeneity}

A system is \emph{heterogeneous} if it comprises 
 different kinds of components.
\Macrop{} a system of multiple heterogeneous components or individuals
 is challenging because 
 making use of the different capabilities of these
 requires an individual-level viewpoint.
Vice versa, \macrop{} homogeneous collectives 
 (such as swarms of homogeneous robots)
 tends to be simpler 
 as any robot is assimilable to another.
In principle, 
 heterogeneity may be abstracted 
 at the programming level
 and encapsulated at the middleware level,
 or code may be organised such that 
 specific behaviour is modularised.
\revision{We also remark that homogeneity and heterogeneity are not sharp characteristics but form a continuum,
 and that abstraction makes things more homogeneous,
 by removing unnecessary details (possibly including differences and peculiarities).
}

Moreover, heterogeneity 
 is not only in shape or capabilities
 but also in aspects
 like autonomy and programmability.
For instance, 
 consider a heterogeneous cyber-physical collective
 made of smart city components (e.g., smart traffic lights, cloudlets, autonomous vehicles)
 and augmented human operators (e.g., through smartphones, smart watches or glasses),
 which may be programmed to support decentralised crowdsensing applications;
 the digital devices worn by those humans
 will move according to those humans' deliberation,
 and hence their mobility could not be programmed
 (but only ``requested'', at best).
Among the surveyed approaches, only
 the SmartSociety platform~\cite{scekic2017hybridcas}
 provides some support for human orchestration,
 where humans and machines are considered \emph{peers}.

\revision{
While collectives tend to be homogeneous, 
 heterogeneity is typically more present in 
 \emph{composites}, namely collections of entities
 related by a notion of \emph{componenthood}~\cite{brodaric2020pluralities-collectives-composites}.
An example is a car, which builds on components such as engine, wheels, etc. However, 
it would be very hard to imagine the possibility of \emph{programming} a car as a whole.
}

To conclude this reflection,
 \macrop{} does not need to assume homogeneity,
 but it does need to take heterogeneity into account
 at some level of its engineering stack (middleware, application, model).
Moreover, we also observe that 
 \macrop{} is not to be thought as a comprehensive approach
 meant to define all aspects of a system behaviour,
 which also leads to the following challenge.

\subsubsection{Integration with other programming paradigms and toolchains}
As discussed in previous sections,
 \macrop{}
 embodies a particular viewpoint of system development,
 which may not capture all the relevant 
 functional and non-functional requirements.
Indeed, a complex system
 may involve the solution of multiple different problems,
 each one best addressed by a specific paradigm.
This is the idea of \emph{multi-paradigm programming}~\cite{DBLP:conf/plsa/SpinellisDE94,DBLP:journals/jsc/AlbertHHOV05}.
On a more pragmatic side,
 supporting \macrop{}
 on top of existing development platforms
 (such as the JVM or .NET)
 may enable quick prototyping
 as well as reuse of features and tools from the host platform.
This has fostered the emergence of \emph{internal DSLs}~\cite{DBLP:books/daglib/0030751} for \macrop{},
 which are embedded as expressive APIs
 on top of existing general-purpose languages:
 this is the case of 
 PyoT (Python)~\cite{Azzara2014pyot-macroprogramming-iot},
 Chronus (Ruby)~\cite{wadaa2010chronus-spatiotemporal-macroprogramming-wsn},
 jDEECo (Java)~\cite{bures2013deeco},
 ScaFi for aggregate programming (Scala)~\cite{scafi},
 Dolphin (Groovy)~\cite{lima2018dolphin},
 D'Artagnan (Haskell)~\cite{mizzi2018dartagnan}, and
 AErlang (Erlang)~\cite{denicola2018aerlang}.
However, 
 this aspect of integration of paradigms 
 poses \emph{architectural} challenges,
 especially considering that \macrop{}
 tends to permeate various dimensions 
 of the system---including structure, behaviour, and interaction.
In summary,
 multi-paradigm programming 
 is appealing
 but must be carefully 
 analysed
 at the level
 of models, 
 architecture, 
 and development practice.

\section{Related Work}
\label{s:rw}

This work integrates, extends upon, and 
 differentiates with respect to other survey papers.
The main difference is that the secondary studies presented in the following,
 while similarly rich and detailed,
 adopt a narrower perspective (spatial computing, WSN, microelectromechanical systems, and swarm robotics, respectively).
By contrast, this survey aims to relate various \macrop{} approaches across disparate domains, and adopts a general software engineering viewpoint.
Moreover, due to their publication time, other surveys only cover works published before 2012.
Indeed, by analysing the twenty-year time-frame from early 2000s to 2020,
 we can also make considerations about trends (see \Cref{sec:domains}).

The most related survey is \cite{beal2012organizing-the-aggregate}, which however focusses on \emph{spatial computing} languages.
It proposes a conceptual framework where 
 spatial computation can be described
 in terms of constructs for
 \emph{(i) measuring space-time} (sensors);
 \emph{(ii) manipulating space-time} (actuators);
 \emph{(iii) computation}; 
 and
 \emph{(iv) physical evolution} (inherent spatiotemporal dynamics).
The device model accounts for the way devices are \emph{discretised} in space-time (distinguishing between discrete, cellular, and continuous models),
 the way they are programmed (e.g., by giving them a uniform programs, heterogeneous programs, or leveraging mobile code),
 their communication scope (e.g., through local, neighbourhood, global regions), and
 their communication granularity (e.g., unicast, multicast, or broadcast).
The survey classifies languages in the following groups:
 (i) amorphous computing (including pattern languages and manifold programming languages);
 (ii) biological modelling;
 (iii) agent-based modelling (including multi-agent and distributed systems modelling);
 (iv) wireless sensor networks
   (distinguishing between region-based, dataflow-based, database abstraction-based, centralised-view, and agent-based languages);
 (v) pervasive computing;
 (vi) swarm and modular robotics;
 (vii) parallel and reconfigurable computing
 (including dataflow, topological, and field languages);
 (viii) formal calculi for concurrency and distribution
 (i.e., process algebras/calculi).
Languages are further analysed based on:
 characteristics of the language (type, DSL implementation pattern, platform, layers),
 supported spatial computing operators,
 and
 abstract device characteristics. Language type ranges over functional, imperative, declarative, graphical, process calculus, and any.

Very related is also \cite{mottola2011programming-wsn},
 a 2011 survey that covers programming approaches for wireless sensor networks.
In their taxonomy, 
 the \emph{interaction pattern} 
 is classified into (i) \emph{one-to-many},
 (ii) \emph{many-to-one},
 and
 (iii) \emph{many-to-many}.
Moreover,
 the extent of distributed processing in space
 can be
 (i) \emph{global}, e.g., in environment monitoring applications; or
 (i) \emph{regional}, e.g., in intrusion detection or HVAC systems in buildings.
Other dimensions include \emph{goal} (sense-only or sense-and-react),
 \emph{mobility} (static, mobile),
 \emph{time} (periodic or event-driven).
Regarding WSN programming abstractions, they define a taxonomy as follows.
\emph{Communication} aspects cover:
\emph{scope} (system-wide, physical neighbourhood-based, or multi-hop group);
\emph{addressing} (physical or logical);
and \emph{awareness} (implicit or explicit).
\emph{Computation} aspects include \emph{scope} of computation (local, group, or global).
The \emph{model of data access} could be database, data sharing, mobile code, or message passing.
Finally, the \emph{paradigm} could be: 
\emph{imperative} (sequential or event-driven);
\emph{declarative} (functional, rule-based, SQL-like, special-purpose); or \emph{hybrid}.

The survey~\cite{DBLP:journals/csur/LiangCLL16}
 on distributed intelligent microelectromechanical systems (MEMS) programming also provides a classification based on the
 distinction between \emph{device-level} and \emph{system-level} programming models.
The surveyed programming models are then analysed w.r.t. the characteristics of
real-time support, application range (general-purpose vs. domain-specific),
syntax complexity, scalability, mobility support, and uncertainty tolerance.

The review \cite{brambilla2013swarm-robotics-review-engineering-perspective}
 of swarm robotics from an engineering perspective
 neglects the programming viewpoint.
However, they provide a taxonomy where
 collective behaviour is classified into
 behavior for 
 (i) \emph{spatial organisation} (e.g., pattern formation, morphogenesis),
 (ii) \emph{navigation and mobility} (e.g., coordinated motion and transport),
 (iii) \emph{collective decision making} (e.g., consensus achievement and task allocation),
 and 
 (iv) \emph{other}.
\emph{Design methods} are categorised into \emph{behaviour-based} (e.g., finite state machines, virtual physics-based) 
and \emph{automatic} (e.g., evolutionary robotics and reinforcement learning-based methods).
\emph{Analysis methods} are categorised into \emph{microscopic models}, \emph{macroscopic models} (e.g., via rate/differential equations or control/stability theory), and \emph{real-robot analysis}.

Finally, certain works proposed concepts useful for 
 classifying and understanding \macrop{} approaches.
These elements have been considered and integrated into the taxonomy provided in \Cref{sec:taxonomy}.
A possible classification of macroprogramming approaches~\cite{choochaisri2012logic-macroprogramming-wsn-sense2p} distinguishes between
\begin{enumerate}
\item \emph{node-dependent} macroprogramming---where the nodes (or, more generally, the components of the micro-level) and their states are referred to explicitly by the macroprogram; and
\item \emph{node-independent} macroprogramming---where the  underlying nodes are not visible at all to the programmer.
\end{enumerate} 
As per the discussion of \Cref{s:macro-not},
 node-dependent approaches tend to enact a weak form of \macrop{}.
Examples of node-independent approaches include, e.g., those that abstract a WSN as a database.
Another distinction can be made between:
\begin{enumerate}
\item \emph{data-driven} macroprogramming~\cite{Pathak2010energy-efficient-data-driven-macroprog}---where macro-programs define tasks consuming and producing data; and
\item \emph{control-driven} macroprogramming~\cite{DBLP:conf/iwdc/BakshiP05}---where macro-programs specify control flow and instructions operating on distributed memory.
\end{enumerate}
The classification in data-driven and control-driven approaches has been applied in other fields such as coordination~\cite{DBLP:journals/ac/PapadopoulosA98}, where the latter are also known as \emph{task-} or \emph{process-oriented} coordination models.

\section{Conclusion}
\label{s:wrap-up}

This paper provides, for the first time,
 an explicit, integrated view of research
 on \macrop{} languages.
It discusses what \macrop{} is,
 its core application domains,
 its main concepts,
 and analyses and classifies 
 a wide range of works
 addressing system development
 by a more-or-less macroscopic perspective.
We argue that such a high-level stance
 could be beneficial 
 for software engineering
 in forthcoming distributed computing scenarios (cf. IoT, CPS, smart ecosystems)
 and for promoting language-based solutions
 to collective adaptive behaviour and intelligence.
In particular,
 the macro-level perspective could represent a complementary viewpoint
 for addressing structure, behaviour, and interaction in complex systems.
\Macrop{} approaches tend to be 
 domain-specific, because domain assumptions
 are generally instrumental to properly and efficiently
 map high-level abstractions to activity on the low-level platform.
However, there is arguably margin for 
 recovering general principles
 through inter-domain discussion and sharing of ideas,
 but this requires a more integrated and structured view
 of \macrop{} as a field, which this article aims to cultivate.

\backmatter
\printbibliography

\end{document}